\documentclass[preprint2]{aastex}

\usepackage{graphicx}
\usepackage{amsmath}
\usepackage{color}

\numberwithin{equation}{section}

\newcommand{\be}{\begin{equation}}
\newcommand{\ee}{\end{equation}} 
\newcommand{\lb}{\label}

\newtheorem{Prop}{Proposition}

\newcommand{\bk}{{\textit {\textbf k}}}
\newcommand{\br}{{\textit {\textbf r}}}
\newcommand{\bu}{{\textit {\textbf u}}}
\newcommand{\bw}{{\textit {\textbf w}}}
\newcommand{\bx}{{\textit {\textbf x}}}
\newcommand{\by}{{\textit {\textbf y}}}
\newcommand{\bz}{{\textit {\textbf z}}}
\newcommand{\bB}{{\textit {\textbf B}}}
\newcommand{\bE}{{\textit {\textbf E}}}
\newcommand{\bJ}{{\textit {\textbf J}}}
\newcommand{\bR}{{\textit {\textbf R}}}

\newcommand{\bell}{\hbox{\boldmath $\ell$}}

\newcommand{\balpha}{\hbox{\boldmath ${\mathit \alpha}$}}
\newcommand{\boeta}{\hbox{\boldmath $\eta$}}
\newcommand{\btau}{\hbox{\boldmath $\tau$}}
\newcommand{\bxi}{\hbox{\boldmath ${\mathit \xi}$}}

\newcommand{\barepsilon}{\hbox{\boldmath ${\mathit \varepsilon}$}}
\newcommand{\bSigma}{\hbox{\boldmath ${\mathit \Sigma}$}}
\newcommand{\bomega}{\hbox{\boldmath ${\mathit \omega}$}}

\newcommand{\btimes}{\hbox{\boldmath $\times$}}
\newcommand{\bdot}{\hbox{\boldmath $\cdot$}}
\newcommand{\grad}{\hbox{\boldmath ${\mathit \nabla}$}}
\newcommand{\bzed}{\hbox{\boldmath $0$}}

\shorttitle{Turbulent Reconnection}
\shortauthors{Eyink}

\begin{document}


\title{Turbulent General Magnetic Reconnection}


\author{G. L. Eyink}
\affil{Department of Applied Mathematics \& Statistics and 
Department of Physics \& Astronomy, The Johns Hopkins University, 
Baltimore, MD 21218}
%\email{eyink@jhu.edu}


%\altaffiltext{1}{Visiting Astronomer, Cerro Tololo Inter-American Observatory.
%CTIO is operated by AURA, Inc.\ under contract to the National Science


\begin{abstract}
Plasma flows with an MHD-like turbulent inertial range, such as the solar wind, vitiate many assumptions 
of standard theories of magnetic reconnection.  In particular, the ``roughness'' of turbulent velocity 
and magnetic fields implies that magnetic field-lines are nowhere ``frozen-in'' in the usual sense. 
This situation demands an essential generalization of the so-called ``General Magnetic Reconnection'' 
(GMR) theory. Following ideas of Axford and Lazarian \& Vishniac, we identify magnetic field-lines by ``tagging" 
them with  plasma fluid elements and then determine their slip-velocity relative to the plasma fluid 
by integrating in arc-length along the wandering field-lines. The main new concept introduced here 
is the {\it slip-velocity source vector}, which gives the rate of development of slip-velocity per unit 
arc-length of field line. The slip-source vector is the ratio of the curl of the non-ideal electric 
field $\bR$ in the Generalized Ohm's Law and the magnetic field strength. It diverges at magnetic nulls,
unifying GMR with theories of magnetic null-point reconnection. Only under restrictive assumptions is the 
slip-velocity related to the gradient of the line-voltage or ``quasi-potential'' obtained by integrating  
the parallel electric field $R_{\|}$ along field-lines.
In an MHD turbulent inertial-range $\grad\btimes \bR$ 
becomes extremely large while $R_{\|}$ is tiny, so that line-slippage occurs freely even as a description
by ideal MHD becomes accurate. This ``paradox'' is resolved by the understanding that ideal MHD 
is valid for a turbulent inertial-range not in the standard sense but in a ``weak'' sense which does 
not imply magnetic line-freezing. The mathematical notion of ``weak solution'' is here explained 
in physical terms of spatial coarse-graining and renormalization-group (RG) theory.  We give a new
first-principles argument for the ``weak'' validity of the ideal Ohm's law in the inertial range, via 
rigorous estimates of the terms in the Generalized Ohm's Law for an electron-ion plasma. 
Particular attention is paid to the conditions in the solar wind, a collisionless, magnetized plasma.  
Coarse-grained to inertial-range lengths, all of the non-ideal terms (from collisional resistivity,  
Hall field, electron pressure anisotropy, and electron inertia) are shown to be irrelevant in the RG sense
and large-scale reconnection is thus governed solely by ideal dynamics. We briefly discuss
some implications for heliospheric reconnection, in particular for deviations from the Parker 
spiral model of interplanetary magnetic field. Solar wind observations show that reconnection in 
a turbulence-broadened heliospheric current sheet, consistent with the \cite{LazarianVishniac99} theory, 
leads to slip velocities that cause field-lines to lag relative to the spiral model. 
\end{abstract}





\section{Introduction}\lb{Intro}

A fundamental assumption of most current theories of magnetic reconnection is that ideal line-freezing 
holds to very good approximation for most of space, except within narrow, sparsely distributed current layers.
This includes the elegant mathematical theories which propose a key role for field-parallel electric fields 
\citep{Schindleretal88, HesseSchindler88}, magnetic flipping \citep{PriestForbes92}, quasi-separatrix layers 
\citep{PriestDemoulin95}, and multi-valued flux-tube velocities \citep{Priestetal03}. More precisely, most 
current theories assume that in the generalized Ohm's law
\be \bE + \bu\btimes \bB= \bR, \lb{gOhm}  \ee
the non-ideal electric field $\bR$ is large in isolated spatial regions of small total volume and that 
outside these ``diffusion regions'' where reconnection solely occurs, ideal MHD equations are valid and 
magnetic field-lines are ``frozen-in'' to a good approximation. One thus finds statements in the literature 
such as  ``reconnection occurs when there is a breakdown of ideal MHD and therefore an electric field component 
($E_{\|}$) along the magnetic field...'' \citep{PriestDemoulin95}. Such concepts are motivated especially by the 
conditions of an initially quiet solar corona, where the underlying assumption of well-localized regions of 
flux-freezing violation may be a good approximation in the early stages of coronal reconnection. 

The basic assumptions of the standard reconnection theories are, however, badly violated in plasmas with 
a turbulent range governed by MHD-like dynamics, such as the solar wind at distances from the sun ranging 
from 0.1 to 100 AU. Here non-ideal 
electric fields $\bR$ are very small in r.m.s. magnitude, ideal Ohm's law is valid down to very small scales within the ``inertial range'' , 
and yet magnetic field-lines are not well ``frozen-in'' anywhere in space (see \cite{Burlagaetal82,KhabarovaObridko12,Richardsonetal13} 
and section \ref{implications} for more discussion).   
The failure of standard flux-freezing in turbulent MHD in the ideal limit (``high magnetic Reynolds-number'') 
has been understood previously from several points of view. A spatial coarse-graining approach similar in spirit  
to renormalization group theory shows that flux-freezing in ideal MHD turbulence can fail to be operationally verifiable,  
even with measurements resolved to increasingly small scales \citep{EyinkAluie06}. From the Lagrangian point 
of view, the turbulent phenomenon of Richardson dispersion of fluid elements explosively amplifies any 
tiny breakdown of line-freezing at plasma microscales rapidly into the inertial-range, at rates non-vanishing 
in the ideal limit \citep{Eyink11,Eyinketal11}. This breakdown of standard flux-freezing has been verified in 
numerical simulations of resistive MHD turbulence at very high conductivities \citep{Eyinketal13}.   

In this work we study turbulent reconnection from yet another point of view, that of magnetic connections 
between plasma elements. In the absence of flux-freezing anywhere in space, the only objectively meaningful 
way to give a magnetic field-line an identity over time is by tagging it with a certain plasma fluid element. As suggested 
by \cite{Axford84}, we understand the crucial feature of magnetic reconnection to be the ``disconnection'' of fluid 
elements that start on the same field line. We thus study how the field line anchored (by convention) to a given 
element moving with the fluid changes its connections to other elements.  The systematic development of this idea 
leads to an essential generalization of so-called ``general magnetic reconnection''  \citep{Schindleretal88, HesseSchindler88}  
and of the notion of slip velocities of lines relative to the plasma  \citep{Priestetal03}. A novel concept introduced 
here is the {\it slip-velocity source vector}, 
\be \bSigma = - \frac{(\grad\btimes \bR)_\perp}{|\bB|},\ee
with $\perp$ denoting the component perpendicular to $\bB.$ {\it Our first main claim is that magnetic reconnection is 
fundamentally related to $\bSigma\neq \bzed$ and not to $R_{\|}\neq 0.$} As we discuss in detail below, the vector field 
$\bSigma$ gives the rate of development of slip velocity per unit arc-length of field-line.  Our analysis thus has a particularly close relation 
to the \cite{LazarianVishniac99} theory of turbulent reconnection, based on ``stochastic wandering'' of magnetic field-lines.  
However, we shall make contact with several other ideas, including the work of \cite{Albright99} on fractal distributions of magnetic nulls.
Our approach applies equally well in laminar and turbulent flows, but it is especially valuable in the latter case. 

Our discussion shows, in particular, how $\bR$ (and thus $R_{\|}$) can be small in r.m.s. magnitude, so that ideal MHD 
can hold in the ``weak sense'', and yet for turbulent, multi-scale plasmas $\bSigma$ can still be very large and thus 
magnetic field-lines no longer ``frozen-in'' even in an approximate sense. Because there are persistent misunderstandings
about what it means for  ideal MHD to ``hold at large scales'', we devote a section to carefully discussing this issue 
for a turbulent plasma. The solar wind, a nearly collisionless plasma, is the best-studied instance of MHD turbulence in Nature \citep{BrunoCarbone05}, 
so that we use it as the showcase example in this section, but our discussion applies much more generally. It is widely believed that 
the plasma modes ``at scales greater than the ion gyroradius'' in the solar wind  are well-described by ideal MHD or by related 
ideal equations \citep{Schekochihinetal09}. We show that this cannot be true in a naive sense of validity of the ideal Ohm's law,
but only in a ``weak'' or ``coarse-grained'' sense, which does not imply that field-lines will be ``frozen-in'' at length-scales
greater than the ion gyroradius $\rho_i$. By a detailed term-by-term analysis of the Generalized Ohm's Law for any plasma
of heavy ions and light electrons, we argue that the ideal magnetic induction equation is indeed valid ``weakly" 
or  in ``coarse-grained'' sense under turbulent conditions of the sort observed in the solar wind. {\it Our second main conclusion is thus that
the non-ideal terms of whatever origin (collisional resistivity, Hall field, electron pressure anisotropy,  electron inertia) are irrelevant 
to reconnection processes at scales substantially greater than $\rho_i,$ which are instead governed at those scales 
solely by ideal MHD-like turbulence dynamics.}  
The main physical problem we shall address here is the observed breakdown of the Parker spiral model in the inner heliosphere, which 
we shall argue in some detail to have its origin in the turbulence of the solar wind.

%\newpage 

The detailed contents of this work are as follows: In section \ref{Turb}, we briefly review the current understanding 
of magnetic flux-freezing for resistive MHD turbulence, which is probably the best understood case from numerical 
simulation studies. The results surveyed there motivate the very general approach to magnetic reconnection in 
the following section \ref{TGMR}. In section \ref{slip} we derive the basic equation for the slip-velocity as one follows along  
a magnetic field-line and which contains the slip-velocity source vector. In section \ref{voltage} we exactly integrate 
the equations for the slip-velocity in terms of the ``quasi-potential'' of GMR but show that the conditions required 
to do so are very restrictive in turbulent flow. In section \ref{coarse} we explain the notion of ``weak'' or ``coarse-grained''
solutions and estimate the dependence on length-scale of all of the terms in the Generalized Ohm's Law using a
non-perturbative RG-like approach. In section \ref{implications} we discuss the implications for heliospheric reconnection. 
 Finally, two appendices contain some more technical details, the first analyzing effects of plasma density variations and the second 
deriving mathematical relations for coarse-graining cumulants used in the estimates.    
    
\section{Resistive MHD Turbulence and Flux-Freezing}\lb{Turb} 

Turbulent cascade is described by ideal magnetohydrodynamic equations in a coarse-grained (``weak'') sense, 
for which non-ideal electric fields $\bR$ are vanishingly small in the sense of distributions.  More precisely, 
consider the low-pass filtered or ``coarse-grained'' magnetic field 
\be \bar{\bB}_\ell(\bx) = \int d^3 r  \ G_\ell(\br) \bB(\bx+\br), \,\,\,\, G_\ell(\br) = \ell^{-3}G(\br/\ell) \ee
for a smooth, rapidly decaying filter kernel $G.$ The coarse-grained field satisfies the induction equation 
\be \partial_t\bar{\bB}_\ell = \grad\btimes \left[(\overline{\bu\btimes\bB})_\ell -\bar{\bR}_\ell\right], \ee
which follows from the generalized Ohm's law (\ref{gOhm}) and Faraday's law, after coarse-graining. 
This equation in turbulent flow is usually expressed in terms of the motional electric fields induced by 
the eddies at scales smaller than $\ell$ \citep{Biskamp03}:
\be \bE_\ell^T=-\barepsilon_\ell=-\left[\overline{(\bu\btimes\bB)}_\ell-\bar{\bu}_\ell\times\bar{\bB}_\ell\right]. \ee
Then the statement is that the large-scale contributions $\bar{\bR}_\ell$ of the non-ideal electric fields are 
very tiny compared with $\barepsilon_\ell,$ for $\ell$ in the inertial range. This is, indeed, the very condition 
defining an ``inertial range.'' Under these circumstances, the equation which 
governs the evolution of the coarse-grained magnetic fields at these scales $\ell$ is 
\be \partial_t\bar{\bB}_\ell = \grad\btimes (\overline{\bu\btimes\bB})_\ell =\grad\btimes (\bar{\bu}\btimes\bar{\bB}_\ell+\barepsilon_\ell), \ee
with the tiny $\bar{\bR}_\ell$ term neglected. This is the condition that ideal Ohm's law holds in the turbulent inertial range 
in the coarse-grained or ``weak'' sense.  Because this notion of validity of ideal MHD in the ``weak sense'' is frequently
misunderstood and invoked to make incorrect conclusions, we discuss it and justify it very carefully in section 
\ref{coarse} of this paper.   

Deferring a general discussion to later, we here illustrate the above observations by the example of 
resistive MHD turbulence,  where the classical Ohm's law holds with scalar resistivity $\eta$: 
\be \bR=\eta \bJ=\eta \grad\btimes\bB. \ee 
The ``zeroth law'' of resistive MHD turbulence states that Ohmic dissipation is non-vanishing in the limit of zero resistivity,  
\be  \varepsilon_B = \lim_{\eta\rightarrow 0}\frac{\eta}{|V|}\int_V d^3x \  J^2  >0. \ee
See \cite{MininniPouquet09} for numerical evidence of this result and \cite{Caflischetal97} for mathematical foundations. 
Hence, $J_{rms}\sim (\varepsilon_B/\eta)^{1/2}\rightarrow \infty$ as $\eta\rightarrow 0,$ while $R_{rms}\sim 
(\eta\varepsilon_B)^{1/2}\rightarrow 0$ in the same limit. Because averaging decreases convex functions,
$\bar{R}_{\ell,rms}\leq R_{rms},$ and this easily implies that the smooth field $\bar{\bR}_\ell$ for fixed length-scale $\ell$ 
becomes vanishingly small everywhere in the limit of vanishing $\eta.$ 
%The vanishing of rms values of $\bR$ is equivalent 
%to the mathematical statement that $\bR\rightarrow \bzed$ in the $L^2$ sense and the Cauchy-Schartz inequality 
%for the Euclidean norm 
%\be \|\bar{\bR}_\ell\|_2 \leq \|G_\ell\|_2 \cdot \|\bR\|_2\ee 
%then easily implies that $\bar{\bR}_\ell$ goes to zero everywhere for fixed $\ell$ in the limit of vanishing $\eta.$ 
The smallness of non-ideal Ohmic electric fields at inertial-range length-scales can be expressed also in terms of 3D wavevector 
spectra, as  
\be E_R(\bk) = \eta^2 k^2 E_B(\bk), \ee
which implies rapidly vanishing values of $\hat{\bR}(\bk)$ for decreasing $\eta$ or $k$ in a Kolmogorov-type 
inertial-range. A physical-space version of this spectral estimate (see section \ref{Ohm}) is
\be \bar{R}_{\ell,rms}\leq \eta \frac{\delta B(\ell)_{rms}}{\ell} \ee
for magnetic increments $\delta B(\ell),$ which is likewise vanishing for decreasing $\eta$ or increasing $\ell.$
Although we considered above classical resistive MHD turbulence, similar results hold when the MHD cascade is 
terminated by physics other than collisional resistivity. For example, non-ideal electric fields are produced by gradients 
of the electron pressure in kinetic Alfv\'en-wave turbulence, but these are expected to become progressively 
smaller at scales much larger than the ion gyroradius \citep{BianKontar10,Bianetal10}. For a detailed analysis 
of all of the contributions to $\bR$ in the Generalized Ohm's Law, cf. section \ref{CG-GOL}. 

The vanishing of $\bR$ in the coarse-grained (``weak'') sense or even in the stronger r.m.s. sense, which suffice 
for ideal MHD to hold weakly, does not require that magnetic flux be conserved in the same manner as 
for smooth, laminar solutions of ideal MHD. The necessary and sufficient condition for standard flux-conservation
is that $\grad\btimes\bR=\bzed$ hold at every space point \citep{Newcomb58}. Even if $\bR\rightarrow \bzed$
uniformly everywhere, it is possible for  $\grad\btimes\bR$ to diverge to infinity everywhere in the same limit. This 
is exactly the situation in resistive MHD turbulence, where the identity 
\begin{eqnarray} \langle|\grad\btimes\bR|^2\rangle &=& \int d^3k \ |\bk|^2 E_R(\bk) \cr
 &=& \eta^2 \int d^3k \ |\bk|^4 E_B(\bk) \end{eqnarray} 
shows that $|\grad\btimes\bR|_{rms}\rightarrow\infty$ for increasingly long power-law inertial ranges
of Kolmogorov-type, even as $R_{rms}\rightarrow 0$. The point here is that the righthand side of the above 
equality scales as $\eta k_d^2 \varepsilon_B,$ where $k_d$ is the dissipative cutoff wavenumber, that satisfies 
$k_d>O(\eta^{-3/4})$ as $\eta\rightarrow 0$ in all current theories of MHD turbulence  (see \cite{Iroshnikov64,Kraichnan65,
GoldreichSridhar95,GoldreichSridhar97,Boldyrev05,Boldyrev06} and section \ref{Ohm}). Because of this divergence of r.m.s. values 
of $\grad\btimes\bR$ it is also not true that the magnetic field-lines are ``frozen-in'', even though $R_{rms}\rightarrow 0,$
because the ``frozen-in'' property of field-lines is equivalent to $\hat{\bB}\btimes(\grad\btimes\bR)=\bzed$ 
\citep{Newcomb58}.  Indeed, magnetic flux-conservation in the usual  sense is expected to be violated in the 
presence of turbulent inertial ranges such as described above \citep{EyinkAluie06,Eyink07}.  The inertial-range 
phenomenon of Richardson dispersion of plasma fluid elements accelerates---by unbounded amounts---microscopic 
disconnections between plasma fluid elements and field-lines \citep{Eyink11,Eyinketal11,Eyinketal13}. Thus, field-lines 
are not ``frozen-in'' anywhere in the turbulent plasma, in the sense that any pair of fluid elements initially residing 
on a common field line are in a short (but macrosopic) time later residing on macroscopically well-separated lines.   

The above conclusions depend only on the ``spontaneous stochasticity'' due to Richardson dispersion,
which is an MHD-type inertial-range turbulence phenomenon, and not upon the specific plasma physics of 
collisional resistivity \citep{Eyinketal11}. Any other fluid limit (e.g. small ion gyroradius, etc.) which leads 
to the same universal MHD-like turbulent inertial range will show the same effects. A toy example of this is 
Burgers equation, which is the simplest PDE example of ``spontaneous stochasticity'' \citep{EyinkDrivas14}.  
For Burgers, the velocity field is ``frozen-in'' to fluid trajectories for smooth, laminar solutions. However, 
the zero-viscosity limit yields weak solutions of inviscid Burgers equation with discontinuous shocks. It has been shown 
by \cite{EyinkDrivas14} that the zero-viscosity limit exhibits ``spontaneous stochasticity'' at shock points and 
the velocity is ``frozen-in'' there only stochastically. Furthermore, it has been proved that zero-hyperviscosity 
limits for Burgers yield precisely the same class of weak solutions \citep{Tadmor04} and thus exhibit the same 
``spontaneous stochasticity'', which is an exact feature of the limiting weak solution. The situation with the ideal 
MHD turbulent cascade is expected to be similar, except that, unlike for Burgers, numerical evidence 
\citep{Eyinketal13} suggests that ``spontaneous stochasticity'' occurs at every space point in high-conductivity 
MHD turbulence and not just at very intense current sheets that approximate ideal MHD rotational or tangential discontinuities.  

\section{Turbulent Generalization of ``General Magnetic Reconnection''}\lb{TGMR}  

These facts of MHD turbulence call for an alternative approach to magnetic reconnection which does not assume 
that the breakdown of the ``frozen-in'' condition is spatially localized in ``diffusion regions'' of small total volume. 
We systematically develop such an approach here. 

\subsection{Line Slip-Velocity and Slippage Source}\lb{slip} 

Our basic notion will be that of a magnetic field-line {\it slip velocity} relative to the plasma, which arises 
from a careful analysis of the idea of magnetic connection between plasma elements. 

\subsubsection{Definitions and Fundamental Equation}

Let $\bxi(s;\bx,t)$ be the point on the magnetic field-line at time $t$ which is a distance $s$ from the 
``base point'' or ``anchor point'' $\bx.$ Thus,
\be \frac{d}{ds} \bxi(s;\bx,t)= \hat{\bB}(\bxi(s;\bx,t),t), \,\, \bxi(0;\bx,t)=\bx \ee
where $\hat{\bB}=\bB/|\bB|$ is the magnetic director field. Now let $\bx(t;\bx_0,t_0)$ denote the position at time $t$ 
of the plasma fluid element that starts at $\bx_0$ at time $t_0,$ so that 
\be \frac{d}{dt} \bx(t;\bx_0,t_0)= \bu(\bx(t;\bx_0,t_0),t), \,\,\bx(t_0;\bx_0,t_0)=\bx_0. \ee  
For a smooth, laminar solution of ideal MHD where field-line freezing holds, it must be the case that a suitable 
function $s(t;s_0,x_0)$ exists so that $\bxi(s(t;s_0,\bx_0); \bx(t;\bx_0,t_0),t)=\bx(t;\bxi(s_0;\bx_0,t_0),t_0).$
Differentiating this equation with respect to time $t$ one obtains that $d\bxi/dt=\bu(\bxi,t) \equiv\tilde{\bu}$ holds
if and only if 
\be \dot{s}(t) \hat{\bB}(\bxi,t) + D_t \bxi =  \tilde{\bu} \lb{dxdtisu} \ee
for $D_t=\partial_t+\bu\bdot\grad$. The parallel component gives the equation to determine $s(t)$ as 
\be \dot{s}(t) = (\tilde{\bu}-D_t\bxi)\cdot \hat{\bB} = (\tilde{\bu}-D_t\bxi)_\|, \,\,\,\, s(t_0)=s_0. \lb{dxdtisupar} \ee
When $s(t)$ is determined in this manner, then substituting (\ref{dxdtisupar}) back into (\ref{dxdtisu}) shows finally 
that $d\bxi/dt=\tilde{\bu}$ holds if and only if 
\be (D_t \bxi)_\perp(s;\bx,t) = \bu_\perp(\bxi(s;\bx,t),t) \lb{FF} \ee
holds for all $s,\bx,t,$ and this condition is equivalent to standard field-line freezing. Although  
our derivation was Lagrangian,  the final result (\ref{FF}) is an instantaneous, single-time condition. 

The condition (\ref{FF}) for frozen-in field-lines motivates us to define in general for non-ideal MHD a (perpendicular) 
slip velocity
\be \Delta\bw_\perp(s;\bx,t) = (D_t\bxi-\tilde{\bu})_\perp(s;\bx,t), \ee
which, given a field-line anchored to a base-point $\bx$ at time $t,$ measures its motion relative to the plasma 
fluid with velocity $\tilde{\bu}$ at the point a distance $s$ along the line from $\bx$. It is simple
calculus to derive the following basic equation for the development of slip velocity along 
a field-line:  
\begin{eqnarray} 
\frac{d}{ds} \Delta \bw_\perp &=& \left[(\grad_{\bxi}\hat{\bB})^\top-(\hat{\bB}\hat{\bB})(\grad_{\bxi}\hat{\bB})\right]
\Delta\bw_\perp \cr
&& \,\,\,\,\,\,\,\,\,\,\,\,\,\,\,\,\,\,\,\,\,\,\,\,\,\,\,
- \frac{1}{|\bB|}(\grad\times\bR)_\perp. \lb{slipeq} \end{eqnarray} 
For details, see section \ref{derive} below.   
%Here we have assumed a generalized Ohm's law
% \be {\textbf E}+\bu\times\bB = \bR. \ee
When the non-ideal term vanishes identically, $\bR\equiv {\textbf 0},$ then it is easy to see from (\ref{slipeq}) 
that $\Delta \bw_\perp\equiv {\textbf 0}$ and standard flux-freezing follows (as long as all fields remain smooth as $\bR\rightarrow\bzed$).   
Indeed $D_t\bxi(0;\bx,t)=\bu(\bx,t),$ so that $\Delta\bw(0)={\textbf 0}$ at the base point and the above equation then implies 
that $\Delta \bw_\perp(s)=\bzed$ along the entire length of field-line. This analysis provides a new {\it ab initio} demonstration 
that (\ref{FF}) does indeed hold under the standard assumption, $\hat{\bB}\btimes(\grad\btimes\bR)=\bzed,$ 
required for field-line freezing \citep{Newcomb58}.  

\subsubsection{Derivation of the Equation}\lb{derive}  

We here derive the fundamental equation (\ref{slipeq}). We use the simple results
\be \frac{d}{ds}\tilde{\bu}=(\hat{\bB}\cdot\grad_{\bxi})\tilde{\bu}, \lb{A1} \ee 
which follows directly from the chain rule, and 
\be  \frac{d}{ds} D_t\bxi = \tilde{D}_t \hat{\bB} + \Delta\bw\cdot\grad_{\bxi}\hat{\bB} \lb{A2} \ee
with $\tilde{D}_t=\partial_t+\tilde{\bu}\cdot\grad_{\bxi}$ and $\Delta\bw= D_t\bxi-\tilde{\bu}.$ 
The result (\ref{A2}) follows from 
\begin{eqnarray*}
\frac{d}{ds}D_t\bxi &=& D_t \frac{d}{ds}\bxi \cr
                              &=& D_t \hat{\bB}(\bxi,t)  \cr
                              &=& \partial_t\hat{\bB} + D_t\bxi\cdot\grad_{\bxi}\hat{\bB} \cr
                              &=& \tilde{D}_t\hat{\bB} +(D_t\bxi-\tilde{\bu})\bdot\grad_{\bxi}\hat{\bB}. 
\end{eqnarray*} 
Now subtracting (\ref{A1}) from (\ref{A2}) gives 
\be \frac{d}{ds}\Delta\bw = \left[\tilde{D}_t\hat{\bB}-(\hat{\bB}\cdot\grad_{\bxi})\tilde{\bu}\right]
+ \Delta\bw\cdot\grad_{\bxi}\hat{\bB}. \lb{A3} \ee 
Of course, from the generalized Ohm's law (\ref{gOhm}) it follows that
\be \tilde{D}_t\bB =(\bB\bdot\grad_{\bxi})\tilde{\bu}-\bB(\grad_{\bxi}\bdot\tilde{\bu})-\grad\btimes\bR \ee
and hence
%, using (\ref{gradBhat}) and the similar result $\partial_t |\bB|=\hat{\bB}\bdot\partial_t\bB/|\bB|$, that 
\begin{eqnarray}
 \tilde{D}_t\hat{\bB} &= & \frac{1}{|\bB|}(\tilde{D}_t\bB)_\perp \cr
 &=&\left[(\hat{\bB}\bdot\grad_{\bxi})\tilde{\bu}
-\left(\hat{\bB}^\top\cdot \grad_{\bxi}\tilde{\bu}\cdot \hat{\bB}\right)\hat{\bB}\right] \cr
&& \,\,\,\,\,\,\,\,\,\,\,\,\,\,\,\,\,\,\,\,\,\,\,\,\,\,\,\,\,\,
-\frac{1}{|\bB|}(\grad\btimes\bR)_\perp. \end{eqnarray} 
Here the curl $\grad\btimes\bR$ is always assumed to be evaluated at $\bxi.$ From this we obtain 
an equation for the general component $\Delta \bw$ of the slip velocity: 
\begin{eqnarray}
&&  \frac{d}{ds}\Delta\bw = -\left(\hat{\bB}^\top\cdot \grad_{\bxi}\tilde{\bu}\cdot \hat{\bB}\right)\hat{\bB}
\,\,\,\,\,\,\,\,\,\,\,\,\,\,\,\,\,\,\,\,\,\,\,\, \cr
&& \,\,\,\,\,\,\,\,\,\,\,\, -\frac{1}{|\bB|}(\grad\btimes\bR)_\perp 
+ \Delta\bw\cdot\grad_{\bxi}\hat{\bB}. \lb{A4} \end{eqnarray}  
To extract an equation for only the perpendicular component, we apply the product rule
\begin{eqnarray}
&&  \frac{d}{ds}\left[(\hat{\bB}\bdot\Delta\bw)\hat{\bB}\right] =
\left(\hat{\bB}\bdot  \frac{d}{ds}\Delta\bw\right)\hat{\bB} \,\,\,\,\,\,\,\,\,\,\,\,\,\,\,\,\,\,\,\,\,\,\,\,\,\,\,\,\,\,\,\,\,\,\,\,\cr
&&\,\,\,\,\,\,\,\,\,\,\,\,\, + \left(\frac{d}{ds}\hat{\bB}\bdot\Delta\bw\right)\hat{\bB}
 + (\hat{\bB}\bdot\Delta\bw)\frac{d}{ds}\hat{\bB} \end{eqnarray} 
 and (\ref{A4}) to obtain
\begin{eqnarray} 
&&  \frac{d}{ds}\left[(\hat{\bB}\bdot\Delta\bw)\hat{\bB}\right] =
-\left(\hat{\bB}^\top\cdot \grad_{\bxi}\tilde{\bu}\cdot \hat{\bB}\right)\hat{\bB} 
\,\,\,\,\,\,\,\,\,\,\,\,\,\,\,\,\,\,\,\,\,\,\,\,\,\,\,\,\,\,\,\,\,\,\,\,\cr
&&\,\,\,\,\,\,\,\,\,\,\,\, + \left(\frac{d}{ds}\hat{\bB}\bdot\Delta\bw_\perp\right)\hat{\bB}
 + \Delta w_{\|} \frac{d}{ds}\hat{\bB}. \end{eqnarray}
Subtracting this equation from (\ref{A4}) gives  
\begin{eqnarray} 
\frac{d}{ds}\Delta\bw_\perp &=& 
-\frac{1}{|\bB|}(\grad\btimes\bR)_\perp + \Delta\bw\cdot\grad_{\bxi}\hat{\bB} \cr
&& \,\,\,\,\,\,\,
-\left(\frac{d}{ds}\hat{\bB}\bdot\Delta\bw_\perp\right)\hat{\bB}
-\Delta w_{\|} \frac{d}{ds}\hat{\bB} \cr
&=& -\frac{1}{|\bB|}(\grad\btimes\bR)_\perp + \Delta\bw_\perp\cdot\grad_{\bxi}\hat{\bB} \cr
&& \,\,\,\,\,\,\,\,\,\,
-\left[(\hat{\bB}\bdot\grad_{\bxi})\hat{\bB}\bdot \Delta\bw_\perp\right]\hat{\bB}. 
\lb{A5} \end{eqnarray}  
Here we have used $\Delta\bw\cdot\grad_{\bxi}\hat{\bB}= \Delta w_{\|}\ \frac{d}{ds}\hat{\bB}+
\Delta\bw_\perp \cdot\grad_{\bxi}\hat{\bB}$ and $\frac{d}{ds}\hat{\bB}=(\hat{\bB}\bdot\grad_{\bxi})\hat{\bB}.$ 
However, it is easy to see that (\ref{A5}) is equivalent to (\ref{slipeq}) written previously. 

It is worth emphasizing that all of our analysis is valid for compressible plasma flows. As is well known,
it is the lines of ${\textit {\textbf G}}=\bB/\rho,$ with $\rho$ the ion mass density, which are generally ``frozen-in'' for laminar ideal 
Ohm's law with a compressible velocity. Since $\hat{{\textit {\textbf G}}}=\hat{\bB},$ however, the lines of ${\textit {\textbf G}}$ parameterized 
by arclength are identical to the lines of $\bB$ parameterized in the same fashion, and the derivation above applies  
without change to compressible MHD flows. In fact, our analysis does not assume an MHD-like fluid  description, but only 
a generalized Ohm's law of the form (\ref{gOhm}). In a collisionless but well-magnetized plasma like the solar wind,
such a generalized Ohm's law holds at scales even below the ion gyroradius but with $\bu$ now identified with the electron fluid 
velocity $\bu_e.$ There is observed in the solar wind at scales between the ion and electron gyroradii a regime of kinetic 
turbulence, with many properties similar to MHD turbulence \citep{Sahraouietal13}, and our discussion here applies also 
to the slipping of magnetic field-lines relative to the electron fluid in such kinetic turbulence.  

The above results, to our knowledge, have not appeared in the previous literature. They are implicit, however,
in the founding  works on ``general magnetic reconnection'' of \cite{Schindleretal88} and \cite{HesseSchindler88} 
but hidden by the use of Euler-Clebsch variables $(\alpha,\beta)$ to label magnetic field-lines.  For example, 
the equations (23a-c) of \cite{HesseSchindler88} when transcribed into our notations\footnote{We use notation $V$ for 
the quantity $-\psi$ in \cite{Schindleretal88, HesseSchindler88}. Here $\psi$ is the so-called ``quasi-potential'' discussed 
at length in the following section.}  read 
\be\dot{\alpha} = \frac{\partial V}{\partial \beta} - R^\beta \ee
\be\dot{\beta} = -\frac{\partial V}{\partial \alpha} + R^\alpha \ee
\be \frac{\partial V}{\partial s} = R^s. \ee
Hence, taking the derivative with respect to arc-length $s,$ one obtains 
\be \frac{d\dot{\alpha}}{ds} = \frac{\partial R^s}{\partial \beta} - \frac{\partial R^\beta}{\partial s} \ee
\be \frac{d\dot{\beta}}{ds} = -\frac{\partial R^s}{\partial \alpha} + \frac{\partial R^\alpha}{\partial s} \ee
The reader inclined to do so can check that these are equivalent to our eq. (\ref{slipeq}), but the derivation
is rather more cumbersome than the one we have given above. One of the minor goals of our work is to 
liberate the subject of ``general magnetic reconnection'' from the tyranny of Euler-Clebsch variables. 
One nice feature of those variables is that they make a mathematical connection with Hamiltonian mechanical formalism 
in the case where $\bR$ and all of its space-derivatives vanish for sufficiently large $s.$ This single advantage 
is not present for turbulent flow, as we discussed in section \ref{Turb}, and does not in any case repay for the many 
severe disadvantages. Euler-Clebsch variables exist at most locally in space and away from magnetic nulls, and 
are therefore unsuitable to describe globally complex magnetic topology. Furthermore, these variables completely 
obscure the intuitive picture of magnetic reconnection in physical space. 
%Our simple eq.(\ref{slipeq}) would doubtless have appeared
%earlier in the literature, if the subject had not been burdened with the unnecessary baggage of Euler-Clebsch variables. 
 
\subsubsection{The Slip-Velocity Source}

In contrast to the theories that identify reconnection with non-vanishing $R_\|,$ our approach identifies non-vanishing 
values of the vector field 
\be \bSigma\equiv -\frac{(\grad\times\bR)_\perp}{|\bB|} \lb{source} \ee 
as the necessary and sufficient source of field-line slippage.  We refer to this quantity as the {\it slip-velocity source}. 
Notice it has units of inverse time and its meaning is the slippage velocity vector developed per unit length 
as the field-line is followed in arc-length $s$ from a selected base point. One of the useful features of the slip-velocity source 
is that it is independent of any base point --- unlike the slip-velocity itself --- and is thus an objective feature of the plasma 
in physical space. It is only by intersecting a point with non-vanishing $\bSigma$ that a field-line can slip relative to the 
plasma flow. The slip-velocity source is thus a useful diagnostic to determine where reconnection ``happens'' in both 
laminar and turbulent plasma flows.    

The slippage source $\bSigma$ diverges at magnetic nulls ($\bB=\bzed$) unless also $(\grad\btimes\bR)_\perp=\bzed$ 
there. As well the homogeneous term in (\ref{slipeq}) diverges at non-degenerate nulls, since 
\be \grad_{\bxi} \hat{\bB}=\frac{1}{|\bB|} (\grad_{\bxi}\bB)({\textbf I}-\hat{\bB}\hat{\bB}). \lb{gradBhat} \ee
Our approach thus makes connection with the theories of magnetic null-point reconnection \citep{Greene88,LauFinn90,LauFinn92}.  
Notice that there is no unique way to integrate (\ref{slipeq}) through a magnetic null, in general. If the incoming field line belongs to a ``fan'' of 
incoming lines, then there are two possible ways to continue the integration in $s$ along the outgoing ``spine''.
Likewise, if the incoming line is along the ``spine'', then there are an uncountable infinity of choices of 
outgoing directions in the ``fan.''   The slippage velocity itself may or may not diverge at the null, e.g. depending upon whether
singularities in (\ref{slipeq}) at the null are $s$-integrable or not. See further discussion of this point in the following section. 

The concept of slip-velocity source helps to resolve the ``paradox'' that turbulent MHD reconnection can have $R_{rms}\rightarrow 0$
with increasing Reynolds numbers, while magnetic reconnection persists. Due to the additional space-gradient
in $(\grad\times \bR)_\perp,$ the slip-source may not vanish even as $R_{rms} \rightarrow 0.$ Also the source of line-slippage 
can be non-zero at magnetic nulls, even if  $(\grad\times \bR)_\perp\rightarrow {\textbf 0}$ there. As pointed
out by \cite{Albright99}, magnetic nulls may proliferate in the high-Reynolds limit of MHD turbulence, forming dense fractal clusters.   
Finally, the magnetic field does not remain smooth in the high-Reynolds-number limit, so that $\grad_{\bxi}\hat{\bB}$ diverges 
everywhere in space (not merely at nulls). This is closely related to the ``line-wandering'' with increasing arc-length $s,$  observed 
by \cite{LazarianVishniac99} to play a crucial role in turbulent reconnection. The wandering of the field-lines 
through space makes it likely that they will encounter the very common
regions where the source of slippage $\bSigma$ has large magnitude. 

Our discussion so far in this section has dealt with ``fine-grained'' fields corresponding to a micro-scale description 
of reconnection at lengths below the turbulent inertial range.  The same considerations apply to inertial-range lengths 
$\ell$ by considering the spatially coarse-grained velocities and magnetic fields, $\bar{\bu}_\ell$ and $\bar{\bB}_\ell.$  
In that case, the effective ``non-ideality'' is the motional electric field induced by turbulent eddies of scale $<\ell:$
\be \bR_\ell^T= -\barepsilon_\ell=-[\overline{(\bu\times \bB)}_\ell-\bar{\bu}_\ell\times\bar{\bB}_\ell]. \ee
It can be shown that $|\bR_\ell^T|\sim |\delta \bu(\ell)\btimes \delta \bB(\ell)|$ where $\delta\bu(\ell),\delta\bB(\ell)$ are increments 
across distance $\ell$ and thus this electric field generally vanishes as $\ell$ decreases through the inertial range. 
See \cite{EyinkAluie06} and section \ref{epsilon} of this paper. On the other hand, the curl has a magnitude $|\grad\times \bR_\ell^T|
\sim \frac{1}{\ell}|\delta \bu(\ell)\btimes\delta \bB(\ell)|$ that instead grows for $\ell$ decreasing through the 
inertial range, until it becomes comparable to the genuinely non-ideal term $\grad\btimes\bR$ at the turbulence
micro-scale $\ell_d.$ But it is important to emphasize that $|\grad\btimes\bR_\ell^T|\gg |\grad \btimes \bar{\bR}_\ell |$ for $\ell\gg \ell_d,$
since $|\grad \btimes \bar{\bR}_\ell | \sim |\hat{\bell}\btimes\bar{\bR}_\ell |/\ell$ in the inertial range (see section \ref{curlE}). 
Thus, the slippage of field lines of coarse-grained $\bar{\bB}_\ell$ in the inertial-range is due to the turbulence-induced 
slip source $(\grad \btimes \bR^T_{\ell})_\perp/|\bar{\bB}_\ell| $ rather than the slip source arising from the true non-ideality 
at scales below $\ell_d.$  

The choice between ``coarse-grained'' and ``fine-grained'' descriptions of turbulent magnetic reconnection 
is purely a matter of convenience. In particular, it should be stressed that existence or not of fast, large-scale reconnection 
for MHD turbulence does not depend in any essential way on coarse-graining. This is, in fact, a general principle in 
physics called ``renormalization group invariance'' \citep{Collins84,Goldenfeld92}.  A standard example is block-spins 
in the Ising model. The statistics of the block-spins must be the same whether they are calculated from the original Ising 
Hamiltonian for the microscopic spins or from an effective Hamiltonian for the block-spins obtained by integrating out the  
microscopic spin degrees of freedom. The same is true in turbulent magnetic reconnection, where reconnection will 
occur for coarse-grained magnetic fields $\bar{\bB}_\ell$ at inertial-range scales $\ell$ when considered either by  
the coarse-grained dynamics or by the fine-grained dynamics. This statement of ``renormalization-group invariance'' seems 
to be almost a triviality, but it can be exploited to obtain nontrivial consequences. \cite{EyinkAluie06} used 
this invariance to derive the necessary conditions for fast magnetic reconnection in ideal MHD.   

\subsection{Flux-Tube Slip Velocities and Line-Voltage}\lb{voltage} 

As we shall now show, the ``slip velocities'' of the previous section generalize the multi-valued flux-tube velocities 
introduced by \cite{Priestetal03}. 

\subsubsection{Basic Definitions}

Let $T$ be any open surface at time $t$ which is everywhere transversal to the magnetic 
field and such that a magnetic field-line intersecting the surface does so at exactly one point. Then for all points $\bx\in T,$
the quantity $\bw_\perp(s;\bx,t)$ defines an instantaneous slip velocity everywhere in the ``magnetic flux tube'' with base 
$T$ and at a distance $s$ along the field-lines inside the tube at time $t$. See Fig.~1. The magnetic field lines are 
now regarded, by convention, as ``frozen-in'' to the entire \\

\vspace{20pt}
\begin{figure}[!ht]
\begin{center}
\includegraphics[width=130pt,height=100pt]{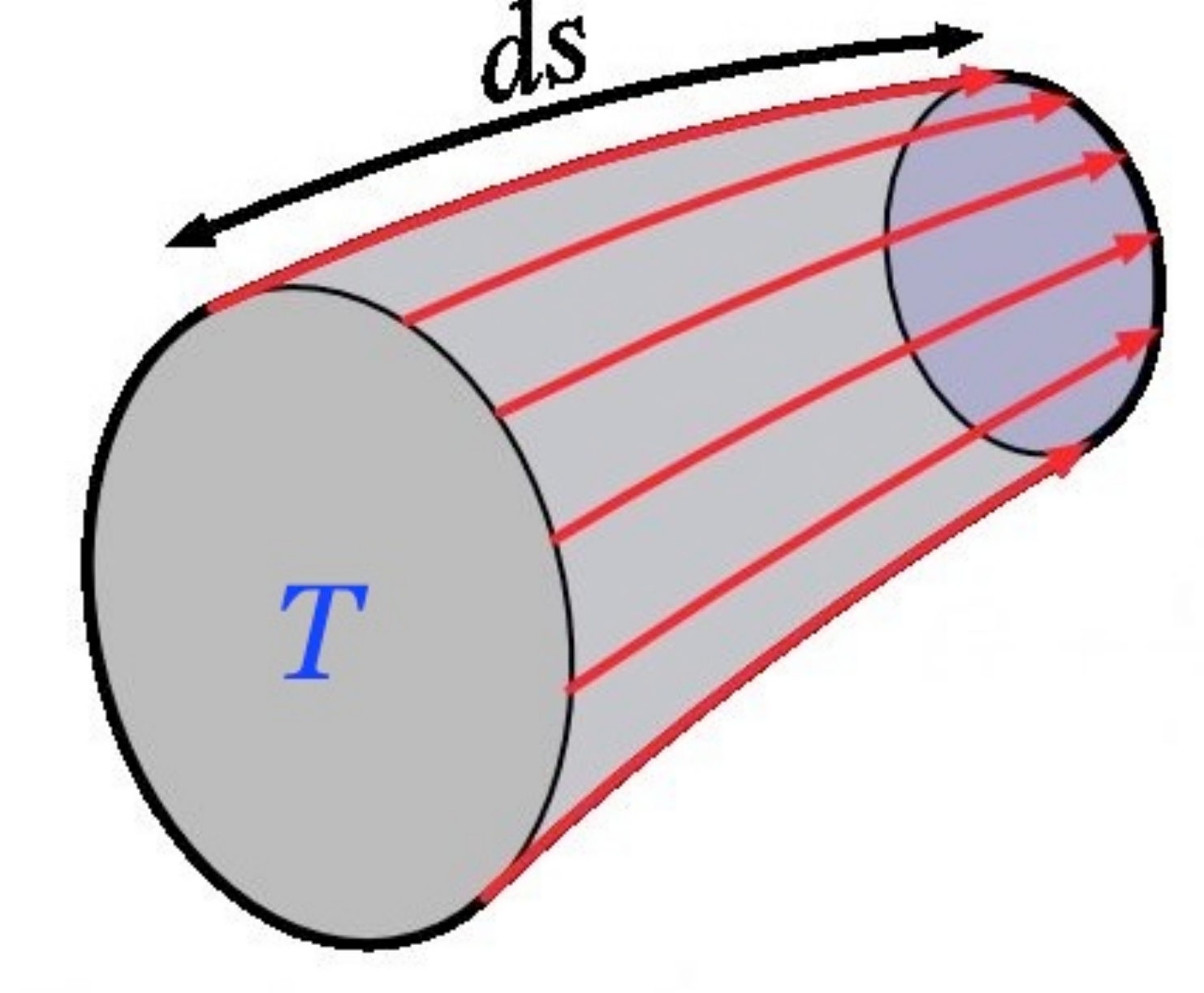}
\end{center}
\caption{{\it Magnetic Flux Tube.} Magnetic flux tube transversal to an open surface $T.$ The magnetic 
field lines are plotted in \textcolor{red}{red}. Arclength element $ds$ along the field-lines is indicated.}
\end{figure}\lb{flux-tube}

\noindent surface $T$ of plasma fluid elements. This slip-velocity within the 
flux tube does, of course, depend upon the particular cross-sectional surface $T$ that is chosen.  
The properties of $T$ required for this definition of flux-tube velocities will be generally preserved by turbulent flow only for a short time. 
That is, the surface $T_{t'}=\bxi(t';T,t)$ advected by the fluid flow to time $t'>t$ will rapidly become wrinkled and folded 
by turbulent advection. Since the ``frozen-in'' property does not hold in the turbulent flow, the magnetic field-lines 
at the later time $t'$ may intersect the warped surface $T_{t'}$ at multiple points or tangentially. In that case, the slip-velocity 
of lines anchored to $T_{t'}$ is no longer well-defined throughout the tube (although slip-velocities can still be defined for field-lines 
anchored to individual plasma fluid elements of $T_{t'}$, as in the previous section). Only for very narrow flux-tubes, with dimensions 
of the initial surface $T$ very small (diameters well below the ``inner'' length $\ell_d$), can the flux-tube velocity be defined over 
extended periods of time.    

Under the above restrictive assumptions and with a very special choice of $T,$ it is possible to integrate exactly our equation 
(\ref{slipeq}) for slip-velocities within the flux-tube, in terms of the so-called ``quasi-potential'' of General Magnetic Reconnection 
theory \citep{Schindleretal88,HesseSchindler88,Priestetal03}. This quantity $V$ is defined for points $\bx'$ within the flux-tube 
by integrating the parallel electric field
$E_{\|}=R_{\|}$ a distance $s$ along the line starting at the unique point  $\bx$ where field-line $L$ through $\bx'$ intersects $T$:
\begin{eqnarray}
 V(\bx') &=& \int_0^s E_{\|}(\bxi(s';\bx,t),t) ds' \cr
 &=& \int_{L:\bx\rightarrow \bx'} \bE(\bxi,t) \bdot d\bxi.  \end{eqnarray} 
We prefer to use the more descriptive term ``line-voltage'' for this quantity. Notice in this setting that the line-voltage $V$ depends 
upon the distance $s$ along the field-line from $T.$ Now, under the stated assumptions, it is not hard to show (see section \ref{derive2}
immediately below) that 
\be \Delta \bw_\perp = \frac{ \hat{\bB}\btimes (\grad V-\bR)}{|\bB|}. \lb{exact} \ee 
gives an exact solution inside the flux tube to the basic equation (\ref{slipeq}). The solution depends upon the particular 
cross-section $T$ of the flux tube which is adopted. The natural choice for $T$ is a normal surface to the vector field $\bR$,
which is transversal to the magnetic field if $R_{\|}\neq 0$ everywhere on the surface. If $T$ is chosen to be such a normal 
surface of $\bR$ then it can be seen from (\ref{exact}) that $\Delta\bw_\perp=\bzed$ on $T,$ and hence (\ref{exact}) inside the 
tube indeed gives the slip velocity for field-lines anchored in $T.$  Of course, such a choice of $T$ can only be made instantaneously, 
since the advected surface $T_{t'}$ for $t'>t$ will not usually be a normal surface for $\bR(\bx,t').$ The condition of normality 
becomes vacuous, on the other hand,  in a space region where $\bR\equiv \bzed$ identically and then any transversal surface 
$T$ may be selected. We thus recover the results of \cite{Priestetal03}  for the ``standard'' situation where the flux tube begins 
and ends in regions with $\bR\equiv\bzed.$  

\subsubsection{Derivations and Proofs}\lb{derive2} 

Here we derive the exact formula (\ref{exact}) for the slip-velocity in terms of the line-voltage $V$ within a flux-tube and 
prove the various statements of the preceding paragraph.  The essential observation is that 
\be \hat{\bB}, \,\,\,\, \bz_\perp\equiv \grad V-\bR, \,\,\,\, \Delta\bw_\perp \equiv \frac{\hat{\bB}\btimes \bz_\perp }{|\bB|} \lb{ortho} \ee
form a right-handed orthogonal (but not orthonormal) coordinate system. In particular note 
\be \hat{\bB}\bdot \bz_\perp = \frac{dV}{ds} - R_{\|} =0. \ee
Our strategy shall be to show that $\Delta \bw_\perp$ defined as above satisfies the basic equation (\ref{slipeq}),
by verifying the projection of that equation onto each of the above coordinate directions. 

For example, differentiating $\hat{\bB}\bdot\Delta\bw_\perp=0$ gives 
\be \hat{\bB}\bdot \frac{d}{ds}(\Delta\bw_\perp) = -\Delta\bw_\perp \bdot \frac{d}{ds}\hat{\bB}
= -\Delta\bw_\perp \bdot (\hat{\bB}\bdot\grad)\hat{\bB}, \ee
which agrees with the $\hat{\bB}$-projection of (\ref{slipeq}). This result shows that the term in 
(\ref{slipeq}) proportional to $(\hat{\bB}\hat{\bB})(\grad\bB)$ has a purely geometric origin and is required 
for that equation to preserve orthogonality with $\hat{\bB}$ as $\Delta\bw_\perp$ is evolved along the $s$-direction.  

Similarly, differentiating $\bz_\perp\bdot\Delta\bw_\perp=0$ gives
\be \bz_\perp\bdot \frac{d}{ds}(\Delta\bw_\perp) = -\Delta\bw_\perp \bdot \frac{d}{ds}\bz_\perp. \lb{zwperp} \ee
To calculate $d\bz_\perp/ds$ we note that 
\be \frac{d}{ds}\bR = (\hat{\bB}\bdot\grad)\bR = \partial_{\|}\bR, \ee
and the relation $ \frac{d}{ds}V = (\hat{\bB}\bdot\grad)V = R_{\|}$ upon taking the space-gradient yields 
\be \frac{d}{ds}\grad V = -(\grad\hat{\bB})\bdot\grad V +\grad R_{\|}. \ee
Hence, taking the difference of these last two equations, 
\be \frac{d}{ds}\bz_\perp  = -\grad\hat{\bB}\bdot \bz_\perp 
+(\grad R_{\|}-\partial_{\|}\bR-\grad\hat{\bB}\bdot\bR). \ee
On the other hand, the gradient of the definition $R_{\|}=\hat{\bB}\bdot\bR$ gives
\be \grad R_{\|} = \grad\hat{\bB}\bdot\bR + \grad\bR\bdot\hat{\bB}, \ee
so that
\begin{eqnarray*}
\grad R_{\|}-\partial_{\|}\bR-\grad\hat{\bB}\bdot\bR &=& \grad\bR\bdot \hat{\bB} -\partial_{\|}\bR \cr
         &=& \left[\grad\bR-(\grad\bR)^\top\right]\cdot\hat{\bB} \cr
         &=& \hat{\bB}\btimes (\grad\btimes\bR).
\end{eqnarray*}         
It follows that 
\be  \frac{d}{ds}\bz_\perp  = -\grad\hat{\bB}\bdot \bz_\perp  + \hat{\bB}\btimes (\grad\btimes\bR). \lb{dzds} \ee
Substituting this into (\ref{zwperp}) and using the definition of $\Delta\bw_\perp$ we conclude that 
\be \bz_\perp\bdot \frac{d}{ds}(\Delta\bw_\perp) = \bz_\perp\bdot\left[
(\Delta\bw_\perp \bdot  \grad)\hat{\bB} - \frac{(\grad\btimes\bR)_\perp}{|\bB|}\right]. \ee
This is the $\bz_\perp$-projection of equation (\ref{slipeq}). 

Finally, differentiation of $|\Delta\bw_\perp|^2=|\bz_\perp|^2/|\bB|^2$ gives 
\be (\Delta\bw_\perp)\bdot \frac{d}{ds}(\Delta\bw_\perp) = \frac{1}{|\bB|^2}\bz_\perp\bdot \frac{d\bz_\perp}{ds}
- \frac{|\bz_\perp|^2}{|\bB|^2} \frac{d}{ds}\ln |\bB|. \ee
Using (\ref{dzds}) for $d\bz_\perp/ds$ gives
\begin{eqnarray} 
&& (\Delta\bw_\perp)\bdot \frac{d}{ds}(\Delta\bw_\perp) = -\frac{1}{|\bB|^2}\bz_\perp\bdot (\grad\hat{\bB})\bdot\bz_\perp \cr
&& \,\,\,\,\,\,\,\,\,\,\,\,\,\,\,\,\,\,\,\,\,\,\,\, - |\Delta\bw_\perp|^2 \frac{d}{ds}\ln |\bB| \cr
&& \,\,\,\,\,\,\,\,\,\,\,\,\,\,\,\,\,\,\,\,\,\,\,\,+\frac{1}{|\bB|^2}\bz_\perp\bdot\hat{\bB}\btimes(\grad\btimes\bR). \lb{dnormds} \end{eqnarray} 
We next note that a relation on the trace of $\grad\hat{\bB}$ follows from the solenoidal condition ${\rm Tr}(\grad\bB)=\grad\bdot\bB=0,$
by substituting $\bB=|\bB|\hat{\bB}$ to obtain 
\be {\rm Tr}(\grad\hat{\bB}) = -\frac{d}{ds}\ln|\bB|. \ee
Expressed in terms of the orthogonal coordinate system (\ref{ortho}), this condition on the trace becomes
\begin{eqnarray}
&&  \frac{1}{|\bB|^2} \bz_\perp\bdot (\grad\hat{\bB})\bdot \bz_\perp + (\Delta\bw_\perp)\bdot (\grad\hat{\bB}) \bdot (\Delta\bw_\perp) \cr
&& \,\,\,\,\,\,\,\,\,\,\,\,\,\,\,\,\,\,\,\,\,\,\,\,   = -|\Delta\bw_\perp|^2 \frac{d}{ds}\ln|\bB|, \end{eqnarray} 
noting that $(\grad\hat{\bB})\bdot\hat{\bB}=0.$  Using the above trace condition and the definition of $\Delta\bw_\perp,$
the equation (\ref{dnormds}) becomes
\begin{eqnarray}
&& (\Delta\bw_\perp)\bdot \frac{d}{ds}(\Delta\bw_\perp) = \,\,\,\,\,\,\,\,\,\,\,\,\,\,\,\,\,\,\,\,\,\,\,\,\,\,\,\,\,\,\,\,\,\,\,\, \cr
&&  \Delta\bw_\perp\bdot\left[
(\Delta\bw_\perp \bdot  \grad)\hat{\bB} - \frac{(\grad\btimes\bR)_\perp}{|\bB|}\right]. \,\,\,\,\,\,\,\,\,\,\,\,\,\,\,\,\,\,\,\end{eqnarray} 
This is the $\Delta\bw_\perp$-projection of equation (\ref{slipeq}). 

The formula (\ref{exact}) thus solves the equation (\ref{slipeq}) for the line-voltage
$V$ developed from any transversal surface $T$ of the flux tube. In order to represent the actual 
slip velocity $\Delta\bw_\perp$ for field-lines anchored to $T,$ the condition $\Delta\bw_\perp=\bzed$
must hold on the surface $T.$ This condition is equivalent to $\grad V = \bR$ on $T$. Since 
$T$ is a zero level-surface of $V,$ by the very definition of $V$, $\grad V$ on $T$ is normal to the surface.
Hence, the condition $\grad V = \bR$ can hold on $T$ only if that surface $T$ is chosen to be everywhere 
normal to $\bR.$ As a matter of fact, this normality condition is not only necessary but also sufficient 
to guarantee that $\grad V = \bR$ on $T.$ If the flux tube starts in a region where $\bR\equiv\bzed,$
then $\grad V = \bR$ trivially. Thus, assume instead that the tube starts in a neighborhood where $\bR$ is 
everywhere nonzero, so that the unit vector $\hat{\bR}$ and its normal surface $T$ are well-defined. 
In that case, the line-voltage $V$ defined for such a choice of $T$ satisfies
\be \grad V=|\grad V| \hat{\bR} \,\,\,\,\,\,\,\, {\rm on} \,\,\,\, T. \ee
Dotting this equation with $\hat{\bB}$ and using $dV/ds=R_{\|}$ gives
\be R_{\|}=\hat{\bB}\bdot\grad V = |\grad V| (\hat{\bR})_{\|}= |\grad V|\frac{R_{\|}}{|\bR|}, \ee
which implies $|\grad V|=|\bR|$ on $T$. It follows that 
\be \grad V=\bR \,\,\,\,\,\,\,\, {\rm on} \,\,\,\, T \ee
and thus $\Delta \bw_\perp=\bzed$ on $T,$ as claimed. 

\subsubsection{Discussion of the Results} 

It is sometimes stated, loosely, that ``reconnection occurs when there is a breakdown of ideal MHD and therefore
an electric field component ($E_{\|}$) along the magnetic field...'' \citep{PriestDemoulin95}. This is an incorrect 
statement for turbulent reconnection. All current evidence supports the idea that Ohmic electric fields vanish 
(distributionally) for MHD turbulence in the infinite-conductivity limit, that ideal MHD holds (in the coarse-grained
or ``weak'' sense) and yet reconnection occurs unabated. A more precise statement is that a non-vanishing gradient 
$\grad V$ is required for reconnection \citep{Schindleretal88,HesseSchindler88}.  It is quite possible that $\bR\rightarrow \bzed$ 
pointwise everywhere, that $V$ vanishes along every field-line segment of finite length, and yet $\grad V$ remains 
finite or diverges to infinity.  

The line-voltage $V$ is sometimes interpreted as a ``reconnection rate'' or rate of change of magnetic flux 
due to reconnection. The rationale for this is roughly as follows. Consider points $\bx'$ and $\by'$ inside the flux tube 
which are connected by a curve $C'$ everywhere orthogonal to $\bB$ and which have base-points $\bx$ and 
$\by$ in $T.$ Then the quantity defined by 
\be \dot{\Phi}_{{\rm slip}} \equiv \int_{C':\bx'\rightarrow \by'} \bB\bdot(\Delta\bw_\perp\btimes d\br) \ee 
is naturally interpreted as the rate of transfer across $C'$ of the magnetic flux in the tube, due to slippage motion 
of field-lines.  Using (\ref{exact}) it is easy to see that
\be \dot{\Phi}_{{\rm slip}} = V(\bx')-V(\by') + \int_{C':\bx'\rightarrow \by'} \bR\bdot d\br. \ee
In the ``standard situation'' where $C'$ is outside an assumed well-localized ``diffusion region'',  so that $\bR=\bzed$ on $C',$ 
the rate of transfer of flux by slippage is simply the difference of the line-voltages, $V(\bx')-V(\by'),$ and $V(\bx'),V(\by')$ become
independent of distance $s$ along the field line.  If furthermore $\by'$ is on a line that lies entirely outside the ``diffusion region'', 
then $V(\by')=0$. In that case,  $\dot{\Phi}_{{\rm slip}} = V(\bx'),$ so that the rate of change of flux may be identified with $V(\bx').$

None of these statements hold obviously in a turbulent regime with $\bR\rightarrow\bzed$. Both $V$ and the line-integral 
of $\bR$ along $C'$ may vanish. This is not required, of course. Even if $\bR\rightarrow\bzed$ almost everywhere 
in space (with respect to Lebesgue measure) as conductivity goes to infinity, there may be uncountably many field-lines, 
densely distributed in space, where the line-voltages do {\it not} vanish, due to spatial intermittency effects. But in that 
case, there is no reason that the line-integral of $\bR$ along $C'$ necessarily vanishes either! It is not clear 
whether $V$ may remain non-zero in high-conductivity MHD turbulence for some subset of lines and, if so, whether 
these values represent ``reconnection rates'' along the corresponding segment of these lines. It is one of the open 
issues in turbulent reconnection whether line-voltages may remain non-vanishing at infinite conductivity for certain lines
due to spatial intermittency effects. 

It is quite possible, however, that $V\rightarrow 0$ for all field-lines and yet flux-conservation is violated throughout the turbulent flow. 
Notice that the transfer of magnetic flux within the tube by slippage can be rewritten as  
\be\dot{\Phi}_{{\rm slip}}=\oint_\Gamma \bR\bdot d\br, \ee
where $\Gamma:\bx \stackrel{L_x}{\rightarrow}\bx'\stackrel{C'}{\rightarrow}\by' \stackrel{-L_y}{\rightarrow}\by \stackrel{-C}{\rightarrow}\bx$
is the closed loop obtained by following up along field-line $L_x$ from $\bx\rightarrow \bx',$ across along $C'$ from $\bx'\rightarrow \by',$
down along line $L_y$ from $\by'\rightarrow \by,$ and back along $C$ from $\by\rightarrow \bx$ inside $T.$ Here $\int_{C} \bR\bdot d\br=0$ 
because $T$ is normal to $\bR.$ Thus, $\dot{\Phi}_{{\rm slip}}=\left.\frac{d}{dt}\Phi(\Gamma,t)\right|_{t=0}$ represents instantaneous 
non-conservation of flux through the loop $\Gamma$ due to the non-ideality. However, flux-conservation is a Lagrangian statement 
and requires that $\frac{d}{dt}\Phi(\Gamma,t)=\oint_{\Gamma(t)}\bR(\br,t)\bdot d\br$ vanish for the advected loop $\Gamma(t)$ moving 
with the plasma fluid, at all times $t$. As discussed in \cite{EyinkAluie06}, the loop $\Gamma(t)$ for any $t>0$ is expected to approach 
a non-rectifiable (fractal) curve in the limit of very long inertial ranges. Thus, it is possible that $V\rightarrow 0$ and $\dot{\Phi}_{{\rm slip}}\rightarrow 0$ but that 
$\frac{d}{dt}\Phi(\Gamma,t)\neq 0$ at positive times $t>0,$ because the vanishing of $\bR$ is compensated by the unbounded 
growth of the length of $\Gamma(t)$ as the inertial range increases in extent. 

A final remark concerns the behavior of the slip velocity at magnetic nulls. It follows from (\ref{exact}) that $\Delta\bw_\perp$ will generally 
diverge at nulls, unless it happens that the magnitude of $(\grad V-\bR)$ vanishes at an equal or faster rate than $|\bB|$ as the null is 
approached. A small flux-tube around a field-line that enters a null along a ``spine'' will impinge on the ``fan'' plane and define $V$, 
at most, in a narrowing layer on that side of the ``fan''. $\grad V$ is then defined only in a one-sided sense on the ``fan'' plane.  
On the other hand, a small flux-tube around a field-line entering a null inside a ``fan'' plane will generally lead to a $V$ which is 
discontinuous at the null and multi-valued on the outgoing ``spine'', since each flux-tube line that belongs to the ``fan'' will usually 
enter the null with a different voltage.  

\section{Weak Solutions and Coarse-Grained Generalized Ohm's Law}\lb{coarse}

We have argued in the previous sections that reconnection is associated fundamentally to the slip-source 
$\bSigma$ and not to the non-ideal electric field $\bR.$ In particular, the example of resistive MHD turbulence 
shows that it is possible for $\bR$ to go to zero (in r.m.s.) and for $\bSigma$ to become simultaneously unboundedly large. 
This fact implies that situations can exist where ideal MHD is valid, but magnetic reconnection will occur freely.
This may sound contradictory, but it is only because the sense of validity of ideal MHD for turbulent solutions 
is in a ``weak''  or ``coarse-grained'' sense, which is quite different than the usual notion of validity for smooth, 
laminar solutions. In this section, we shall explain ``weak solutions'' in a non-technical manner, relating them to 
the more physically familiar ideas of spatial coarse-graining and renormalization-group theory.   

Furthermore, we shall argue in detail for the validity of ideal Ohm's law in this ``weak'' or ``coarse-grained'' sense,
for the modes at length-scales within an MHD-like turbulent inertial-range. We do so by means of an analysis 
of the Generalized Ohm's Law for an electron-ion plasma. The plasma (ion) momentum equation could be 
treated similarly, but, since our primary  interest is reconnection, we do not consider it here. The principal
example we have in mind is the solar wind, where detailed empirical evidence is available to support our argument.   

\subsection{Weak Solutions and Spatial Coarse-Graining} 

The MHD equations and related hydromagnetic equations such as 2-fluid models are conservations laws 
(for mass, momentum, energy and magnetic field). Thus, they have a standard {\it weak formulation} \citep{Evans10}. 
This notion is perhaps most familiar to space scientists and astrophysicists in the context of discontinuous solutions 
of ideal MHD, such as fast/slow shocks and rotational/tangential discontinuities, and numerical methods to solve for 
such solutions \citep{Levequeetal98}. It is also understood that turbulence is described by such weak solutions of 
ideal fluid equations, as suggested originally by Onsager for hydrodynamic turbulence \citep{Onsager49,EyinkSreenivasan06,
Eyink08,DeLellisSzekelyhidiJr12} and later by others for MHD turbulence and reconnection \citep{Caflischetal97,EyinkAluie06}. 
Here we give a brief self-contained discussion. 

Using as an example the magnetic induction equation 
\be \partial_t\bB= -\grad \btimes \bE, \ee 
the weak formulation corresponds to smearing with a smooth test function $\varphi(\br,s)$ and moving all derivatives
to the test function: 
\begin{eqnarray} 
&&  \int d^3r \int ds \, \big[\partial_s\varphi(\br,s)\bB(\br,s) \,\,\,\,\,\,\,\,\,\,\,\,\,\,\,\,\,\,\,\,\,\,\,\,\,\,\,\,\,\,\,\,\,\,\,\,\,\,\,\,\,\,\,\,\,\,\,\,\cr
&& \,\,\,\,\,\,\,\,\,\,\,\,\,\,\,\,\,\,\,\,\,\,\,\,\,\,\,\,\,\,\,\,\,\,\,\,+\grad\varphi(\br,s)\btimes\bE(\br,s)\big] =0. \end{eqnarray}  
This notion extends the meaning of ``solution'' to singular fields $\bE,\bB$ for which ordinary classical derivatives do not 
exist.  As a matter of fact, it is often sufficient to smear only in the space-variable. Using a sequence of test 
functions $\phi(\br,s)$ that approximate $\psi(\br)\chi_{[0,t]}(s),$ with $\chi_{[0,t]}(s)$ the characteristic function 
of the time-interval $[0,t],$ one obtains   
\begin{eqnarray} 
&&  \int d^3r \ \psi(\br) \bB(\br,t) = \int d^3r\, \psi(\br)\bB_0(\br) \,\,\,\,\,\,\,\,\,\,\,\,\,\,\,\,\,\,\,\,\,\,\,\, \cr
&& \,\,\,\,\,\,\,\,\,\,\,\,\,\,\,\,\,\,\,\,\,\,\,\,- \int_0^t ds \int d^3r\, \grad\psi(\br)\btimes\bE(\br,s) \end{eqnarray} 
and for almost every time $t$
\be \frac{d}{dt}\int d^3r\, \psi(\br)\bB(\br,t)=- \int d^3r\, \grad\psi(\br)\btimes\bE(\br,t), \ee
if the righthand-side in the last equation above is Lebesgue-integrable in time.

The above considerations may seem rather technical, but they can be interpreted in a standard physical picture 
of {\it spatial coarse-graining}, if one takes test functions $\psi(\br)$ of the form
\be  \psi_{\bx,\ell}(\br) = \ell^{-3} G((\br-\bx)/\ell) \equiv G_\ell(\br-\bx) \ee
 for some smooth, rapidly decaying, non-negative function $G$ with unit integral and $\ell>0$. In that case, the weak 
 formulation implies the equation 
 \be \partial_t\bar{\bB}_\ell(\bx,t)= -\grad \btimes \bar{\bE}_\ell(\bx,t), \ee 
 for the coarse-grained fields at length-scale $\ell$:
\be \bar{\bB}_\ell(\bx,t) = \int d^3r\ G_\ell(\br) \bB(\bx+\br,t), \ee
and so forth for $\bar{\bE}_\ell,$ etc. As a matter of fact, the above coarse-grained equations for all $\bx$ and $\ell$
are equivalent to the usual weak formulation. We briefly explain this standard fact without elaborate mathematical detail. 
Note that for a general smooth test function $\psi$
\be \lim_{\ell\rightarrow 0} \int d^3x \, \psi(\bx) G_\ell(\br-\bx) = \psi(\br). \ee
On the other hand, one can also approximate the convolution integral by a finite Riemann sum,
\be \int d^3x \, \psi(\bx) G_\ell(\br-\bx) \simeq  \sum_{i=1}^N (\Delta V)_i \psi(\bx_i) G_\ell(\br-\bx_i)  \ee
converging as $N\rightarrow\infty.$ Hence, by taking a linear combination of the coarse-grained equations 
at points $\bx_i$ with coefficients $(\Delta V)_i \psi(\bx_i)$ and taking the limits first $N\rightarrow\infty$
and then $\ell\rightarrow 0,$ one recovers the standard weak formulation with an arbitrary smooth test function $\psi.$   

The coarse-graining point of view is the more important one for practical applications, because it gives a physical relevance 
to weak solutions as valid descriptions {\it over a certain range of length-scales}. For example, it is widely believed 
that the ideal Ohm's law $\bE= -\bu\btimes \bB$ is valid in the solar wind at lengths $\ell$ much greater than the ion 
gyro-radius $\rho_i$ (e.g. see \cite{Schekochihinetal09} for a detailed discussion). However, the ideal magnetic induction equation 
cannot hold in the classical sense of partial-differential equations, since the solar wind is observed to have a Kolmogorov-type 
turbulent inertial range with magnetic energy spectrum $E_B(k,t)\sim k_\perp^{-5/3}$ down to ion scales, and kinetic turbulence 
at smaller scales. Thus, fine-scale magnetic  gradients $\grad_\perp\bB$ have a spectrum $k_\perp^2 E_B(k,t)\sim k_\perp^{1/3}$ 
with increasing r.m.s. contributions up to wave-numbers of order $1/\rho_i,$ and continuing to grow up to electron scale wavenumbers
\cite{Sahraouietal13}. Thus, fine-scale magnetic-field gradients will be dominated by modes at electron scales and 
there is no sense in which the ideal induction equation can hold for the standard sense of derivatives. What is plausible and consistent with 
observations is instead that the {\it coarse-grained ideal induction equation} 
\be \partial_t\bar{\bB}_\ell = \grad \btimes \overline{(\bu\btimes \bB)}_\ell  \lb{coarse-ideal} \ee
is valid to a very good approximation for $\ell\gg \rho_i.$ In this precise sense, the ideal induction equation is valid in the 
``weak sense'' for length-scales greater than the ion gyroradius.\footnote{Note that physical derivations of MHD-like equations 
at large scales in the solar wind, such as \cite{Schekochihinetal09} via gyrokinetics, generally impose conditions on wavenumbers $k_\perp,$
$k_{\|}$ and perhaps also on frequencies $\omega.$ This is equivalent to a weak formulation in which one uses test functions
of the form $\psi_{\bk_\perp,k_{\|}}(\br)=\exp(i\bk_\perp\bdot\br_\perp+ik_{\|}r_{\|}),$ or perhaps $\varphi_{\bk_\perp,k_{\|},\omega}(\br,s)
= \exp(i\bk_\perp\bdot\br_\perp+ik_{\|}r_{\|}-i\omega s)$ if one wishes to select for frequencies as well. In fact, this is the original
approach of \cite{Onsager49} to define weak Euler solutions describing infinite Reynolds-number turbulent flow. 
For a careful mathematical discussion, see \cite{DeLellisSzekelyhidiJr12}. This Fourier approach is mathematically 
equivalent to our filtering method and amounts to using a sharp spectral filter/Fourier truncations to define the effective 
equations in a given range of wavenumbers and frequencies. Note, however, that the analogue of the turbulent electric field 
$\barepsilon_\ell$ is present also for a sharp Fourier filter, but was omitted without justification in \cite{Schekochihinetal09}.}   

This sense of validity has, however, a quite different meaning than the naive validity of the ideal 
induction equation for the coarse-grained variables $\bar{\bu}_\ell,\bar{\bB}_\ell$. In fact, we can rewrite 
the above equation in terms of the turbulent electric field induced by motions at scales $<\ell$ as 
\be \partial_t\bar{\bB}_\ell = \grad \btimes (\bar{\bu}_\ell\btimes \bar{\bB}_\ell +\barepsilon_\ell),  \ee 
which differs from the naive ideal equation by the apparent ``non-ideal'' term 
\be \barepsilon_\ell = \overline{(\bu\btimes \bB)}_\ell - \bar{\bu}_\ell\btimes \bar{\bB}_\ell. \ee
Yet note that this apparently ``non-ideal'' electric field is obtained just by coarse-graining the ideal Ohm's law.
The physical meaning is that the turbulent cascade of magnetic energy is governed entirely by ideal MHD-like 
dynamics\footnote{In fact, the flux of magnetic energy from scales $>\ell$ to scales $<\ell$ due to turbulent 
energy cascade is given exactly by the expression $\Pi_\ell^B=-\bar{\bJ}_\ell\bdot \barepsilon_\ell,$ where
$\bar{\bJ}_\ell$ is the coarse-grained electric current. See \cite{AluieEyink10}.}. 
The difference with the naive sense of validity of ideal MHD is, however, the source of many misunderstandings 
and erroneous conclusions. For example, one often encounters statements like ``At $k_\perp \rho_i \ll 1$, ions ... are
magnetized and the magnetic field is frozen into the ion flow'' \citep{Schekochihinetal09},  but they are fundamentally incorrect.  
Instead, the induced electric field $\barepsilon_\ell$ in a turbulent environment leads to magnetic reconnection, as  
was first pointed out, to our knowledge, by \cite{MatthaeusLamkin86}. 

In the next section below we argue in detail for the validity of (\ref{coarse-ideal}) at length-scales $\ell$ much greater 
than than the relevant ``inner'' length-scale $\ell_d$ of astrophysical plasma turbulence (the precise length-scale 
involved depending upon the microscopic plasma properties). Assuming a general, abstract form of Ohm's law, 
$\bE+\bu\btimes\bB=\bR,$ it follows that 
\be \partial_t\bar{\bB}_\ell = \grad \btimes [\overline{(\bu\btimes \bB)_\ell}+\bar{\bR}_\ell]. \ee
To obtain (\ref{coarse-ideal}) it is enough for $\bR$ to vanish in r.m.s  magnitude as $\ell_d\rightarrow 0.$ 
To see this, note by an integration by parts that 
\be \grad\btimes\bar{\bR}_\ell(\bx) = \frac{1}{\ell} \int d^3r\ (\grad G)_\ell(\br) \btimes \bR(\bx+\br).  \ee
Hence, by the Cauchy-Schwartz inequality\footnote{This yields an estimate pointwise in $\bx$. One can also obtain estimates 
for $p$th-order moments in space-averages over $\bx,$ which in mathematics is called an ``$L_p$-estimate,'' 
with $\|\bR\|_p=\left(\int d^3x\, |\bR(\bx)|^p\right)^{1/p}$ the $L_p$-norm. For example,
an application of the (continuous) Minkowski inequality gives $\|\grad\btimes\bar{\bR}_\ell\|_p\leq (1/\ell)
\int d^3r \, |\grad G(\br)| \cdot \|\bR\|_p$ for any $p\geq 1.$ Any of the estimates that we obtain here and below 
can be interpreted either in a pointwise sense or for $pth$-order moments in space averages. See discussions 
in \cite{Eyink05}. To avoid burdening our presentation with excessive mathematical detail, we will not specify 
in any of our estimates below the various possible senses of validity.}   
\be |\grad\btimes\bar{\bR}_\ell(\bx)|\leq  \frac{1}{\ell} \| (\grad G)_\ell\| \cdot \|\bR\| \ee
where $\|\bR\|=\sqrt{\int d^3x \ |\bR(\bx)|^2}$ is the space $L^2$-norm. This coincides with the r.m.s. magnitude,
$R_{rms}=\|\bR\|$, when the space-average $\int d^3x \,\bR(\bx)=\bzed.$  Hence, if $R_{rms}\rightarrow 0$
as $\ell_d\rightarrow 0,$ then $\grad\btimes\bar{\bR}_\ell\rightarrow 0$ for every fixed $\ell$ as $\ell_d\rightarrow 0.$
Note it is not required here that $\bR(\bx)$ vanish for all $\bx:$ \\ 
due to  spatial intermittency, there could be 
non-empty sets of zero volume where $\bR(\bx)\neq \bzed$ as $\ell_d\rightarrow 0.$ The ideal induction equation 
would still hold in the limit in the ``weak'' sense. We argued in Section \ref{Turb} 
that this situation occurs for resistive
MHD with $\bR=\eta\bJ$, in the limit as $\eta\rightarrow 0.$ Below we argue that the same result holds for 
more general forms of plasma non-ideality. 
 
\subsection{Coarse-Grained Generalized Ohm's Law}\lb{CG-GOL} 

The ``weak'' validity of the ideal Ohm's law in an MHD-like inertial range such as the solar wind can be understood starting
from the ``Generalized Ohm's Law'' of plasma physics, which is, we recall, a  rewriting of the electron momentum equation 
that ignores terms of order $O(m_e/m_i),$ the electron-ion mass ratio. We have so far in this paper used dimensionless 
variables with an MHD scaling, but we now write the generalized Ohm's law in dimensional cgs units, as  
\begin{eqnarray} 
&&  \bE +\frac{1}{c}\bu\btimes \bB= \eta\bJ + \frac{1}{nec}\bJ\btimes\bB -\frac{1}{ne}\grad\bdot{\textbf P}_e
\,\,\,\,\,\,\,\,\,\,\,\,\,\,\,\,\,\, \cr
&& +\frac{m_e}{ne^2}\left[\frac{\partial \bJ}{\partial t}
+\grad\bdot\left(\bJ\bu+\bu\bJ-\frac{1}{ne}\bJ\bJ\right) \right]. \lb{Gohm} \end{eqnarray} 
This equation has often been used in discussions of magnetic reconnection \citep{Vasyliunas75,Bhattacharjeeetal99,
CraigWatson03}, where the final term in the square bracket is frequently omitted under the assumption that $J\ll ne u.$  
On the other hand, a recent paper of \cite{Ohiaetal12} studies magnetic reconnection using a fluid model that 
retains only the first and fourth terms in the square bracket and which gives results in good agreement with 
PIC simulations of kinetic reconnection. We therefore keep all four bracketed terms here. 
The coarse-grained version of this equation is 
\begin{eqnarray}
&& \bar{\bE}_\ell +\frac{1}{c}\bar{\bu}_\ell\btimes \bar{\bB}_\ell= -\barepsilon_\ell + \eta\bar{\bJ}_\ell 
\,\,\,\,\,\,\,\,\,\,\,\,\,\,\,\,\,\,\,\,\,\,\,\,\,\,\,\,\,\,\,\,\,\,\,\, \,\,\,\,\,\,\,\,\,\,\,\,\,\,\,\,\,\,\cr
&& \,\,\,\,\,\,\,\,\,\,\,\,\,\,\,\,\,\,+ \frac{1}{nec}\overline{(\bJ\btimes\bB)_\ell} -\frac{1}{ne}\grad\bdot{\overline{\textbf P}}_{e,\ell} \,\,\,\,\,\,\,\,\,\,\,\,\,\,\,\,\,\,\cr
&& +\frac{m_e}{ne^2}\left[\frac{\partial \bar{\bJ}_\ell}{\partial t}
+\grad\bdot\overline{\left(\bJ\bu+\bu\bJ-\frac{1}{ne}\bJ\bJ\right)_\ell} \right]. 
\end{eqnarray} 
Here we assume for simplicity of presentation that the density $n$ is spatially constant, otherwise density-weighted Favre-averging
is required \citep{Favre69,Aluie13}. For that analysis, see  Appendix \ref{density}. 
As we shall now show in detail, the dominant term on the righthand side at length-scales 
$\ell$ in an MHD-like inertial range is the first term $-\barepsilon_\ell$ from ideal Ohm's law and all of the other terms are negligible.  
The reason is that all of the genuinely non-ideal terms involve at least one overall space-gradient. This is obvious 
for all of the terms except the Hall electric field, where it follows from the non-relativistic Ampere's law 
$\bJ=\frac{c}{4\pi} \grad\btimes\bB,$ the standard vector calculus identity
\be (\grad\btimes \bB)\btimes \bB= (\bB\bdot\grad)\bB -\frac{1}{2}\grad(|\bB|^2), \lb{vec-calc-id} \ee
and the solenoidality condition $\grad\bdot\bB=0.$
Each overall space-gradient $\grad$ brings in a factor $\ell^{-1}$ upon coarse-graining at scale $\ell,$ and thus
terms with overall gradients give diminishing contributions for increasing $\ell.$ These terms are thus essentially ``irrelevant'' 
variables in the renormalization-group sense as the generalized Ohm's law is coarse-grained to successively larger length-scales. 
To make more quantitative estimates, one must compare $\ell$ with other relevant length-scales, as we now do term-by-term.  
Because there is no small parameter on which to base an expansion, our analysis is non-perturbative and exploits 
an exact cumulant or linked-cluster expansion of coarse-graining averages. See Appendix \ref{cumulant} for details. 
 
 \subsubsection{Turbulent subscale electric field}\lb{epsilon}  
 
The leading order term in an MHD-like inertial range is the electric field induced by motion of eliminated turbulent eddies
at scales $<\ell$, which (see \cite{EyinkAluie06} and Appendix \ref{cumulant}) can be written in terms of field increments
$\delta \bB(\br;\bx)=\bB(\bx+\br)-\bB(\bx),$ etc.  as    
\begin{eqnarray} 
&& \!\!\!\!\!\!\!\! \!\!\!\!\!\!\!\!   \barepsilon_\ell = \frac{1}{c} \left[\int d^3r \, G_\ell(\br) \, \delta \bu(\br)\btimes \delta \bB(\br)\right. 
\,\,\,\,\,\,\,\,\,\,\,\,\,\,\,\,\,\,\,\,\,\,\,\,\,\,\,\,\,\,\,\,\,\,\,\cr
&& \!\!\!\!\!\!\!\!\!  \!\!\!\!\!\!\!\! -\left. \int d^3r  \int d^3r' \, G_\ell(\br) G_\ell(\br') \, \delta \bu(\br)\btimes \delta \bB(\br') \right]. 
\lb{CET} \end{eqnarray} 
In this expression and following ones the $\bx$-variable for simplicity is not written explicitly. 
It is easily seen from this expression for a spherically-symmetric filter kernel $G$ that 
\be \barepsilon_\ell \sim \frac{1}{c} \langle \delta \bu(\bell)\btimes \delta \bB(\bell)\rangle_{{\rm ang}}, \lb{ETmag} \ee
where $\langle\cdot\rangle_{{\rm ang}}$ denotes a spherical average over the direction vector $\hat{\bell}.$
This will be the same order of magnitude as 
\be \barepsilon_\ell \sim \frac{1}{c} \delta u(\ell)\cdot \delta B(\ell), \lb{ETmagB} \ee
unless the vectors $\delta \bu(\bell)$, $\delta \bB(\bell)$ exhibit ``dynamic alignment,'' which may cause them 
to be nearly parallel \citep{Boldyrev05,Boldyrev06}. There is some evidence for this phenomenon in 
the solar wind \citep{Podestaetal09,Hnatetal11,Wicksetal13b}, but dynamic alignment, if it really occurs for $\ell$ 
much smaller than the turbulent outer scale $L,$ only leads to a reduction by some factor $(\ell/L)^\beta,$ 
with $\beta=1/4$ a popular value \citep{Boldyrev06}. As we shall see below, this modest reduction does not affect 
our conclusion that $\barepsilon_\ell$ is the leading term. 

\subsubsection{Ohmic electric field}\lb{Ohm}  

Next consider the Ohmic electric field $\bE^{{\rm Ohm}}=\eta\bJ$. This contribution is quite tiny in the nearly 
collisionless solar wind, but it can be the leading non-ideal term in other cases, such as the solar photosphere. 
The Ohmic field coarse-grained to length $\ell$ can be written in terms of the magnetic increment as   
\be \bar{\bE}_\ell^{{\rm Ohm}} = 
\frac{\lambda}{\ell c} \int d^3r (\grad G)_\ell(\br) \btimes \delta \bB(\br), \ee 
with $\lambda=\eta c^2/4\pi$ the magnetic diffusivity.  The magnitude of the Ohmic electric field is thus
\be \bar{E}_\ell^{{\rm Ohm}}\sim \lambda \frac{\delta B(\ell)}{\ell c}. \ee 
Taking into account scale-dependent anisotropy as proposed by \cite{GoldreichSridhar95}, the $\ell$ in the prefactor 
of this and other estimates should be interpreted as $\ell_\perp,$ since the increments that make the 
largest contributions are those with $\br\perp \bar{\bB}_\ell.$ In absence of dynamic alignment, the coarse-grained
Ohmic field thus matches the turbulent electric field when $\ell_\perp \delta u(\ell)\sim \lambda,$ which is the condition 
defining the resistive dissipation length $\ell_\eta.$ For example, assuming Goldreich-Sridhar scaling $\delta u(\ell)
\sim (\varepsilon \ell_\perp)^{1/3},$ with $\varepsilon$ the energy cascade rate, the resistive dissipation length is 
$\ell_{\eta,\perp} \sim (\lambda^3/\varepsilon)^{1/4}.$ At length-scales $\ell_\perp\gg \ell_{\eta,\perp}$ the turbulent 
electric field is therefore much larger than the Ohmic field. Dynamic alignment changes the velocity scaling to 
$\delta u(\ell)\sim (\varepsilon \ell_\perp^{1-\beta}L^\beta)^{1/3},$ and thus the resistive dissipation length to 
$\ell_{\eta,\perp}\sim (\lambda^3/\varepsilon L^\beta)^{1/(4-\beta)}.$ The situation is qualitatively unchanged, 
with the turbulent field again dominating for $\ell_\perp\gg \ell_{\eta,\perp}.$

\subsubsection{Hall electric field} 

The coarse-grained Hall electric field has two contributions, a resolved part  
\be \frac{1}{nec} \bar{\bJ}_\ell\btimes \bar{\bB}_\ell \ee
and a Hall contribution to the subscale electric field
\be \barepsilon_\ell^{{\rm Hall}} = \frac{1}{nec} \left[\overline{(\bJ\btimes\bB)_\ell}-\overline{\bJ}_\ell\btimes\overline{\bB}_\ell\right]. \ee 
The resolved Hall field can be neglected relative to $\frac{1}{c}\bar{\bu}_\ell\btimes\bar{\bB}_\ell,$ since 
$ \bar{\bu}_\ell$ is much larger than $\bu_\ell^{{\rm Hall}}\equiv \bar{\bJ}_\ell/ne.$ In fact, using the previous 
estimate that $\delta \bar{J}_\ell\sim c \delta B(\ell)/{4\pi\ell_\perp},$ one finds that
\be \bar{u}_\ell^{{\rm Hall}}/\bar{u}_\ell \sim \left(\frac{v_A}{\bar{u}_\ell}\right) \left(\frac{\delta_i}{\ell}\right)
\left(\frac{\delta B(\ell)}{B_0}\right), \ee
where $v_A=B_0/\sqrt{4\pi m_i n}$ is the Alfv\'en speed based on the mean magnetic field $B_0$ and $\delta_i=
c(m_i/4\pi ne^2)^{1/2}$ is the ion skin depth.  In the turbulent inertial range $\delta B(\ell)/B_0\ll1.$ The 
ratio $v_A/\bar{u}_\ell$ is a constant factor, which is less than one in super-Alfv\'enic flow such 
as the solar wind at 1 AU. Finally, at inertial-range lengths $\ell\gg \delta_i,$ the other factor is also very small.

The subscale Hall term can be estimated as in \cite{Eyinketal11} by using the previously stated vector calculus 
identity (\ref{vec-calc-id}) to write 
\be\barepsilon_\ell^{{\rm Hall}} = \frac{1}{4\pi ne}\left[ \grad\bdot \btau^{{\rm Max}}_\ell -\frac{1}{2}\grad({\rm tr}\, \btau^{{\rm Max}}_\ell)\right]\ee
where $\btau^{{\rm Max}}_\ell= \overline{(\bB\bB)_\ell}-\overline{\bB}_\ell\overline{\bB}_\ell$ is the turbulent Maxwell stress tensor. 
Note that $\btau^{{\rm Max}}_\ell$ can be easily written in terms of magnetic field increments as
\begin{eqnarray} 
&& \!\!\!\!\!\!\!\!\!\!\!\!\!\!\!\!\!\!\!\!\!\!\!\! \tau^{{\rm Max}}_{\ell,ij} =  \int d^3r \, G_\ell(\br) \delta B_i(\br)\delta B_j(\br) \cr
&& \!\!\!\!\!\!\!\!\!\!\!\!\!\!\!\!\!\!\!\!\!\!\!\!\! -\int d^3r \, G_\ell(\br) \delta B_i(\br)\cdot  \int d^3r' \, G_\ell(\br') \delta B_j(\br'), \end{eqnarray} 
but furthermore so can its space-gradient:
\begin{eqnarray}
&&  \!\!\!\!\!\!\!\!\!\!\!\!\!\!\!\!\partial_k \tau^{{\rm Max}}_{\ell,ij} =  -\frac{1}{\ell} \left[\int d^3r \, (\partial_kG)_\ell(\br) \delta B_i(\br)\delta B_j(\br)\right.\cr
&& \!\!\!\!\!\!\!\!\!\!\!\!\!\!\!\!- \int d^3r \, (\partial_kG)_\ell(\br) \delta B_i(\br)\cdot  \int d^3r' \, G_\ell(\br') \delta B_j(\br')  \cr
&& \!\!\!\!\!\!\!\!\!\!\!\!\!\!\!\!\left. - \int d^3r \, G_\ell(\br) \delta B_i(\br)\cdot  \int d^3r' \, (\partial_kG)_\ell(\br') \delta B_j(\br')\right]. \cr
&& \,\,\,\,\,\,\, \,\,\,\,\,\,\, \,\,\,\,\,\,\, \,\,\,\,\,\,\, \,\,\,\,\,\,\, \,\,\,\,\,\,\, \,\,\,\,\,\,\, \,\,\,\,\,\,\, \,\,\,\,\,\,\,  \end{eqnarray}  
See Appendix \ref{cumulant}. This last identity gives the estimate 
\be \varepsilon_\ell^{{\rm Hall}} \sim \frac{1}{4\pi ne} \frac{(\delta B(\ell))^2}{\ell_\perp}. \ee
If possible dynamic alignment is ignored, then $ \varepsilon_\ell^{{\rm Hall}}\ll \varepsilon_\ell $ when 
\be \delta u(\ell) \gg  \frac{c}{4\pi ne} \frac{\delta B(\ell)}{\ell_\perp} \sim \frac{\delta_i}{\ell_\perp} \cdot \delta b(\ell), \ee
where ${\textit {\textbf  b}}=\bB/\sqrt{4\pi m_in}$ is the magnetic field in Alfv\'en velocity units. Since weakly compressible MHD-like 
turbulence in the solar wind consists mainly of shear-Alfv\'en waves with $\delta u(\ell)\sim \delta b(\ell),$ the above inequality  
is well satisfied for $L\gg \ell_\perp\gg \delta_i.$  Dynamic alignment, if it occurs, changes the result only slightly, with   
$\ell_\perp\gg \delta_i^{1/(1+\beta)} L^{\beta/(1+\beta)}\equiv \delta_i^*$ now required. While $\delta_i^*>\delta_i,$
nevertheless $\delta_i^*/L=(\delta_i/L)^{1/(1+\beta)}\ll 1,$ if $\beta$ is small as expected. We conclude that the Hall electric 
field contributions can all be neglected for $\ell_\perp$ in the MHD inertial range $\delta_i^*\ll \ell_\perp\ll L.$. This 
conclusion is in agreement with numerical simulations of Hall MHD turbulence, which find that the Hall term has 
negligible effects at scales greater than the ion skin depth \citep{DmitrukMatthaeus06}. 
 
\subsubsection{Electron pressure-tensor electric field} 

It is more straightforward to analyze the electric field arising from the electron pressure tensor because 
this contribution has an explicit overall gradient. Thus, 
\be \frac{1}{ne}\grad\bdot\bar{{\textbf P}}_{e,\ell}(\bx) = -\frac{1}{ne \, \ell} \int d^3r\, (\grad G)_\ell(\br) \bdot {\textbf P}_e(\bx+\br) \ee
contains at least one factor $\ell^{-1}.$ Notice that we have not replaced ${\textbf P}_e(\bx+\br)$ with the increment
${\textbf P}_e(\bx+\br)-{\textbf P}_e(\bx),$ because the electron pressure is a sub-inertial-range quantity with possible rapid 
variations on electron scales. Thus, ${\textbf P}_e(\bx+\br)-{\textbf P}_e(\bx)$ need not be any smaller than
${\textbf P}_e(\bx+\br)$. The size of the latter term is often estimated in reconnection discussions     
\citep{Vasyliunas75,Bhattacharjeeetal99,CraigWatson03} from the assumption that 
 \be P_e\sim P_i\sim n m_i u_{th}^2 \sim\frac{ne}{c} \delta_i u_{th}B_{th} \ee
with $B_{th}=(4\pi m_i n)^{1/2} u_{th}$, which implies that 
\be \frac{1}{ne}\grad\bar{P}_{e,\ell}(\bx) 
\sim \frac{P_e}{ne \, \ell} \sim \left(\frac{\delta_i}{\ell}\right) \frac{u_{th}B_{th}}{c}. \ee
The electric field from electron pressure gradient is thus suppressed by one factor of $\delta_i/\ell$ in an 
MHD-like inertial-range where $\ell\gg \delta_i,$ and can be neglected relative to $\barepsilon_\ell\sim 
\delta\bu(\bell)\btimes\delta\bB(\bell)/c.$ Note that the electron pressure tensor need not be isotropic
and the above estimate then applies to each individual component of the tensor. 

The above estimate may however be improved by more refined evaluation of the electron pressure, 
using the assumption of a strong ambient magnetic field $\bB_0$. In this limit simple gyrofluid models can be derived,
the lowest-order model of which type has often been employed in simulations of magnetic reconnection 
\citep{LoureiroHammett08,Klevaetal95,Grassoetal00}. Using this lowest-order gyrofluid model, 
\cite{BianKontar10} and \cite{Bianetal10} have derived for kinetic Alfv\'en  wave turbulence as in the solar wind that 
\be P_e = \frac{ne B_0}{c} \rho_s^2 \omega_{\|}  \ee
where $\rho_s=c_s/\Omega_{c,i}$ is the ion gyroradius calculated from the sound speed $c_s$ and the 
ion cyclotron frequency $\Omega_{c,i},$ and $\omega_{\|}$ is the component of the ion fluid vorticity
$\bomega=\grad\btimes\bu$ along the direction of the magnetic field $\bB_0$.  In this case, the coarse-grained electric field 
contribution becomes   
\begin{eqnarray} 
&& \!\!\!\!\!\!\!\!\!\!\!\!
\frac{1}{ne}\grad \bar{P}_{e,\ell} = -\frac{1}{c}\left(\frac{\rho_s}{\ell}\right)^2 
  \int d^3r\, \delta\bu(\br)\btimes \bB_0\bdot(\grad \grad G)_\ell(\br)\cr
&&  \,\,\,\,\,\,\, \,\,\,\,\,\, \sim\left(\frac{\rho_s}{\ell}\right)^2 
  \frac{1}{c}B_0\delta u(\ell). \end{eqnarray} 
Thus, the gyrofluid model predicts that the electron pressure contribution is actually suppressed by 
the factor $(\rho_s/\ell)^2$ and is even smaller than implied by the first estimate.   

It is worth emphasizing here that, except for the last estimate, all of our coarse-graining analysis is more 
general than gyrokinetics, because we do not assume a mean magnetic $B_0$ stronger than the fluctuations. 
We have ignored  in this section fluctuations in the density (and temperature) for simplicity of presentation, but the analysis
of Appendix \ref{density} includes those effects. The main assumption that we have made which is not required
in gyrokinetics is that the electron-ion mass ratio is small, $m_e/m_i\ll 1,$ but this condition is always satisfied 
in the solar wind.
\footnote{Even this assumption is not strictly required for our analysis. The exact Generalized Ohm's Law 
for any two-species plasma of oppositely-charged ions and electrons, retaining terms of all orders in $m_e/m_i,$ is 
\begin{eqnarray}
&& \!\!\!\!\!\!\!\!\!\!\!\!\!\!\!\!\!\!\!\!\!\!\! \bE +\frac{1}{c}\bu\btimes \bB = \frac{1}{en}{\textit {\textbf  F}} + \frac{m_i}{m_i+m_e}\frac{\bJ\btimes\bB}{nec}\cr
&& \,\,\,\,\,\,\,\,\,\,-\frac{1}{ne}\grad\bdot \left( \frac{m_i {\textbf P}_e -m_e {\textbf P}_i}{m_i+m_e}\right) \cr
&&\!\!\!\!\!\!\!\!\!\!\!\!\!\!\!\!\!\!\!\!\!\!\!+\frac{m_e m_i}{(m_i+m_e) ne^2} \left[\frac{\partial \bJ}{\partial t}
+\grad\bdot\left(\bJ\bu+\bu\bJ-\frac{1}{ne}\bJ\bJ\right) \right], \end{eqnarray}
where $\bu$ is the ion fluid velocity and ${\textit {\textbf  F}}$ is the total drag force density between the two fluid species (including 
frictional drag due to relative motion and thermal drag). Our analysis can be applied to this equation even in an extreme 
limit of a non-relativistic electron-positron plasma where the ``ion'' is the positron with $m_i=m_e.$ Note that we have
generally neglected the thermal drag, since it is expected to be small for large $B_0$, as in the solar wind. See
\cite{Schekochihinetal09}, section 4.2.} 

\subsubsection{First electron inertia contribution}

The first contribution to the coarse-grained electric field arising from electron inertia effects can be evaluated 
using the Maxwell equations to be 
\begin{eqnarray} \frac{m_e}{ne^2}\frac{\partial \bar{\bJ}_\ell}{\partial t} &=& c\frac{m_e}{4\pi ne^2}\grad\btimes \frac{\partial \bar{\bB}_\ell}{\partial t} \cr
&=& -c^2 \frac{m_e}{4\pi ne^2}\grad\btimes (\grad \btimes\bar{\bE}_\ell) \cr
&=& -\delta_e^2 \grad\btimes(\grad\btimes \bar{\bE}_\ell) \end{eqnarray} 
with $\delta_e$ the electron skin depth. Assuming quasi-neutrality, $\grad\bdot \bar{\bE}_\ell=0,$ and then one obtains the further 
simplification that $-\grad\btimes(\grad\btimes \bar{\bE}_\ell)=\triangle \bar{\bE}_\ell.$ In either case, there are now two 
space-gradients, leading to a suppression by $\ell^{-2}.$ More precisely, 
\begin{eqnarray} 
\frac{m_e}{ne^2}\frac{\partial \bar{\bJ}_\ell}{\partial t} &=& \left(\frac{\delta_e}{\ell}\right)^2
\int d^3r\, (\triangle G)_\ell(\br) \bE(\bx+\br) \cr
&\sim&  \left(\frac{\delta_e}{\ell}\right)^2 \bar{\bE}_\ell 
\end{eqnarray} 
and this is much smaller than $\bar{\bE}_\ell$ itself for $\ell\gg\delta_e.$

\subsubsection{Second electron inertia contribution}  

The second contribution  to the coarse-grained electric field arising from electron inertia effects has two terms, one 
factorized term and one cumulant term: 
\begin{eqnarray}
&&  \frac{m_e}{ne^2}\grad\bdot\overline{\left(\bJ\bu+\bu\bJ\right)_\ell} =
\frac{m_e}{ne^2} \grad\bdot\left(\bar{\bJ}_\ell\bar{\bu}_\ell+\bar{\bu}_\ell\bar{\bJ}_\ell\right) \cr
&& \,\,\,\,\,\,\,\,\,\,\,\,\,\, + \frac{m_e}{ne^2}\grad\bdot\left(\tau_\ell(\bJ,\bu)+\tau_\ell(\bu,\bJ)\right) \end{eqnarray} 
We denote by $\tau_\ell(f,g)=\overline{(fg)_\ell}-\bar{f}_\ell\bar{g}_\ell$ the 2nd-order cumulant from coarse-graining. 

The factorized term is easily estimated by writing 
\begin{eqnarray}
&&  \!\!\!\!\!\!\!\!\!\!\!\!\!\!
\grad\bdot\left(\bar{\bJ}_\ell\bar{\bu}_\ell+\bar{\bu}_\ell\bar{\bJ}_\ell\right) = 
(\bar{\bJ}_\ell\bdot\grad)\bar{\bu}_\ell \cr
&&  \,\,\,\,\,\,\,\,\,\,\,\,\,\,    + (\grad\bdot\bar{\bu}_\ell)\bar{\bJ}_\ell + (\bar{\bu}_\ell\bdot\grad)\bar{\bJ}_\ell \end{eqnarray}
and by using methods like those previously to show that 
\be \grad\bar{\bu}_\ell \sim \frac{\delta u(\bell)}{\ell}, \,\,\,\, \grad\bar{\bJ}_\ell \sim \frac{c}{4\pi}\frac{\delta B(\bell)}{\ell^2}. \ee 
Hence, 
\begin{eqnarray} 
&& \!\!\!\!\!\!\!\!\!\!\!\!\!\!\frac{m_e}{ne^2} \grad\bdot\left(\bar{\bJ}_\ell\bar{\bu}_\ell+\bar{\bu}_\ell\bar{\bJ}_\ell\right)
\,\,\,\,\,\,\,\,\,\,\,\,\,\,\,\,\,\,\,\,\,\,\,\,\,\,\,\,\cr
&& \!\!\!\!\!\!\!\!\!\!\!\!\!\!\sim \left(\frac{\delta_e}{\ell}\right)^2 \left[ \frac{2}{c}\delta u(\ell)\delta B(\ell)+ \frac{1}{c} u \, \delta B(\ell) \right] 
\end{eqnarray} 
and this electric field is small relative to $\barepsilon_\ell\sim \delta\bu(\bell)\btimes\delta\bB(\bell)/c$ for $\ell\gg \delta_e.$
 
The cumulant term is more complex.  As shown in Appendix \ref{cumulant} 
\begin{eqnarray}
&&  \!\!\!\!\!\!\!\!\!\!\!\!\!\!\grad\bdot\tau_\ell(\bJ,\bu) = -\frac{1}{\ell} \left[\int d^3r \ (\grad G)_\ell(\br) \bdot \delta \bJ(\br)\delta\bu(\br)\right. \cr
&& \!\!\!\!\!\!\!\!\!\!\!\!\!\!- \int d^3r \ (\grad G)_\ell(\br) \int d^3r' \ G_\ell(\br') \ \bdot \delta \bJ(\br)\delta\bu(\br') \cr
&& \!\!\!\!\!\!\!\!\!\!\!\!\!\!\left. -\int d^3r \ G_\ell(\br) \int d^3r' \ (\grad G)_\ell(\br') \bdot \delta \bJ(\br)\delta\bu(\br') \right]. \cr
&& \end{eqnarray} 
However, this yields only one factor $\ell^{-1}.$ Furthermore, $\bJ$ is a very rough field on MHD inertial-range 
scales and, in collisionless plasmas like the solar wind, even down to electron scales. The lack of any smoothness
at inertial-range separations $\br$ means that current increments are not small, but instead take on large 
values $\delta \bJ(\br)\sim \bJ\sim (c/4\pi) \delta B(\delta_e)/\delta_e$ for a collisionless plasma with magnetic fields 
rough down to the length-scale $\delta_e.$\footnote{There are both theoretical arguments \citep{Schekochihinetal09}
and empirical evidence in the solar wind \citep{Sahraouietal13} that the true ``inner scale'' for kinetic turbulence
in a well-magnetized but collisionless plasma is the electron gyroradius $\rho_e$ rather than the electron 
skin-depth $\delta_e$. In that case, we should really estimate $\bJ\sim (c/4\pi) \delta B(\rho_e)/\rho_e$. Since 
$\rho_e=\sqrt{\beta_e}\delta_e,$ use of this estimate will change our results in the text by some factors of 
$\beta_e.$ At least for the solar wind, $\beta_e\simeq 1$ rather generally and thus it is largely immaterial whether 
one uses $\delta_e$ or $\rho_e$ as an estimate of the inner length.} One thus obtains an estimate 
\be \frac{m_e}{ne^2}\grad\bdot\left(\tau_\ell(\bJ,\bu)+\tau_\ell(\bu,\bJ)\right) \sim 
\left(\frac{\delta_e}{\ell}\right) \cdot \frac{1}{c}\delta u(\ell)\delta B(\delta_e). \ee
This term is suppressed by only a single factor of $(\delta_e/\ell).$ On the other hand, it is also true in the 
solar wind and other similar cases of collisionless plasma turbulence that $\delta B(\delta_e)\ll \delta B(\ell)$
for $\ell\gg \delta_e,$ since $\delta B(\ell)\sim \ell_\perp^h$ with $h\simeq 1/3$ for $L\gg \ell\gg \delta_i$ and 
$h\simeq 2/3$ for $\delta_i\gg\ell\gg\delta_e$ \citep{Sahraouietal13}. Thus, there is suppression relative to $\barepsilon_\ell\sim 
\delta\bu(\bell)\btimes\delta\bB(\bell)/c$ by the total factor $\delta_e\delta B(\delta_e)/\ell\delta B(\ell)\ll1.$

%However, the two fields being coarse-grained over regions of radius $\ell$ live on different scales, with $\delta \bJ(\br)$
%varying with $\br$ over lengths $\delta_e$ and having the previous stated magnitude, whereas $\delta \bu(\br)$ 
%is varies slowly on scales $\delta_e$ and has maximum magnitude $\delta u(\ell).$ Thus, there should be large 
%cancellations in the integral over $\br.$ Using a heuristic approach borrowed from turbulence theory \cite{EyinkAliue09}, 
%one can estimate the cancellations by introducing a ``decorrelation factor'' $\delta_e/\ell$ associated to the ratios 
%between the scales. In this way one gets a heuristic estimate 
%\be \frac{m_e}{ne^2}\grad\bdot\left(\tau_\ell(\bJ,\bu)+\tau_\ell(\bu,\bJ)\right) \sim 
%\left(\frac{\delta_e}{\ell}\right)^2 \cdot \frac{1}{c}\delta u(\ell)\delta B(\delta_e) \ee
%which is likewise small relative to $\barepsilon_\ell\sim \delta\bu(\bell)\btimes\delta\bB(\bell)/c$ for $\ell\gg \delta_e.$   

\subsubsection{Third electron inertia contribution} 

The third contribution from electron inertia, 
$ -\frac{1}{n^2e^3c}\grad\bdot\overline{\left(\bJ\bJ\right)_\ell}, $
can be understood in a similar fashion to the cumulant term from the second contribution above. Noting that 
\begin{eqnarray} \grad\bdot\overline{\left(\bJ\bJ\right)_\ell} &=&  -\frac{1}{\ell} \int d^3r \ (\grad G)_\ell(\br) \bdot \bJ(\bx+\br)\bJ(\bx+\br) \cr
&\sim &\frac{1}{\ell} J^2 \end{eqnarray} 
and using the estimate $J\sim (c/4\pi) \delta B(\delta_e)/\delta_e,$ one finds that
\be -\frac{1}{n^2e^3c}\grad\bdot\overline{\left(\bJ\bJ\right)_\ell}\sim
\left(\frac{\delta_e}{\ell}\right) \cdot  \frac{1}{c}\frac{(\delta B(\delta_e))^2}{\sqrt{4\pi m_i n}}. \ee
If one assumes approximate equipartition of magnetic and velocity fields down to electron scales 
\citep{BianKontar10},  then
$\delta u(\delta_e)\sim \delta b(\delta_e)=\delta B(\delta_e)/\sqrt{4\pi m_i n}$ and the estimate becomes
\be -\frac{1}{n^2e^3c}\grad\bdot\overline{\left(\bJ\bJ\right)_\ell}\sim
\left(\frac{\delta_e}{\ell}\right) \cdot  \frac{1}{c} \delta u(\delta_e)\delta B(\delta_e)\ee
and this is smaller than $\barepsilon_\ell\sim 
\delta\bu(\bell)\btimes\delta\bB(\bell)/c$ by the factor $\delta_e \delta u(\delta_e)\delta B(\delta_e)
/\ell\delta u(\ell)\delta B(\ell)\ll1.$

\subsection{The contributions to $\grad\btimes\bar{\bE}_\ell$}\lb{curlE}  

Having completed our estimate of all the terms in the coarse-grained Generalized Ohm's Law, we now 
make a similar estimate of all the terms in the coarse-grained Faraday's law
\be \partial_t\bar{\bB}_\ell = -c\grad\btimes\bar{\bE}_\ell.\ee 
Fortunately, most of the work has already been done in our estimate of the contributions to $\bar{\bE}_\ell.$ 
As we shall now show, the contributions to $\grad\btimes\bar{\bE}_\ell$ differ from the corresponding 
contributions to $\bar{\bE}_\ell$ only by an additional factor of $\ell^{-1}$

For example, the resolved induction gives   
\be \grad\btimes(\bar{\bu}_\ell\btimes \bar{\bB}_\ell) = -(\bar{\bu}_\ell\bdot\grad)\bar{\bB}_\ell+
(\bar{\bB}_\ell\bdot\grad)\bar{\bu}_\ell -\bar{\bB}_\ell (\grad\bdot\bar{\bu}_\ell), \ee 
which yields the estimate
\be \grad\btimes(\bar{\bu}_\ell\btimes \bar{\bB}_\ell) \sim \frac{1}{\ell}u \delta B(\ell) + \frac{2}{\ell}B\delta u(\ell). \ee 
The subscale turbulent induction contribution is estimated from the identity 
\begin{eqnarray}
&& \!\!\!\!\!\!\!\!\!\!\!\!
c\grad\btimes \barepsilon_\ell = \frac{1}{\ell} \left[\int d^3r \ (\grad G)_\ell(\br) \btimes (\delta \bu(\br)\btimes \delta\bB(\br))\right. \cr
&& \!\!\!\!\!\!\!\!\!\!\!\!
- \int d^3r \ (\grad G)_\ell(\br) \int d^3r' \ G_\ell(\br') \ \btimes (\delta \bu(\br)\btimes \delta\bB(\br'))  \cr
&& \!\!\!\!\!\!\!\!\!\!\!\!
\left. -\int d^3r \ G_\ell(\br) \int d^3r' \ (\grad G)_\ell(\br')  \ \btimes (\delta \bu(\br)\btimes \delta\bB(\br')) \right]. \cr
&&
\end{eqnarray}
For a spherically symmetric filter kernel, $\grad G(r) =G'(r)\hat{\br}$ and thus 
\be c\grad\btimes \barepsilon_\ell \sim \frac{1}{\ell} 
\langle\hat{\bell}\btimes (\delta \bu(\bell)\btimes\delta \bB(\bell))\rangle_{{\rm ang}}. \ee
One can see that these are the dominant contributions to $c\grad\btimes\bar{\bE}_\ell.$

Indeed, all of the contributions to $c\grad\btimes\bar{\bE}_\ell$ from the non-ideal terms in the Generalized Ohm's Law
can be estimated with the help of the following general identities:
\be \partial_{i_1}\cdots\partial_{i_p}\bar{f}_\ell = \frac{(-1)^p}{\ell^p} \int d^3r 
\ (\partial_{i_1}\cdots \partial_{i_p} G)_\ell(\br) \, \delta f(\br), \lb{multiple} \ee
and 
 \begin{eqnarray} 
 && \!\!\!\!\!\!\!\!\!\!\!\!\partial_i\partial_j\tau_\ell(f,g) =  \frac{1}{\ell^2} \left[\int d^3r \, (\partial_i\partial_j G)_\ell(\br) \delta f(\br)\delta g(\br)\right.\cr
&&  \!\!\!\!\!\!\!\!\!\!\!\!- \int d^3r \int d^3r' \,(\partial_i \partial_jG)_\ell(\br)G_\ell(\br') \, \delta f(\br) \delta g(\br')  \cr
&&  \!\!\!\!\!\!\!\!\!\!\!\!- \int d^3r \int d^3r'  \, G_\ell(\br) (\partial_i\partial_jG)_\ell(\br') \, \delta f(\br) \delta g(\br')\cr
&& \!\!\!\!\!\!\!\!\!\!\!\!-\int d^3r  \int d^3r'  \, (\partial_i G)_\ell(\br) (\partial_jG)_\ell(\br') \, \delta f(\br) \delta g(\br') \cr
%&& \,\,\,\,\,\,\,\,\,\,\,\,\,\,\,\,\,\,\,\,\,\,\,\,\,\,\,\,\,\,\,\,\,\,\,\,\,\,\,\,
&& \!\!\!\!\!\!\!\!\!\!\!\! \left.  -\int d^3r \int d^3r'  \, (\partial_jG)_\ell(\br) (\partial_i\partial_jG)_\ell(\br') \, \delta f(\br) \delta g(\br')\right]. \cr
&&
\lb{twice} \end{eqnarray} 
These and other such identities invoked previously can be checked by tedious calculation or derived more 
conveniently by a cumulant generating function technique (\cite{Eyink-Notes} and Appendix \ref{cumulant}). 
We leave it to the reader to check  with these identities that the curl of each of the non-ideal contributions to 
$\bar{\bE}_\ell$ is modified in magnitude simply by an additional factor of $\ell^{-1}.$ Hence, the relative magnitude 
of all of the terms is unchanged and the ideal contributions remain the largest to the curl $c\grad\btimes\bar{\bE}_\ell.$ 

From this fact we draw two important conclusions. 
First, the ideal induction equation is the leading-order dynamical description in the coarse-grained or ``weak'' sense for length-scales
$\ell$ in the turbulent inertial range. This is a very different statement, however, than saying that the coarse-grained 
variables  $\bar{\bu}_\ell, \bar{\bB}_\ell$ satisfy the ideal equation in the naive sense. The latter statement overlooks the 
contribution of the turbulent subscale electric field $\bR^T_\ell=-\barepsilon_\ell$ which appears as an apparent 
``non-ideal'' term in the coarse-grained Ohm's law but which arises in fact from ideal turbulence dynamics. 
Second, the leading-order contribution to the slip source $-(\grad \btimes \bar{\bR}_\ell)_\perp/|\bar{\bB}_\ell |$ for lengths in the 
turbulent inertial range $\ell$, is the ideal MHD term $(\grad \btimes \barepsilon_\ell)_\perp/|\bar{\bB}_\ell |.$ Despite ideal Ohm's
law holding at those length-scales $\ell$ in the ``weak'' or coarse-grained sense, field-lines of the coarse-grained magnetic field 
$\bar{\bB}_\ell$ are {\it not} ``frozen-in'' and the line-slippage at those scales is due to ideal turbulence physics 
rather than to non-ideal plasma effects.  

At length-scale $\ell$ in the inertial range, the slip-velocity acquired along a length 
$\ell$ of field-line will have magnitude $~|\delta\bu(\bell)\btimes\delta\bB(\bell)|/|\bar{\bB}_\ell|$. Note that this is 
consistent on order of magnitude with a ``turbulent $\bE\btimes\bB$-drift velocity'' given by ${\textit {\textbf v}}_{{\rm slip},\ell}=-c\barepsilon_\ell\btimes
\widehat{\bar{\bB}}_\ell/|\bar{\bB}_\ell|$ \citep{EyinkAluie06}, which will allow lines of the coarse-grained magnetic field $\bar{\bB}_\ell$
to drift relative to the coarse-grained plasma fluid velocity $\bar{\bu}_\ell.$  Here let us just note the exact decomposition 
of the turbulent emf 
\be c \barepsilon_\ell = \alpha_\ell \bar{\bB}_\ell + {\textit {\textbf v}}_{{\rm slip},\ell}\btimes\bar{\bB}_\ell, \ee
with $\alpha_\ell=c\barepsilon_\ell\bdot\widehat{\bar{\bB}}_\ell/|\bar{\bB}_\ell |$ a pseudo-scalar field with dimensions
of velocity that measures inverse cascade of magnetic helicity/dynamo action. When $\alpha_\ell\ll v_{{\rm slip,\ell}},$
one can assign a velocity $\bar{\bu}_\ell+{\textit {\textbf v}}_{{\rm slip},\ell}$ to magnetic field-lines which is 
dynamically consistent with the induction equation \citep{Newcomb58}. This approach to turbulent-induced slip 
of field-lines is therefore less general than the slip-velocity source in section \ref{slip}, but it is more practical to
apply to the limited (one-dimensional) information available from most single-spacecraft observations of the solar wind,
as we see below.  

\section{Implications for Heliospheric Reconnection}\lb{implications}

There are important physical implications of the results presented in this paper. We consider here briefly 
just a few of them.  

\subsection{Breakdown of the Parker Spiral} 

We begin with the \cite{Parker58} spiral model of the interplanetary magnetic field, 
which is one of the most famous applications in astrophysics and space science of the ``frozen-in'' principle 
for magnetic field-lines. The model has been shown to be approximately valid when taking into account 
solar cycle variations in source magnetic field strength and latitude/time variation in solar wind speeds.
Using yearly-averaged magnetic field strengths calculated from observations of Voyager 1 and 2 in the period 
1978-2001 at solar distances 1-81 AU and comparing with Parker's model for estimated magnetic field strength 
and wind speed on a source surface at 1 AU, \cite{Burlagaetal02} found a mean deviation of only -2\% between 
observations and model. On the other hand, typical yearly deviations were around $\pm 20$\% and the 
maximum error (in 1993) was about -40\%. These sizable yearly deviations may be partly due to inaccuracies 
in inputs or errors in observations of the magnetic field, but some part is also clearly due to inaccuracy of the model. For example, there is a clear correlation 
of the relative error with the solar cycle, with the greatest deviations at years of solar maximum (see Figure 4 of 
\cite{Burlagaetal02}). One should recall that \cite{Parker58} concluded his paper with a ``warning to the reader 
against taking too literally any of the smooth idealized models which we have constructed in this paper". 

Whereas the study of \cite{Burlagaetal02} considered only magnetic field strengths, the earlier work 
of \cite{Burlagaetal82} had studied the magnetic vector geometry and found ``notable deviations" from the 
spiral model. \cite{Burlagaetal82} studied daily averages of magnetic field observations of Voyager 1 and 2
in the ecliptic plane at solar distances $R=$1-5 AU during a period of increasing solar activity in the years 1977-1979.
In contrast to the Parker predictions for radial magnetic field component radial dependencies $B_R\sim R^{-2}$ and azimuthal 
component $B_T\sim R^{-1},$ \cite{Burlagaetal82} found $B_R\sim R^{-1.56}$ and $B_T\sim R^{-1.20}.$ 
They contrasted their findings with earlier Pioneer 10, 11 investigations between 1 and 8.5 AU during the period 
1972-1976 when the sun was less active, which confirmed the spiral model modulo 
variability on time scales shorter than a solar rotation period. \cite{Burlagaetal82} attributed the observed deviations
``to temporal variations associated with increasing solar activity, and to the effects of 
fluctuations of the field in the radial direction.''

These early observations were recently confirmed by \cite{KhabarovaObridko12}, who presented evidence 
on the breakdown of the Parker spiral model for time- and space-averaged values of the magnetic field 
from several spacecraft (Helios 2, Pioneer Venus Orbiter, IMP8, Voyager 1) in the inner heliosphere at solar 
distances 0.3-5 AU and in the years 1976-1979. This study thus significantly overlapped with that of 
\cite{Burlagaetal02}, but used time averages over longer periods (from 76 days to 2 years) coupled with 
space averages over intervals of width up to 1 AU. For details of the averaging, see Table 1
of  \cite{KhabarovaObridko12}. This more extensive averaging eliminated sizable fluctuations still
observed in the daily averages of \cite{Burlagaetal82} (see their Figure 1). The study of \cite{KhabarovaObridko12}
is in essential agreement with \cite{Burlagaetal82}, finding dependencies $B_R\sim R^{-1.66}$ and $B_T\sim R^{-1.10},$ 
just a bit closer to the Parker predictions. \cite{KhabarovaObridko12} interpret their observations as due to 
``a quasi-continuous magnetic reconnection, occurring both at the heliospheric current sheet and 
at local current sheets inside the IMF sectors''. They present extensive evidence that most 
nulls of $B_R$ and $B_T,$ where reconnection may occur, are not associated to the 
heliospheric current sheet. They as well observe a rapid disappearance of the regular sector structure 
at distances past 1 AU, which they attribute to ``turbulent processes in the inner heliosphere.''
See also \cite{Robertsetal05,Roberts10}. 

This interpretation is 
consistent with our results, since line-slippage due to pervasive turbulence in the near-ecliptic solar wind 
will lead to a less tightly wound spiral and a stronger radial field-strength than in the Parker model, as observed 
by \cite{KhabarovaObridko12}. This is essentially the same phenomenon as the ``reconnection diffusion'' 
proposed by \cite{Lazarian05} and \cite{Santos-Limaetal10}, as a mechanism of removal of magnetic fields 
from collapsing molecular clouds in star formation, and equivalent to the ``turbulent $\bE\btimes\bB$-drift'' of \cite{EyinkAluie06}.  
The magnitude and direction of the drift velocity at length-scale $\ell$ is given by 
\be {\textit {\textbf v}}_{{\rm slip},\ell}= -c\barepsilon_\ell\btimes
\widehat{\bar{\bB}}_\ell/|\bar{\bB}_\ell|. \lb{EA06} \ee
As discussed in section \ref{epsilon}, this can be estimated on order of magnitude from the approximate 
expression 
\begin{eqnarray}
 {\textit {\textbf v}}_{{\rm slip},\ell} &\sim &-(\delta\bu(\bell)\btimes\delta\bB(\bell))\btimes \widehat{\bar{\bB}}_\ell/|\bar{\bB}_\ell|^2 \cr
&\sim& \delta u(\ell)\delta B(\ell)/B_0\sim 0.5 (\delta B(\ell)/B_0)^2 v_A, \cr
&& \lb{vslip-LV99} 
\end{eqnarray}
with $v_A$ the Alfv\'en speed based on $B_0.$  The second line follows by ignoring any suppression due to dynamic alignment effects.
It is worth observing that the latter speed $v_{{\rm slip},\ell}$ 
given by equation (\ref{vslip-LV99}) is numerically identical\footnote{The physical meaning is somewhat different, however, 
because  ${\textit {\textbf v}}_{{\rm slip},\ell}$ does not correspond to a plasma inflow speed into a reconnection region.
Instead it describes the speed of effective motion of magnetic field-lines relative to the resolved plasma velocity.}  
to the turbulent reconnection speed at perpendicular scale $\ell$ in the theory of \cite{LazarianVishniac99}. 
Here we have assumed that $\delta u(\ell)/v_A \sim 0.5 \ \delta B(\ell)/B_0,$ as is often observed in the inertial range. 
If $\delta B(\ell)/B_0\simeq 1$ for large $\ell$, one can expect $v_{{\rm slip}}\sim v_A$.

To give more precise estimates, we appeal to observations. We use two sets of {\it Ulysses} data from the ``COHOWeb''  
data sets compiled by J. King at the National Space Science Data Center (NSSDC). One set of data is from polar-latitude 
fast wind near solar minimum in 1995, on days 100-200 when {\it Ulysses} moved between distances 1.36-2.02 AU and heliospheric 
latitudes $21.3^\circ - 80.2^\circ$. The other set is from near-ecliptic pure slow wind near solar maximum in 1992, on days 
151-159 when the spacecraft passed from 5.34 AU to 5.38 AU and from latitude $-10.11^\circ$ to $-13.14^\circ.$ 
These two extreme cases should give a good idea of the range of variation of the turbulent drifts. The {\it Ulysses} 
magnetic field data has 1 sec resolution but the plasma data has only 4 min resolution. This is sufficient 
for our purposes, since we are primarily interested in large lengths $\ell.$ As usual, we can employ Taylor's hypothesis 
to interpret time-increments $\tau$ as space-increments $\ell,$ by the formula $\ell=\bar{u}\tau$ where $\bar{u}$
is the mean solar wind speed.  This should work well for the very slowly evolving large-scale features that are our 
principal concern. As a matter of fact, all of our formalism carries over also to time-averaging \citep{Germano92}, so 
that we will present the coarse-grained quantities in terms of time variables. The bulk-averaged properties of the two 
solar wind cases are presented in Table 1.

\vspace{10pt}

%\begin{minipage}
\hspace{-27pt} \scalebox{.9}{
\begin{tabular}{llll}
\multicolumn{4}{c}{{\textbf Table 1}} \\
\multicolumn{4}{c}{Parameters for the Two Solar Wind Datasets} \\
\hline
\hline
Parameter & high-speed &  low-speed  \\
\hline
velocity magnitude $u$ (km$\, {\rm sec}^{-1}$) & 770 & 430 \\
radial velocity $u_R$  (km$\, {\rm sec}^{-1}$) & 769 & 429 \\
tangential velocity $u_T$ (km$\, {\rm sec}^{-1}$) & 19.4  & 5.69  \\
normal velocity  $u_N$ (km$\, {\rm sec}^{-1}$) & -0.657 & 4.57  \\ 
\hline
field strength $B$ (nT)  & 2.41 & 0.951 \\
radial field $B_R$  (nT) & 1.24 & 0.110  \\
tangential field $B_T$ (nT) & -0.386 & 0.093 \\
normal field $B_N$ (nT) & -0.0438 & -0.135\\ 
\hline
ion density $n_i\,\,\,\, ({\rm cm}^{-3})$ & 0.973 & 0.341 \\
ion temperature $T_i \,\,\,\, (10^4 \,\, {\rm K})$ & 20.5 & 3.78 \\
Alfv\'en speed $v_A$ (km$\, {\rm sec}^{-1}$) & 53.2 & 43.1 \\
ion gyroradius $\rho_i$  (km) & 252 & 274 \\
ion plasma beta $\beta_i$ & 1.20 & 0.493 \\
\hline
& & & \\
\end{tabular}
}
%\end{minipage}
\newpage

\begin{figure}[!ht]
\begin{center}
%\begin{tabular}{cc}
%{\it high-speed} & {\it low-speed} \\
%\includegraphics[width=100pt,height=100pt]{Bstrucfun-High.pdf} & 
%\includegraphics[width=100pt,height=100pt]{Bstrucfun-Low.pdf}\\
%\includegraphics[width=100pt,height=100pt]{Vstrucfun-High.pdf} & 
%\includegraphics[width=100pt,height=100pt]{Vstrucfun-Low.pdf}\\
%\end{tabular} 
\includegraphics[width=210pt,height=210pt]{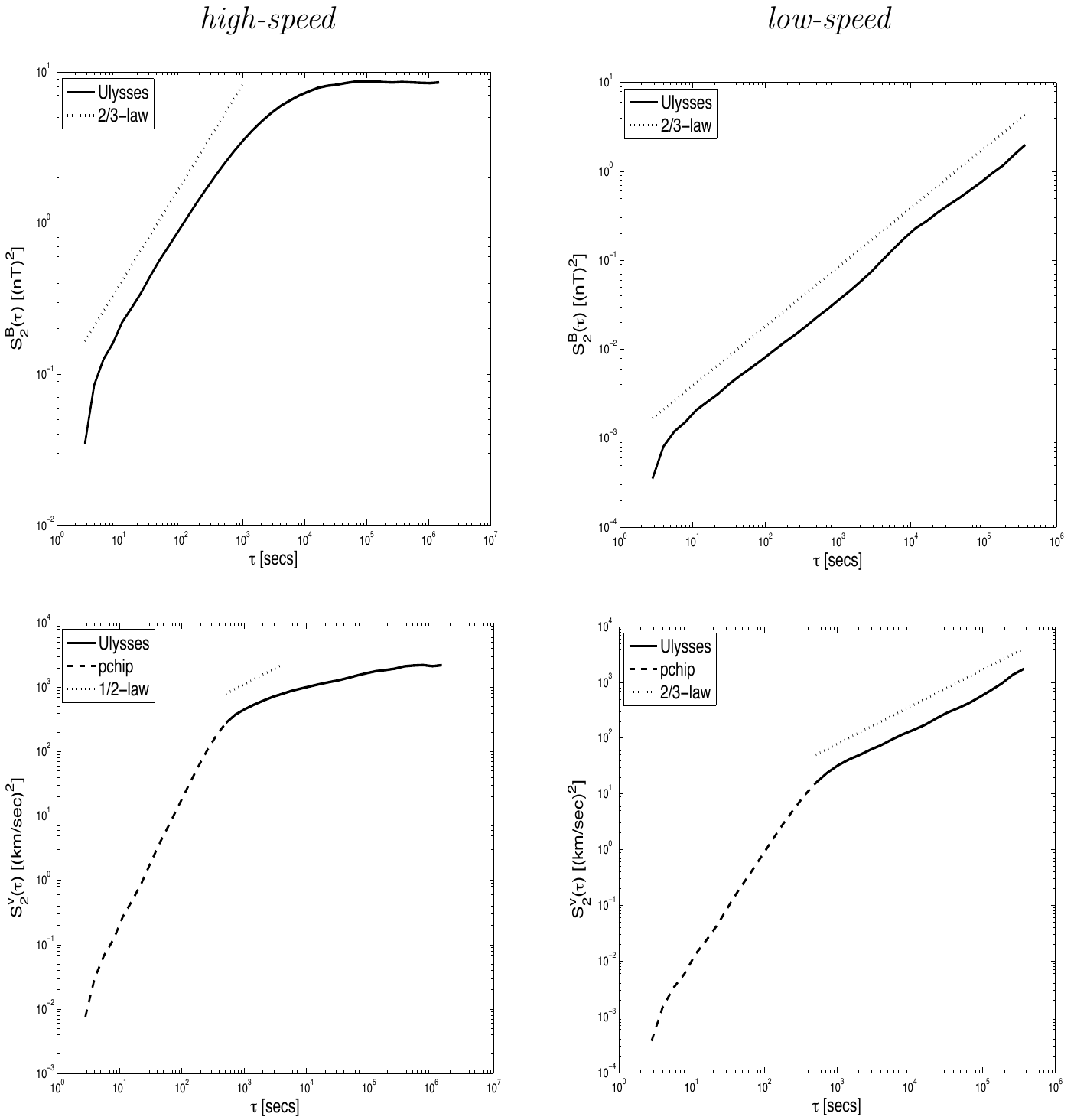}
\end{center}
\caption{{\it Second-Order Structure Functions}. Top Panels: Magnetic structure functions.
The solid line plots the {\it Ulysses} data, the dotted line is a 2/3 power law. Bottom Panels:
Velocity structure functions. The solid line is again the raw data. The dotted line for the 
high-speed case on the left is a 1/2 power, while the dotted line for the low-speed case on the 
right is a 2/3 power. The dashed lines are the results below 4 mins resulting from the 
pchip interpolation of the velocity field.}
\end{figure}\lb{strucfns}

We next present results for the second-order structure-functions of time increments
\be  S_2^X(\tau) = \overline{ |{\textit {\textbf  X}}(t+\tau)-{\textit {\textbf  X}}(t)|^2} \ee 
with ${\textit {\textbf  X}}=\bB,\bu$ and overline $\overline{()}$ denoting time-average over the 
entire interval of data. For the purposes of later evaluation of the slip velocity via 
formula (\ref{EA06}) we interpolate the 4 min plasma velocity data to the 1 sec 
grid of the magnetic data using piece-wise cubic Hermite interpolating polynomials
(pchip). Although there is no physical significance of the interpolated velocity data at times less
than 4 min, we plot their structure functions down to 1 sec time separations in order 
to show the effects of the \ interpolation. The structure functions plotted in Figure 2 show the typical 
features of high-speed and low-speed solar wind around 1 AU. 
The low-speed solar wind exhibits a Kolmogorov-type 2/3 power-law scaling for both 
magnetic and velocity fields over the entire range of available times, \ consistent with 
the theory of \cite{GoldreichSridhar95}. The magnetic field data begin to bend over 
to a steeper scaling at $\tau$ a few seconds, around the ion gyro-period. 
$\,$ The velocity field structure function if accurately measured to 1 sec resolution would presumably
have similar behavior, but  the interpolated data lead to a steeper decay with a power of 2
due to the smooth polynomial.  The high-speed wind structure functions are quite similar to 
those for the low-speed solar wind at times less than about $10^4$ sec. The primary difference
in that range is that the scaling exponent for the velocity structure function appears closer to 1/2
than 2/3, as often observed in the high-speed wind near 1 AU \citep{Podestaetal07}, although the precise 
determination here is impossible since only about a decade exists between 4 minutes and the 
outer time, which is somewhat less than $10^4$ sec. The main difference between high-speed and 
low-speed wind appears above that outer time, where the high-speed wind shows a flattening 
of both the magnetic and the velocity structure functions. This corresponds to the so-called
``1/{\it f} range'' of scales \citep{MatthaeusGoldstein86}, which is believed to be largely a mixture 
of a noninteracting, antialigned population of Alfv\'en waves and magnetic force-free structures
ejected from the sun \citep{Wicksetal13a,Wicksetal13b}. Hereafter we discuss the low- and
high-speed cases separately. 

\subsubsection{Low-Speed Case}\lb{low}

We first plot the time series for the magnetic and velocity fields in the slow-speed case. See 
Fig.~3. From examination of the data for tangential component of the magnetic field, one can
see that there are sector crossings during day 152, late on day 153 and a major crossing 
during the whole of day 157. The last is an example of a complex, broadened heliospheric current 
sheet (HCR), of the type which has been shown to occur with increasing likelihood at greater distances 
from the sun and under conditions of solar maximum \citep{Robertsetal05}. Note that, not only does the 
tangential component of the magnetic field undergo a large reversal in sign on day 157, but also
the other two components reverse sign. Fig.~4, which plots other plasma properties, shows additional 
signatures of the HCS on day 157, such as sizable increases in density and temperature and 
different levels of magnetic field strength be- 

\begin{figure}[!ht]
\begin{center}
%\begin{tabular}{c}
%\includegraphics[width=400pt,height=125pt]{Bcomponents-Low.pdf}\\
%\includegraphics[width=400pt,height=125pt]{Vcomponents-Low.pdf}\\
%\end{tabular} 
\includegraphics[width=210pt,height=135pt]{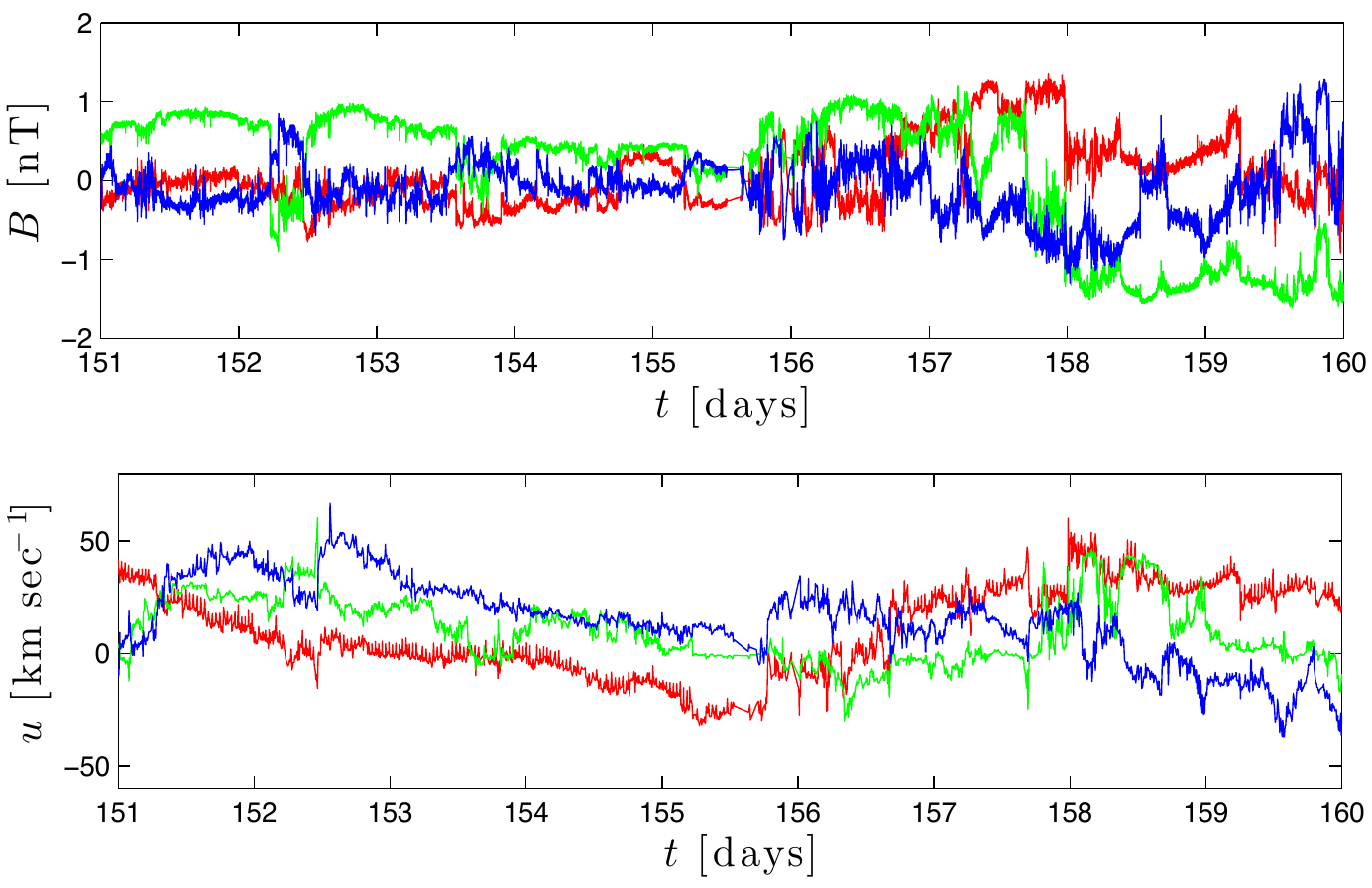}
\end{center}
\caption{{\it Magnetic and Velocity Components of Slow Wind Time-Series}. The upper panel plots the
magnetic field components and the lower panel velocity field components. The \textcolor{red}{red} line 
represents radial component, \textcolor{green}{green} tangential, and \textcolor{blue}{blue} normal,
in the standard RTN coordinate system. The plotted radial component of the velocity is the fluctuation 
field $u_R'=u_R-\bar{u}_R,$ with the average value $\bar{u}_R=$769 km ${\rm sec}^{-1}$ subtracted.}
\end{figure}

\begin{figure}[!ht]
\begin{center}
%\begin{tabular}{c}
%\includegraphics[width=400pt,height=125pt]{Bstrength-Low.pdf}\\
%\includegraphics[width=400pt,height=125pt]{density-Low.pdf}\\
%\includegraphics[width=400pt,height=125pt]{temperature-Low.pdf}\\
%\end{tabular} 
\includegraphics[width=210pt,height=200pt]{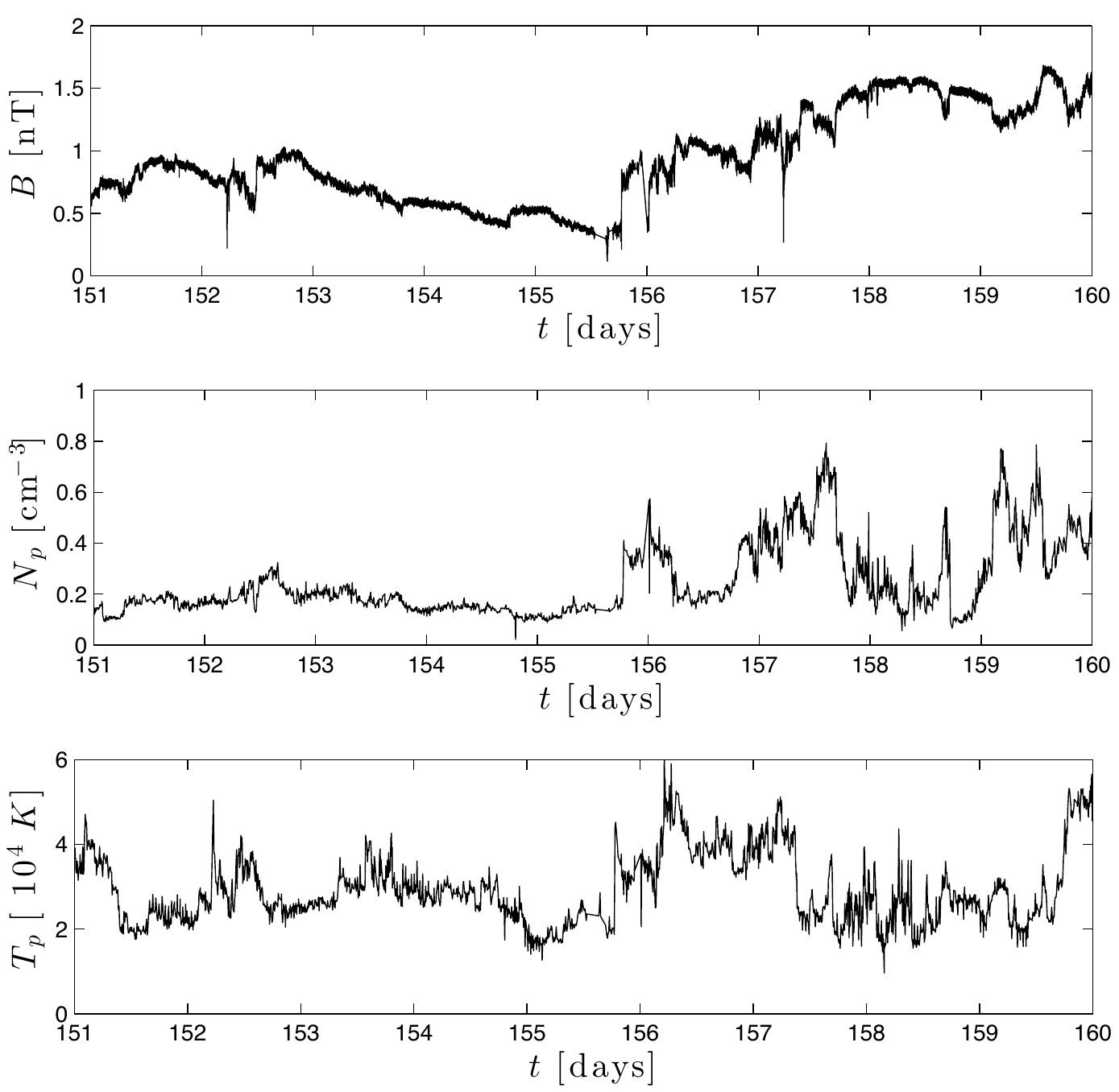}
\end{center}
\caption{{\it Field Strength, Density and Temperature of Slow Wind Time-Series.} Upper 
panel: field strength, middle panel: density, and bottom panel: temperature.}
\end{figure}

\noindent fore and after the transition. 

A close examination of the tangential velocity components in Fig.~3 shows also apparent veloc-

\begin{figure}[!ht]
\begin{center}
%\begin{tabular}{c}
%\includegraphics[width=400pt,height=250pt]{BTcomponent-sm.pdf}\\
%\includegraphics[width=400pt,height=250pt]{VTcomponent-sm.pdf}\\
%\end{tabular} 
\includegraphics[width=210pt,height=245pt]{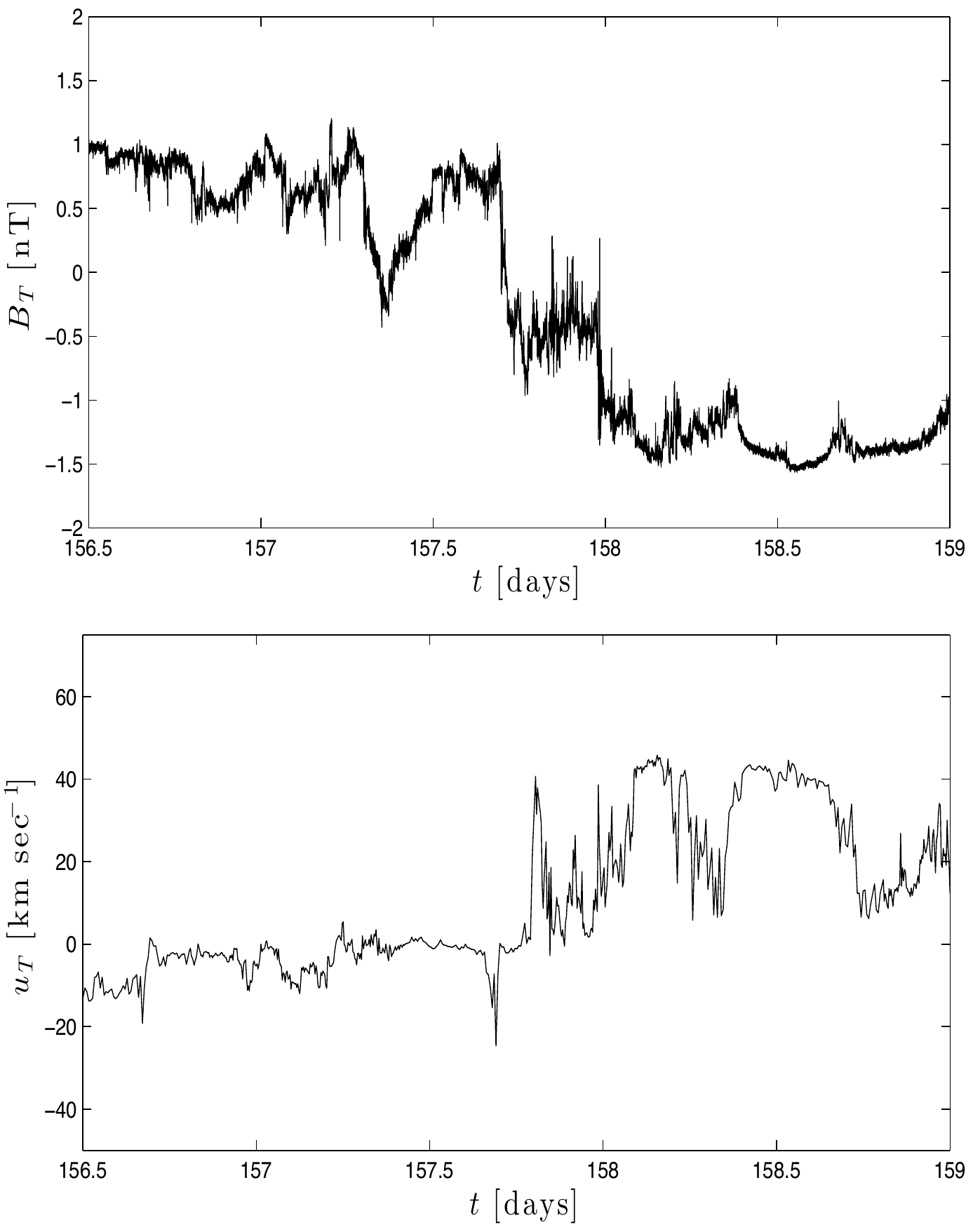}
\end{center}
\caption{{\it Tangential Magnetic Field and Velocity at Broad Heliospheric Current Sheet.} Upper panel:
tangential component of magnetic field. Lower panel: the tangential component of plasma velocity.}
\end{figure}

\noindent 
ity jets associated with each of the three sector crossings, in  particular at the HCS. To examine 
the latter more carefully, we show a closer view of the tangential components of magnetic and 
velocity fields across the HCS in Fig.~5. The reversal of the tangential component of the magnetic 
field is particularly broad, extending over days 157-158.5. The associated exhaust 
is a bit narrower, covering days 157.75-158.75. Notice that the speed of this broad outflow is 
about 40 km ${\rm sec}^{-1}$, just a bit less than the local upstream Alfv\'en speeds. This seems
to be a likely example of \cite{LazarianVishniac99} turbulent reconnection, associated to the 
embedding of the HCS in a strongly turbulent environment. Notice that the observed value of 
$B_{rms}/\bar{B}$ away from the HCS itself is about 0.5. Furthermore, the plasma fluid within the 
HCS is itself strongly turbulent, as required by the \cite{LazarianVishniac99} theory. Fig.~6 plots 
the structure function of the magnetic field inside the HCS, averaging over days 157-158.5. There is a 

\begin{figure}[!ht]
\begin{center}
\includegraphics[width=210pt,height=180pt]{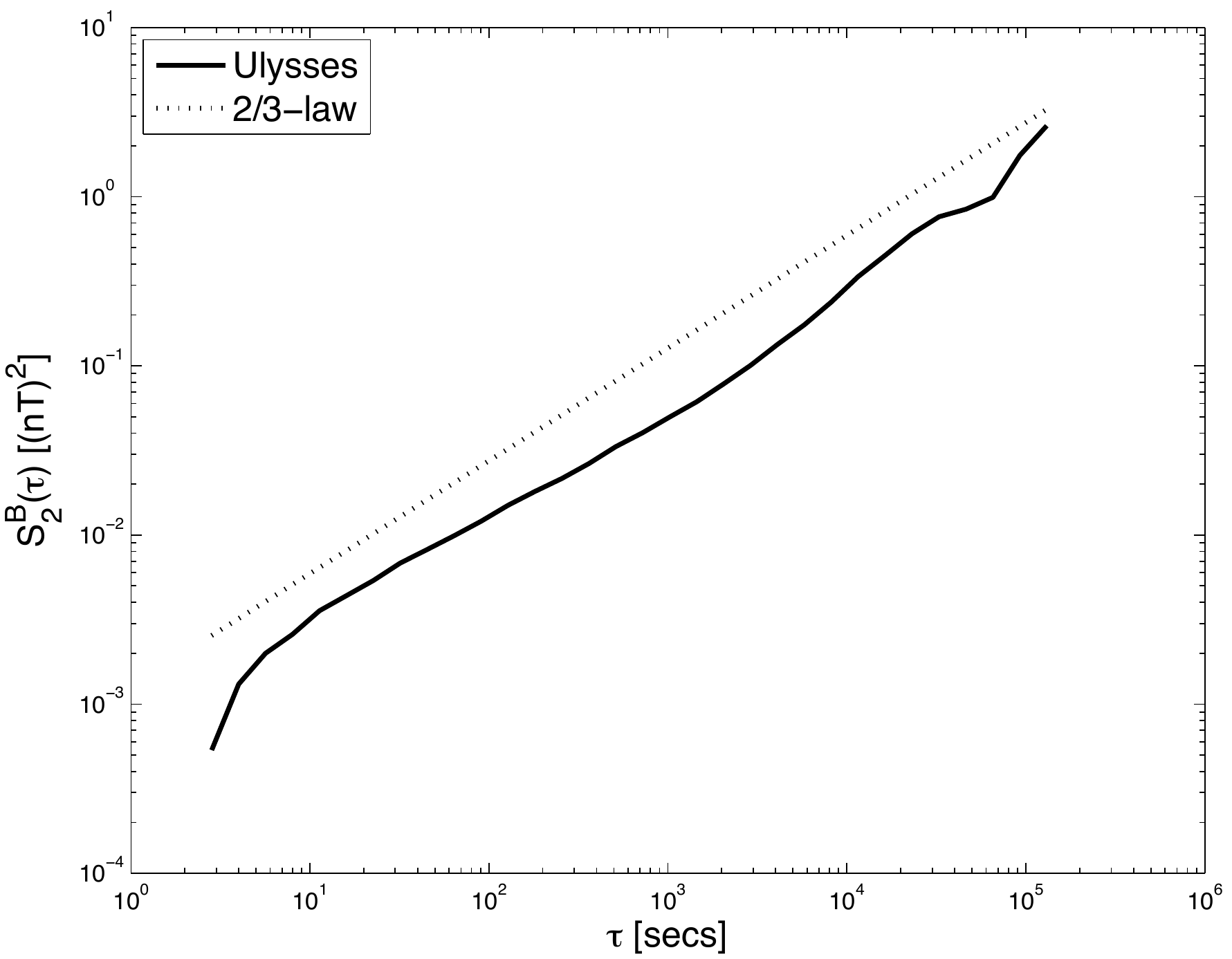}
\end{center}
\caption{{\it Magnetic Structure Function inside the Broad Heliospheric Current Sheet.} Same as in Fig.2,
top right panel, but defined by an average only over days 157-158.5 inside the HCS.}
\end{figure}

\noindent 
clear 2/3-law scaling, covering about 4 decades from $10-10^5$ secs. Finally, we 
note that our coarse-graining analysis in section \ref{CG-GOL} shows that at the length-scales of this event,
of order an AU, all microscopic plasma non-idealities are completely irrelevant. Turbulent reconnection 
seems to be the only plausible explanation.  

We next turn to the evaluation of the formula (\ref{EA06}) for the turbulent slip velocity. Since the full three-dimensional
data is unavailable to perform low-pass filtering, we follow standard practice in experimental turbulence studies 
(e.g. \cite{BrunoCarbone05}, Section 7; \cite{Stolovitskyetal98}) and use filtering on the time axis as a one-dimensional 
surrogate. To carry out the required coarse-graining, we apply a box-filter to the observational time series, with the filter 
half-width identified with coarse-graining time-scale $\tau.$ As discussed earlier, we use pchip interpolation of the velocity  
to define the pointwise cross product $\bu\btimes\bB.$ We plot in Fig.~7 the  instantaneous turbulent drift velocities 
for the filter half-width $\tau=$2 days. It is immediately obvious that there is a large tangential component of the drift 
velocity associated with the HCS, from day 156-158, with a maximum slip speed of nearly 40  km ${\rm sec}^{-1}$. 
The plots for other choices of half-width $\tau$ (not shown) are \ similar, \ as long as $\tau$ is greater than \ 1 day. 

\begin{figure}[!ht]
\begin{center}
%\begin{tabular}{c}
%\includegraphics[width=400pt,height=220pt]{driftV-tau4day-Low.pdf}\\
%\includegraphics[width=400pt,height=220pt]{alpha-tau4day-Low.pdf}\\
%\end{tabular} 
\includegraphics[width=210pt,height=225pt]{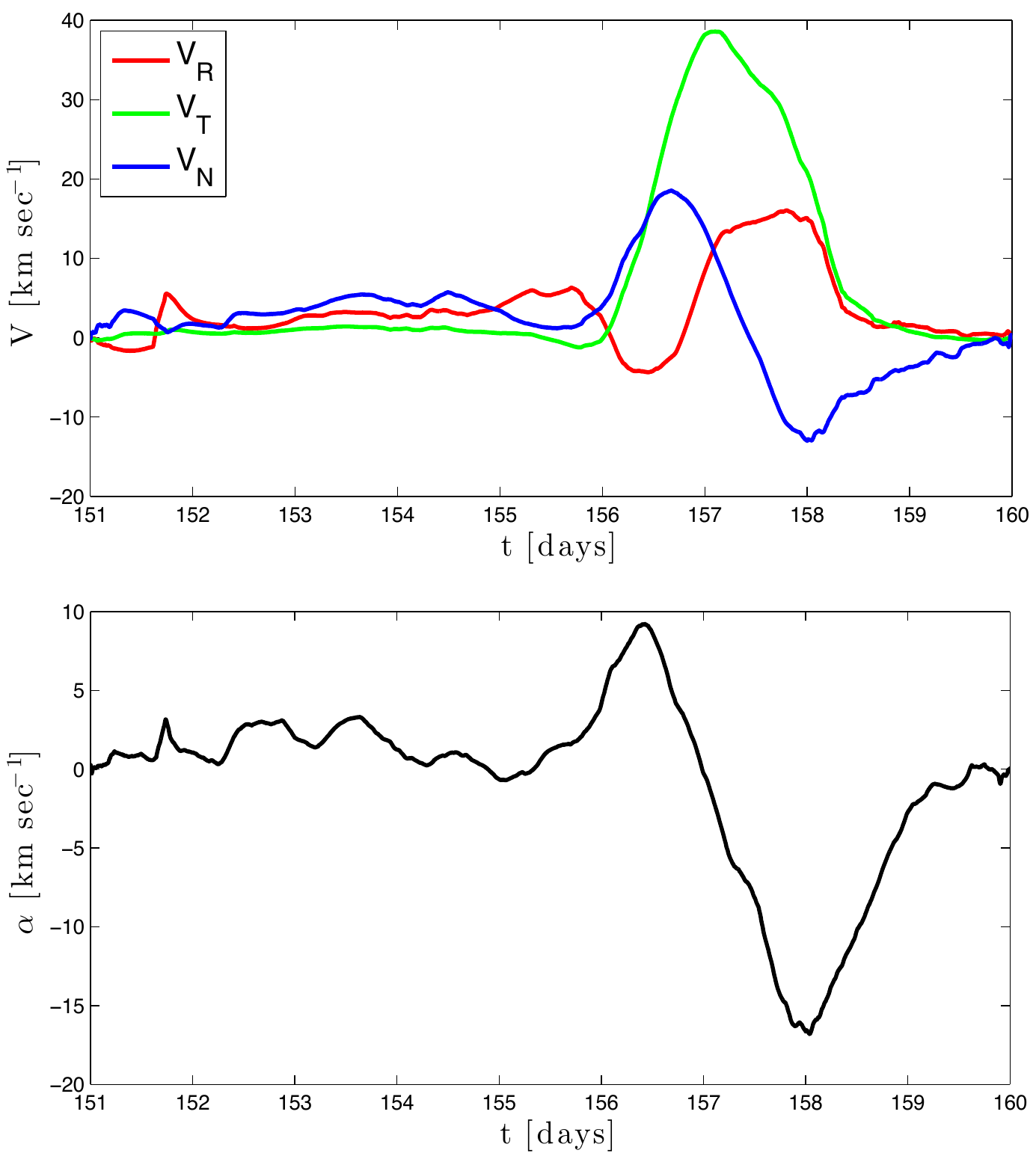}
\end{center}
\caption{{\it Turbulent Drift Velocity and Alpha Parameter for Low-Speed Time-Series.} Top panel:
drift velocity for filter half-width $\tau=2$ days. The color scheme is as in Fig.3. Bottom 
panel: the $\alpha$ parameter for the same half-width $\tau=2$ days.}
\end{figure}

\noindent
The peak tangential drift velocity increases with $\tau$ over the range that we can study\footnote{For $\tau>2,$
we begin to see edge-effects in filtered quantities at the location of the HCS, due to finite duration of the time series.}, 
with peak values of $v_{{\rm slip},T}^{{\max}}=$2.21, 14.9, 29.0, and 38.6 km ${\rm sec}^{-1}$, for $\tau=$0.5, 1.0, 1.5, 2.0 days, 
respectively. There is no observable ``jet" of tangential slip velocity at the HCS 
for $\tau=0.5,$ but the ``jet'' exists and becomes broader for increasing $\tau\geq 1.$
%\footnote{
As remarked at the end of section \ref{coarse}, the net velocity of the field lines is in fact the sum of the 
resolved plasma velocity and the drift velocity at that scale. These total tangential speeds of magnetic field-lines are even a bit larger, with 
maximum values in the HCS of $\bar{u}_{T}+v_{{\rm slip},T}=$28.6, 18.9, 34.8, and 44.0 km ${\rm sec}^{-1}$, 
for $\tau=$0.5, 1.0, 1.5, 2.0 days, respectively.
%}. 

The most important observation is that the largest component of the slip velocity is the tangential component, 
which has a positive sign. With the standard definition of RTN coordinates, this slip velocity is in the direction 
of the solar rotation. Hence, this slippage will tend to make 

\begin{figure}[!ht]
\begin{center}
%\begin{tabular}{c}
%\includegraphics[width=400pt,height=180pt]{driftV-tau2day-High.pdf}\\
%\includegraphics[width=400pt,height=180pt]{driftV-tau4day-High.pdf}\\
%\end{tabular} 
\includegraphics[width=210pt,height=170pt]{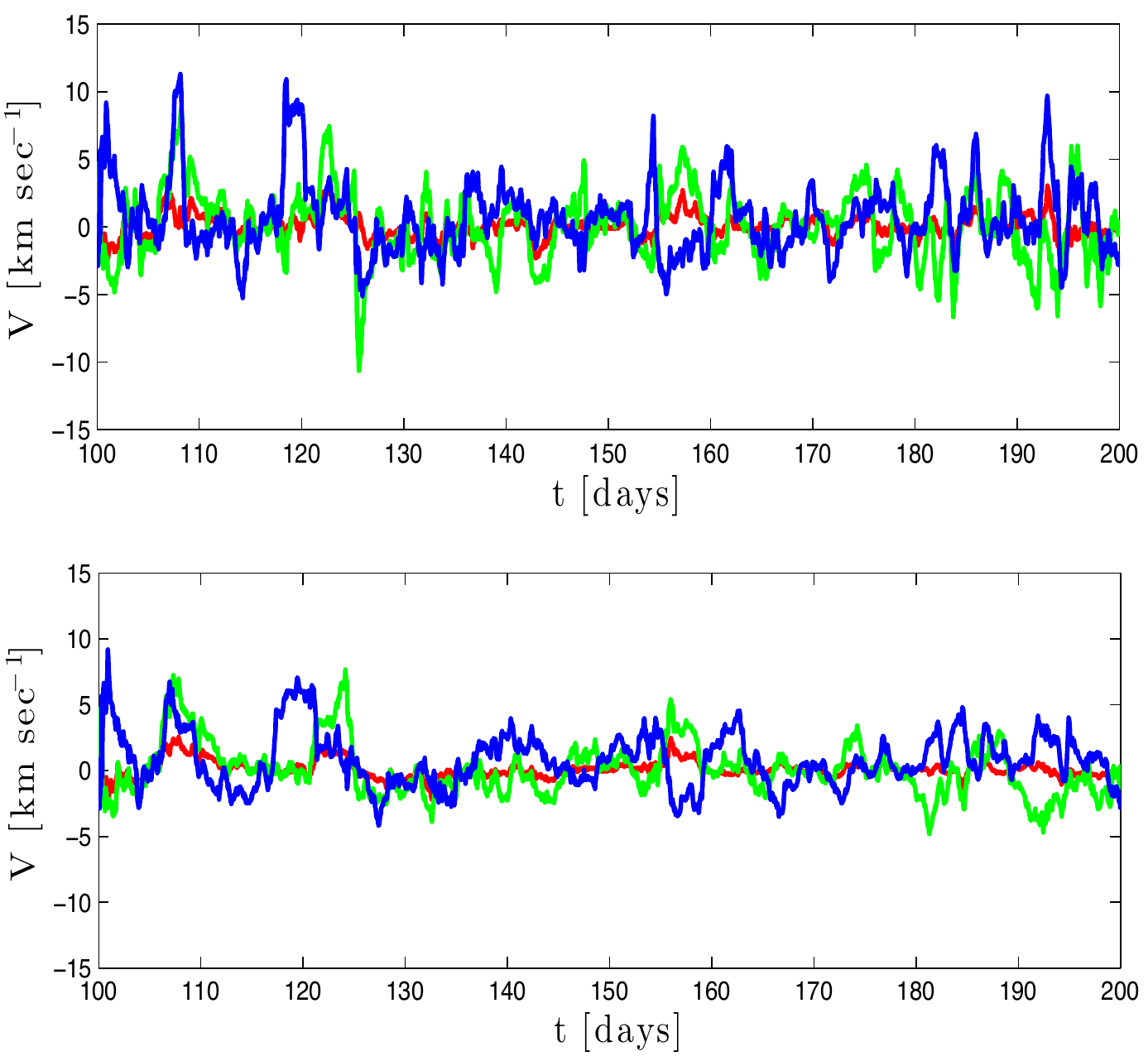}
\end{center}
\caption{{\it Turbulent Drift Velocity for High-Speed Time-Series at Two Scales.} Upper panel: components 
of turbulent drift velocities for filter half-width $\tau=1$ day. Lower panel: components of drift 
velocity for half-width $\tau=2$ days. The color scheme is as in Fig.3.}
\end{figure}

\noindent 
the field-lines less frozen into the plasma fluid and 
rotate more with the sun, so that the spiral will become less tightly wound. This is in qualitative agreement with 
the observations of \cite{Burlagaetal82} and \cite{KhabarovaObridko12}. For consistency, we should check that 
the $\alpha$ parameter is small relative to the drift velocity. As can be seen from  Fig.~7, this is marginally 
true, with $\alpha\sim (1/3)v_{{\rm slip}}$ for $\tau=2$ days. The fact that the $\alpha$ parameter is not 
completely negligible suggests that there is some magnetic dynamo action/inverse cascade of magnetic 
helicity in the HCS. We should note also that the tangential drift velocity is positive away from the HCS
(except briefly at the end of day 155) and takes on typical values of about 1 km ${\rm sec}^{-1}$.  The minor 
sector crossings on days 152 and 153 presumably contribute to such drifts. Such background turbulent slippage
will also contribute to the observed deviations from the Parker spiral model.  

\subsubsection{High-Speed Case}\lb{high}

The situation for the high-speed wind case is quite different. We plot in Fig.~8 the turbulent 

\begin{figure}[!ht]
\begin{center}
%\begin{tabular}{c}
%\includegraphics[width=400pt,height=150pt]{driftV-tau4day-High-sm.pdf}\\
%\includegraphics[width=400pt,height=125pt]{Bcomponents-High-sm.pdf}\\
%\includegraphics[width=400pt,height=125pt]{Vcomponents-High-sm.pdf}\\
%\end{tabular} 
\includegraphics[width=210pt,height=210pt]{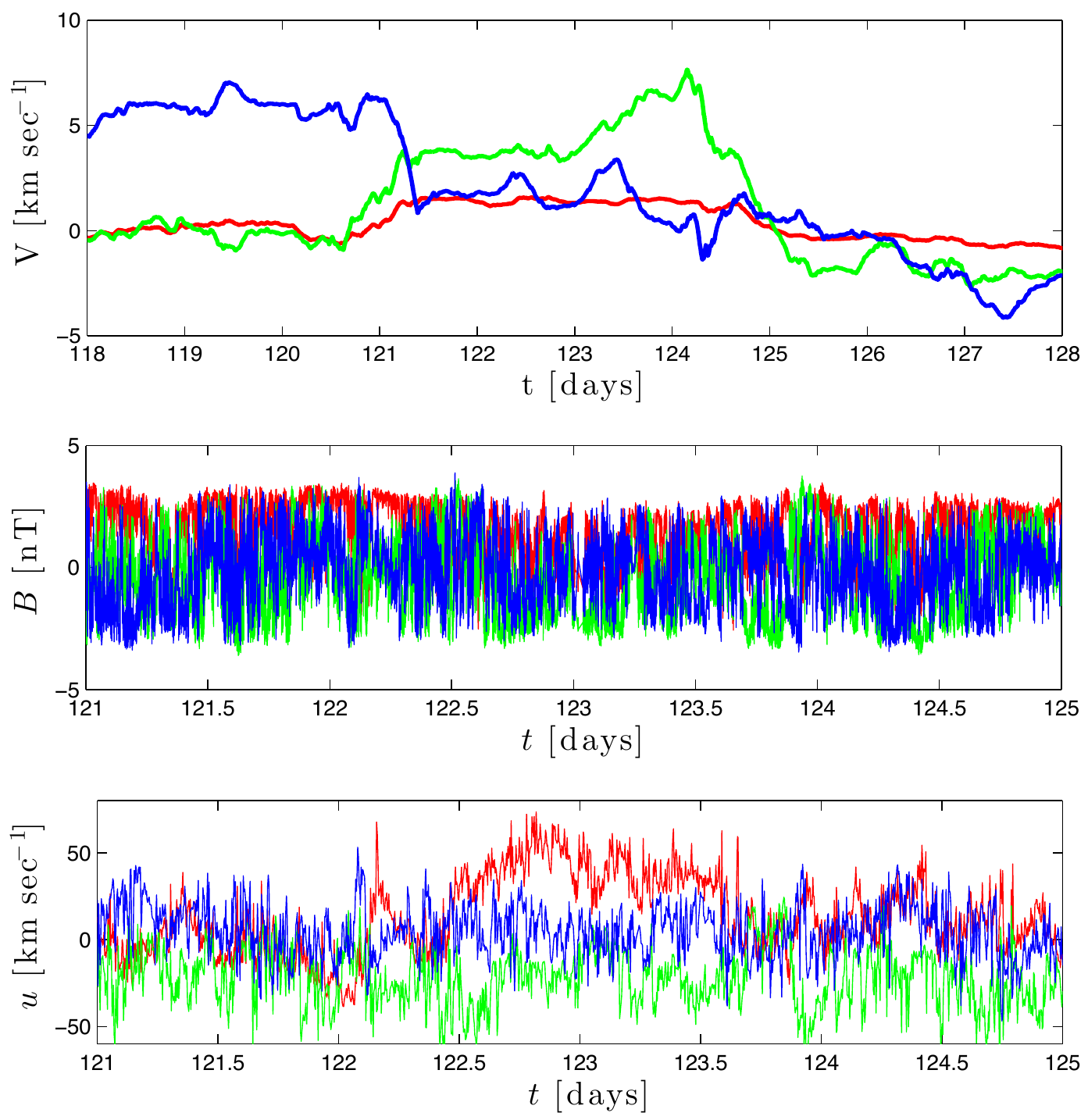}
\end{center}
\caption{{\it Close-up View of High-Speed Wind Data for Large Tangential Drift Region.} Top panel: turbulent 
drift velocities for filter half-width $\tau=2$ days over the smaller interval of days 118-128. Middle panel: 
components of magnetic field for days 121-125. Bottom panel: components of velocity field for days 
121-125. The color scheme for all panels is as in Fig.3, and as there the plotted radial component in the 
bottom panel is $u_R',$ the fluctuation field.}
\end{figure}

\noindent slip velocities
over the entire 100 days, for the two filter half-widths $\tau=1$ day and $\tau=2$ days. The first observation 
is that the peak values are only about 10 km ${\rm sec}^{-1}$, one-quarter of the peak for the slow wind case. 
Second, the largest component is the normal component, not the tangential. Finally, the peak magnitudes 
{\it decrease} going from $\tau=1$ day to $\tau=2$ days, unlike for the slow-wind case where there was a 
sizable increase. There is in fact a slight increase in the mean values over the whole interval, with $[\bar{v}_R,\bar{v}_T,
\bar{v}_N]=[0.0639,\ 0.0878,\ 0.6360]$ km ${\rm sec}^{-1}$ for $\tau=1$ day, but $[\bar{v}_R,\bar{v}_T,\bar{v}_N]
=[0.0981,\ 0.0782,\  0.7841]$ for $\tau=2$ days. However, the extreme values are clearly decreased. 

There are at least two plausible explanations for this decrease. One reason may be the strong alignment of 
$\delta \bB(\bell)$ and $\delta \bu(\bell)$ in the ``1/{\it f}-range'' of the high-speed wind, which leads to a large 
depletion of the slip velocity when $\ell$ increases in that range. Note indeed that \cite{Wicksetal13a,Wicksetal13b} find the non-aligned, nonlinearly 
cascading modes in the ``1/{\it f}-range'' have about an order of magnitude smaller energy than the aligned, 
non-interacting modes.  The second reason has to do with the atypicality of very large-scale reconnection events
of conventional type in the high-speed wind \citep{Gosling07}. In the entire 100 days of high-speed data, we see no 
AU-scale events of the type which occurred in the low-speed wind at the HCS. If we focus on the regions of 
peak tangential drift velocity in the high-speed wind, we see a very different structure. Fig.~9 zooms in on such
a peak around day 123 for $\tau=$ 2 days. As can be seen, the magnetic field components appear to the 
eye rather structureless and stochastic. On the other hand, there is clear structure in the velocity field, with 
the radial velocity fluctuation $u_R'$ around the mean value 769 km ${\rm sec}^{-1}$ increased over a 1-day interval around day 123. 
The peak in the slip velocity at this point appears to result therefore from the interaction between two extensive
high-speed streams, with radial velocities differing by about 50 km ${\rm sec}^{-1}$, and not from any 
large-scale magnetic structure.  

\subsubsection{Discussion} 

Our purpose in this section was to use observational data on the solar wind to make a first estimate of the 
size, directionality, and origin of turbulent slip velocities in that environment. These effects appear promising 
to explain the observed deviations from the Parker spiral in the inner heliosphere, but to do so quantitatively 
will require more extensive studies. We hope to carry these out in the near future. Several open issues remain to be 
clarified, such as the relative importance in the slow wind of background turbulence drift and of the larger turbulent 
slips at the HCS. The accuracy of one-dimensional surrogates of three-dimensional filtering should be 
investigated in numerical simulations of MHD turbulence, etc.  On the other hand, this theory already accounts 
naturally for the smaller deviations from the spiral model observed in the high-speed solar wind than in low-speed,  
and in solar minimum conditions than in solar maximum.  The peak values of slip velocity in the slow wind at solar 
maximum are about 10\% of the mean plasma velocity, whereas for the high-speed wind at solar minimum they are 
only about 1\% (and even less in the tangential direction). Furthermore, our coarse-graining analysis of the Generalized 
Ohm's Law in section \ref{CG-GOL} implies that, for the very long-time averages employed by \cite{Burlagaetal82} 
and \cite{KhabarovaObridko12}, alternative microscopic plasma mechanisms of line-slippage are irrelevant and 
cannot explain the observed deviations. 

\subsection{Other Large-Scale Heliospheric Reconnection}

In addition to the breakdown of the Parker model considered above, another type of deviation has been documented 
in daily averages of Voyager 1 observations in the heliosheath between 2007-2011, when the spacecraft was about 
110 AU from the Sun \citep{Richardsonetal13}. At this distance, the Parker model predicts that the magnetic field 
is almost entirely tangential, $B=B_T,$ and magnetic flux conservation implies that $u_R B R$ should be constant 
in $R$ \citep{Parker63}. \cite{Richardsonetal13} interpreted their observations of a ``flux deficit'' in terms of kinetic-scale magnetic reconnection, 
due to ion-scale current sheets created by compression of magnetic sectors in the heliosheath. 
However, these  observations may also be explained by magnetic reconnection in the heliosheath due to MHD 
turbulence, as earlier suggested by \cite{LazarianOpher09}. This would lead to reconnection 
when the sector widths were still much greater than required by kinetic mechanisms. Note that there is direct evidence
for MHD-like turbulence with a -5/3 energy spectrum of magnetic fluctuations in the sectored region of the heliosheath, 
from Voyager 1 data in the year 2009  (\cite{BurlagaNess10}, Figure 8), within the time period where reconnection
appears to occur. 

Finally, many magnetic reconnection events of conventional nature have been apparently directly observed in the solar wind at solar distances
around 1 AU, as documented in a recent  review of \cite{Gosling12}. These are more extreme events than 
background, gradual slippage and involve a sizable release of magnetic energy. Small-scale current sheets with widths 
of order the ion inertial length that exhibit observable reconnection usually have exhausts at most a few hundred times wider 
and frequently have small shear angles/strong guide fields \citep{Goslingetal07, GoslingSzabo08}. However, there is also 
a sizable number of very large-scale reconnection events in the solar wind, some of them associated with interplanetary coronal 
mass ejections and magnetic clouds or occasionally magnetic disconnection events at the heliospheric current sheet 
\citep{Phanetal09, Gosling12}. These events have reconnection outflows with widths up to nearly $10^5$ ion inertial lengths 
and appear to be in a prolonged, quasi-stationary regime with reconnection lasting for several hours. Gosling interprets
these events in terms of the \cite{Petschek64} reconnection model although, as he points out himself, there are
observations not in strict agreement with that model. This interpretation is also theoretically puzzling since numerical studies \citep{Biskamp86,UzdenskyKulsrud00} 
have shown that Petschek reconnection is not sustainable in laminar, MHD flows, but relaxes to the slow Sweet-Parker type. 
The belief that these events are Petschek-like seems to rest on the fact that  the Hall and electron inertia effects can 
stabilize such X-type reconnection in ion-scale geometries \citep{Shayetal98}. 

The scale of the apparent reconnection events documented by \cite{Gosling12} is so great, however,  
that they will still appear to be reconnection even if coarse-grained on inertial-range scales $\ell$ much greater than the 
ion gyroradius $\rho_i.$  At those scales, our analysis implies that microscopic non-ideal terms from Hall field and 
electron inertia are irrelevant. Such large-scale reconnection events must be due to the ideal turbulence effects 
in the solar wind environment rather than to plasma non-ideality. In fact, these events appear to be very promising 
candidates for turbulent reconnection as in the \cite{LazarianVishniac99} theory. It should be noted that the events observed 
by Gosling often show substantially increased proton densities in the exhausts, as well as enhancements of magnetic field strength 
and proton temperature, effects also seen in the HCS event analyzed in section \ref{low}. 
Whereas the coarse-graining method employed in section \ref{coarse} assumed constant density, 
the more general analysis carried out in Appendix \ref{density} is able to describe such effects.  

\vspace{1in} 

In any case, the discussion of the slip-source in section \ref{TGMR}, which does not assume incompressibility, 
already makes it clear that reconnection in the most general sense must occur ubiquitously throughout the turbulent 
solar-wind.

\acknowledgments

{\bf Acknowledgements}. I wish to thank E. Vishniac and A. Lazarian for many discussions, which have helped to form my understanding 
of turbulent magnetic reconnection. In particular, 
an initial suggestion of E. Vishniac triggered the present study. The solar wind data were all retrieved from the 
NSSDC at http://omniweb.gsfc.nasa.gov/coho, and we acknowledge the many people responsible for the provision of those data sets. The author's 
work was partially supported by NSF grant CDI-II: CMMI 0941530 and also by the Institute for Pure and Applied 
Mathematics at UCLA, where the paper was completed during the fall 2014 long program on ``Mathematics of Turbulence.'' 


%% in parentheses, after the keyword, but they will not be verified.

%{\it Facilities:} \facility{Nickel}, \facility{HST (STIS)}, \facility{CXO (ASIS)}.



\appendix

\section{Density Variations}\lb{density}

Previously we ignored variations of density in our discussion of the Generalized Ohm's Law.
This would be a serious omission for the solar wind, where the observations of apparent reconnection events 
by Gosling and colleagues \citep{Gosling12} show frequently sizable enhancements of density and proton temperature 
in the outflows. The interpretation of Gosling for these events is that they are \cite{Petschek64} reconnection with the 
outflows bounded by slow-mode shocks. In addition to these density variations during reconnection, there are also 
known to be turbulent fluctuations of plasma density in the solar wind. A paper of \cite{Bellamyetal05} 
contains recent observations at solar distances 1-60 AU and a good review of previous studies. 
The spectrum of turbulent density fluctuations has two power-law ranges, a low-wavenumber part with a $k^{-2}$ 
spectrum and a high-wavenumber part with a $k^{-5/3}$ spectrum. \cite{Bellamyetal05} interpret 
the $k^{-2}$ spectrum as due to abrupt discontinuities on the time-scales of a few hours and up, which are 
left over by their data selection procedure, which removes obvious shocks. The high-wavenumber $k^{-5/3}$
spectrum of density fluctuations is in roughly the same range of wavenumbers as the similar inertial-range spectrum 
of magnetic fluctuations. A popular interpretation is that it is a passive scalar cascade of slow magnetosonic
modes driven by the nonlinear cascade of shear-Alfv\'en modes \citep{Schekochihinetal09}. There is 
evidence from observations that density fluctuations in the solar wind consist mainly of slow-modes with
possibly a tiny admixture of fast modes \citep{Howesetal12}. 

As we show now, the extension of our analysis of the Generalized Ohm's Law carries over easily to 
plasma flows with arbitrarily large variations of density (including shocks). The  simplest way to do 
so is to make use of the density-weighted, spatial-coarse-graining {\it Favre average} \citep{Favre69}, 
which is defined by the formula 
\be  \tilde{f}_\ell = \frac{1}{\bar{n}_\ell}\overline{(n f)_\ell}.  \ee
With this density-weighted averaging operation, it is straightforward to obtain the coarse-grained form of the 
Generalized Ohm's Law to be
\be \tilde{\bE}_\ell +\frac{1}{c}\tilde{\bu}_\ell\btimes \tilde{\bB}_\ell = -\tilde{\barepsilon}_\ell +
   \widetilde{\left(\eta\bJ\right)_\ell} 
+ \frac{1}{\bar{n}_\ell ec}\overline{(\bJ\btimes\bB)_\ell}
-\frac{1}{\bar{n}_\ell}\grad\bdot{\overline{\textbf P}}_{e,\ell}+\frac{m_e}{\bar{n}_\ell e^2}\left[\frac{\partial \bar{\bJ}_\ell}{\partial t}
+\grad\bdot\overline{\left(\bJ\bu+\bu\bJ-\frac{1}{ne}\bJ\bJ\right)_\ell} \right], \ee
where the turbulent electric field is now given by
\be \tilde{\barepsilon}_\ell = -\frac{1}{c}\left[\widetilde{(\bu\btimes\bB)_\ell}-\tilde{\bu}_\ell\btimes \tilde{\bB}_\ell\right], \ee
the Favre-average 2nd-order cumulant. 

The first and most important observation is that, other than the Ohmic contribution, all of the non-ideal 
terms in the coarse-grained Generalized Ohm's Law are identical to those obtained in section \ref{CG-GOL},
except that they are multiplied by $1/\bar{n}_\ell$ rather than $1/n.$  The only difference to the old estimates 
is thus that they are multiplied by an additional factor
\be \frac{n}{\bar{n}_\ell} = 1 - \frac{1}{\bar{n}_\ell} \int d^3r\ G_\ell(\br) \delta n(\br). \ee
This is generally $\approx 1$ and is an order unity factor even in the presence of shocks in the density. 
Here is a good point to observe that all of our conclusions in section \ref{CG-GOL} remain valid in 
the presence of shocks. We generally assumed K41 scaling $\delta \bB(\br)\sim r_\perp^{1/3},$
$\delta n(\br)\sim r^{1/3}$ or something similar, but at shock points these increments become nearly 
independent of the separation vector magnitude  $r,$ until $r$ falls below the width of the shock. However,
all of our estimates are pointwise in space position $\bx$ and, even at points where increments are 
independent of $r,$ we always have extra factors of  $\delta_e/\ell$ to make 
the coarse-grained non-ideal terms small for $\ell\gg \delta_e.$  

Another observation is that the Favre-averaged quantities $\tilde{\bE}_\ell,\tilde{\bu}_\ell,\tilde{\bB}_\ell,$
and $\tilde{\barepsilon}_\ell$  are rather close in value to the ordinary coarse-grained quantities
$\bar{\bE}_\ell,\bar{\bu}_\ell,\bar{\bB}_\ell$ and $\barepsilon_\ell$. This can be seen from simple identities \citep{Aluie13},  
for the Favre average 
\be \tilde{f}=\bar{f}+\frac{1}{\bar{n}}\tau(n,f)=\bar{f} +  \frac{1}{\bar{n}}O\left( \delta n\ \delta f\right) \lb{Fav1} \ee
and for the Favre-average 2nd-cumulant $\tilde{\tau}(f,g)=\widetilde{fg}-\tilde{f}\tilde{g},$
\begin{eqnarray}
\tilde{\tau}(f,g)&=&\tau(f,g)+\frac{1}{\bar{n}} \tau(n,f,g) -\frac{1}{\bar{n}^2}\tau(n,f)\tau(n,g) \cr
                      &=& \tau(f,g)+\frac{1}{\bar{n}} O\left(\delta n\ \delta f\ \delta g\right) +
                             \frac{1}{\bar{n}^2} O\left( (\delta n)^2\ \delta f\ \delta g\right).  
\lb{Fav2} \end{eqnarray} 
For the order-of-magnitude estimates, see Appendix \ref{cumulant}. The first identity (\ref{Fav1}) yields
\be \tilde{\bB}_\ell = \bar{\bB}_\ell \left[1+ O\left( \delta n(\ell)\ \delta B(\ell)\right)\right] \simeq
 \bar{\bB}_\ell \left[1+ O\left(\frac{\ell}{L}\right)^{2/3}\right],  \ee
where the final estimate holds at typical points with K41 scaling of the increments and where $L$ is the outer 
scale of the inertial range. Exactly similar results hold for $\tilde{\bu}_\ell$ and $\tilde{\bE}_\ell$
by the same argument\footnote{Note that observations show that the electric field in the high-speed solar wind 
has an inertial-range energy spectrum of power-law form $k^{-3/2},$ similar to that observed for the 
velocity field \citep{Chenetal11}. This is quite reasonable, since the leading contribution to the electric 
field in the presence of a strong mean magnetic field $\bB_0$ is just $\bE=-\frac{1}{c}\bu\btimes\bB_0.$}. 
For the turbulent electric field one can likewise infer from (\ref{Fav2}) that 
\be \tilde{\barepsilon}_\ell = \barepsilon_\ell \left[1+O\left(\frac{\delta n(\ell)}{\bar{n}_\ell}\right)\right]
\simeq \barepsilon_\ell \left[1 + O\left(\frac{\ell}{L}\right)^{1/3}\right], \ee
with the final estimate holding again at typical points. We see that the Favre coarse-grained and ordinary
coarse-grained quantities are very similar when density variations are small relative to the mean, 
and of similar orders of magnitude when density variations are large (e.g. at shocks).   

The most troublesome term is, in fact, the Favre-averaged Ohmic electric field. Although 
the current is a total space-derivative, because of Ampere's law $\bJ=(c/4\pi)\grad\btimes\bB,$ 
the same is not true of the product of density and current, $n \bJ.$ Thus, unlike before, coarse-graining 
this product to obtain the Favre-average of the Ohmic electric field does not yield any powers of $\ell^{-1}$, 
which would make it irrelevant for increasing $\ell.$ Additional complicating $\bx$-dependences 
to the Ohmic electric field arise from the fact that resistivity is a tensor
$\boeta= \frac{1}{\sigma_\|} \hat{\bB}\hat{\bB} + \frac{1}{\sigma_\perp} \left({\textbf I}-\hat{\bB}\hat{\bB}\right) $
with a position-dependent director field $\hat{\bB}(\bx).$ Furthermore, the collisional conductivities 
$\sigma_\perp,\sigma_\|$ have very weak (logarithmic) dependence 
on density but strong dependence $\propto [T_e(\bx)]^{3/2}$ on the electron temperature. Luckily, the Ohmic 
electric field is tiny in the solar wind, even without coarse-graining. Using $\sigma\sim e^2 n\tau_{e,i}/m_e$, 
with $\tau_{e,i}$ the electron collision time with ions, $\lambda_{mfp,e}=v_{th,e} \tau_{e,i}$ the electron mean-free path,
and $J\sim c \delta B(\delta_e)/4\pi\delta_e,$ one finds easily 
\be \eta J\sim \left(\frac{\delta_e}{\lambda_{mfp,e}}\right) \cdot\frac{1}{c}v_{th,e} \delta B(\delta_e). \ee
Since the ratio $\delta_e/\lambda_{mfp,e}$ is $10^{-8}$  or smaller in the solar wind, the Ohmic electric field 
can be neglected even without coarse-graining. 

\newpage

\section{Mathematical Identities for Coarse-Graining Cumulants}\lb{cumulant} 

Suppose that $\{f_i | i=1,2,3\cdots\}$ are any set of space fields. 
Note that
\begin{eqnarray*}
\overline{(f_{i_1} \ldots f_{i_p})}_\ell(\bx) &=& \int d^d r \,\,G_\ell(\br) f_{i_1}(\bx + \br) \ldots f_{i_p}(\bx + \br) \cr
&=& \langle(\sigma f_{i_1}) \ldots (\sigma f_{i_p}) \rangle_\ell(\bx)
\end{eqnarray*}
\noindent where $G$ is any smooth, rapidly decaying, positive function with space integral unity, and   
\be(\sigma f_i)(\bx) = f_i(\bx+\br)\ee
is the {\it{shift operator}} and $\langle\cdot\rangle_\ell$ denotes average over the displacement vector $\br$ with density
$G_\ell(\br).$ We thus see that $\overline{f_{i_1} \ldots f_{i_p}}$ is a correlation function of the ``random variables'' $\sigma f_{i_1}$,...,$\sigma f_{i_p}$.  
(In this formula and the following we omit for simplicity of notations any explicit reference to the coarse-graining 
length $\ell.$) The cumulants of the variables $f_1(\bx + \br), \ldots,f_n(\bx + \br)$ 
for space-averaging with respect to the density $G_\ell(\br)$ on $\br$, denoted by
\be\tau(f_{i_1},\ldots,f_{i_p}) = \langle(\sigma f_{i_1}) \ldots (\sigma f_{i_p})\rangle^c, \ee
are defined as follows:
\begin{eqnarray*}
\bar{f}_1 &=& \tau(f_1) \cr
\overline{f_1 f_2} &=& \tau(f_1, f_2) + \bar{f}_1 \bar{f}_2 \cr
\overline{f_1 f_2 f_3} &=& \tau(f_1, f_2, f_3) + \bar{f}_1 \tau(f_2, f_3) + \bar{f}_2 \tau(f_1, f_3) + \bar{f}_3 \tau(f_1, f_2) + \bar{f}_1 \bar{f}_2 \bar{f}_3
\end{eqnarray*}
and, iteratively,
\be
\overline{f_1 \ldots f_n} =\sum_{I\in \mathcal{P}} \prod^p_{j=1} \tau(f_{i_1^{(j)}},\ldots,f_{i_{n_j}^{(j)}}) \ee
where the sum is  over the set $\mathcal{P}$ of all partitions $I=\{i^{(1)}_1,\ldots,i^{(1)}_{n_1}\}, \ldots, \{i^{(p)}_1,\ldots,i^{(p)}_{n_p}\}$ 
of the set $\{1,2,...,n\}$ with $\sum^p_{j=1} n_j = n$. We thus see that 
\be
\overline{f_1 \ldots f_n} = \tau(f_1,\ldots,f_n) + \mbox{terms defined by lower-order cumulant functions} \ee
so that one may solve successively to obtain 
\begin{eqnarray*}
\tau(f_1) &=& \bar{f}_1, \,\,\,\,\,\,\,\,\,\,\,\, \tau(f_1, f_2) =\overline{f_1 f_2} - \bar{f}_1 \bar{f}_2,  \cr
\tau(f_1, f_2, f_3) &=& 
\overline{f_1 f_2 f_3} - \bar{f}_1 \overline{f_2 f_3} - \bar{f}_2 \overline{f_1 f_3} - \bar{f}_3 \overline{f_1 f_2} + 2 \bar{f}_1 \bar{f}_2 \bar{f}_3, 
\,\,\,\,{\rm etc.} 
\end{eqnarray*}
These cumulants are called ``generalized central moments'' in fluid turbulence literature 
\citep{Germano92}  and ``connected correlation functions" in statistical physics and quantum field theory \citep{Huang87}. 

A very important fact is that the cumulants of the shift fields can be re-expressed as cumulants 
of the {\it difference fields}:
\be\delta_\br f_i(\bx) = \sigma_\br f_i(\bx) - f_i(\bx). \ee
The precise statement is as follows:
\begin{Prop} The connected correlation functions of $\delta f_i$ and $\sigma f_i$ are related for $p = 1$ by 
\be \tau(f_i) = \bar{f}_i =  f_i+\langle\delta f_i\rangle. \ee
and for $p >1$ are equal
\be \tau(f_{i_1},\cdots,f_{i_p})= \,\, \langle\delta f_{i_1} \ldots \delta f_{i_p}\rangle^c \ee
\end{Prop}
Concrete examples are for $p = 2$
\be\tau(f_i,f_j) = \langle\delta f_i \delta f_j\rangle - \langle\delta f_i\rangle \langle\delta f_j\rangle,\ee
and for $p =3$
\begin{eqnarray*}
\tau(f_i,f_j,f_k) \,\,&=&\,\,\langle\delta f_i \delta f_j \delta f_k\rangle - \langle\delta f_i \delta f_j\rangle \langle\delta f_k\rangle \cr
&& -\langle\delta f_i \delta f_k\rangle \langle\delta f_j\rangle - \langle\delta f_j \delta f_k\rangle \langle\delta f_i\rangle \
+2 \langle\delta f_i\rangle \langle\delta f_j\rangle \langle\delta f_k\rangle. 
\end{eqnarray*}
\cite{Constantinetal94} gave a concise proof of Onsager's theorem on dissipative anomaly for turbulent 
Euler solutions using the formula for $p=2$, which our proposition generalizes to all $p$-th order cumulants. 
The result follows from the {\it shift invariance} of the cumulants, that is, the fact that the cumulants
of the  ``random'' variables $\sigma_\br f_i(\bx)=f_i(\bx+\br)$ do not  change under a ``non-random'' 
(i.e. $\br$-independent) shift by $-f_i(\bx).$ Because this property plays a fundamental role in our calculations 
we give, for completeness, a standard proof using a method of generating functions. 

The generating function for correlation functions of shifted fields is
\be Z^\sigma(\balpha) \,= \,\langle\exp (\sum_{i \in I} \alpha_i \,\sigma f_i)\rangle. \ee
In fact, it is easy to check that
\be \overline{(f_{i_1} \ldots f_{i_p})} = \frac{\partial^p}{\partial \alpha_{i_1} \ldots \partial \alpha_{i_p}} Z^\sigma (\balpha) |_{\balpha = {\textbf 0}}\ee
The cumulants are generated by the logarithm of that function
$W^\sigma(\balpha) = \ln Z^\sigma(\balpha), $
i.e.
\be \tau(f_{i_1} \ldots f_{i_p}) = \frac{\partial^p}{\partial \alpha_{i_1} \ldots \partial \alpha_{i_p}} W^\sigma (\balpha) |_{\balpha = {\textbf 0}}. \ee
This is the so-called {\it linked cluster-theorem}. See \cite{Huang87}, section 10.1. 
The correlation functions of the increments $\langle\delta f_{i_1} \ldots \delta f_{i_p}\rangle$ are likewise generated by the function
\be Z^\delta(\balpha) \,= \,\langle\exp (\sum_{i \in I} \alpha_i \,\delta f_i)\rangle, \ee
and the connected correlation functions by the function
$W^\delta(\balpha) = \ln Z^\delta(\balpha), $
i.e.
\be \langle(\delta f_{i_1}) \ldots (\delta f_{i_p})\rangle^c = \frac{\partial^p}{\partial \alpha_{i_1} \ldots \partial \alpha_{i_p}} W^\delta (\balpha) |_{\balpha = {\textbf 0}},\ee
again by the linked-cluster theorem. Now comes the key observation: since $f_i(\bx)$ does not depend on $\br$, it can be taken outside 
the average $\langle.\rangle$. Thus, using $\delta f_i = \sigma f_i - f_i$, 
\begin{eqnarray*} 
Z^\delta(\balpha) \,&=& \,\langle\exp (\sum_{i \in I} \alpha_i \,\delta f_i)\rangle, 
= \langle\exp(\sum_{i \in I} \alpha_i \,\sigma f_i)\rangle \exp(-\sum_{i \in I} \alpha_i f_i) 
= Z^\sigma(\balpha) \exp(-\sum_{i \in I} \alpha_i f_i)
\end{eqnarray*}
Taking the logarithm of both sides then gives
\be W^\delta(\balpha) = W^\sigma(\balpha) - \sum_{i \in I} \alpha_i f_i \ee
The Proposition 1 then follows by multiple differentiations with respect to $\balpha.$

A further property of the cumulants implied by their shift-invariance is a simple transformation law under 
space-translations. We illustrate this property by the following expression for the space-translated 
2nd-order cumulant:   \newpage
\begin{eqnarray}
\tau(f_i,f_j)(\bx + {\textit {\textit {\textbf a}}}, t) &=& \int d^d r \,\,G_{{\textit {\textit {\textbf a}}},\ell}(\br) \delta f_i(\br,t) \delta f_j(\br,t) \cr
&& \,\,\,\,\,\,\,\,-\int d^d r \,\,G_{{\textit {\textit {\textbf a}}},\ell}(\br) \delta f_i(\br,t) \int d^d r' G_{{\textit {\textit {\textbf a}}},\ell}(r') \delta f_j(\br',t)
\label{ident}
\end{eqnarray}
\noindent with kernel $G_{{\textit {\textit {\textbf a}}},\ell}$ centered at point ${\textit {\textit {\textbf a}}}:$ 
\be G_{{\textit {\textit {\textbf a}}}, \ell}(\br) = \ell^{-d} G(\frac{\br-{\textit {\textit {\textbf a}}}}{\ell}). \ee
The general fact is that such expressions for translated cumulants are identical to the corresponding 
expressions for untranslated cumulants in terms of increments, except that the average with respect 
to $G_\ell$ is replaced by average with respect to $G_{{\textit {\textit {\textbf a}}},\ell}.$ To establish this fact, first note 
\begin{eqnarray*}
\tau(f_i,f_j)(\bx + {\textit {\textit {\textbf a}}}) = \int d^d r \,\,G_\ell(\br)[ f_i(\bx + {\textit {\textit {\textbf a}}} + \br)- f_i(\bx + {\textit {\textit {\textbf a}}})][f_j(\bx+{\textit {\textit {\textbf a}}} + \br) - f_j(\bx + {\textit {\textit {\textbf a}}})] \cr
- \int d^d r \,\,G_\ell(\br)[ f_i(\bx + {\textit {\textit {\textbf a}}}+ \br) - f_i(\bx + {\textit {\textit {\textbf a}}}) ] \int d^d r' \,\,G_\ell(r')[f_j(\bx + {\textit {\textit {\textbf a}}} + \br') - f_j(\bx + {\textit {\textit {\textbf a}}})]
\end{eqnarray*}
\noindent Making the change of variables $\br + {\textit {\textit {\textbf a}}} \rightarrow \br,\,\, \br' + {\textit {\textit {\textbf a}}} \rightarrow \br',$ gives
\begin{eqnarray*}
\tau(f_i,f_j)(\bx + {\textit {\textit {\textbf a}}}) = \int d^d r \,\,G_{{\textit {\textit {\textbf a}}},\ell}(\br) [\delta f_i(\br;\bx) - \delta f_i({\textit {\textit {\textbf a}}};\bx)] [\delta f_j(\br;\bx) - \delta f_j({\textit {\textit {\textbf a}}}; \bx)] \cr
- \int d^d r \,\,G_{{\textit {\textit {\textbf a}}},\ell}(\br)[ \delta f_i(\br;\bx)-\delta f_i({\textit {\textit {\textbf a}}};\bx)] \int d^d r' \,\,G_{{\textit {\textit {\textbf a}}},\ell}(r') [\delta f_j(\br';\bx) - \delta f_j({\textit {\textit {\textbf a}}}; \bx)],
\end{eqnarray*}
where use was made of the fact that 
\be
\bu(\bx + \br) - \bu(\bx + {\textit {\textit {\textbf a}}}) = [\bu(\bx + \br) - \bu(\bx)] -[\bu(\bx + {\textit {\textit {\textbf a}}}) - \bu(\bx)] 
= \delta \bu(\br;\bx) - \delta \bu({\textit {\textit {\textbf a}}}; \bx).
\ee
\noindent However, $\delta \bu({\textit {\textit {\textbf a}}}; \bx)$ does not depend upon $\br$ and is thus a ``constant'' with respect to the average over $\br$ with 
density $G_{{\textit {\textit {\textbf a}}},\ell}(\br)$. Since cumulants are invariant to shifts of the random variables by constants, this yields the formula 
(\ref{ident}). The same argument obviously works also for all $p>2.$

An important consequence of (\ref{ident}) is that all space-derivatives of the cumulants with respect to 
$\bx$ can be shifted to space-derivatives of the filter kernels $G_\ell(\br)$with respect to $\br.$
For example,
\begin{eqnarray*}
\partial_k \tau(f_i,f_j) &=& -\frac{1}{\ell} \left\{ \int d^d r (\partial_k G)_\ell(\br) \delta f_i(\br) \delta f_j(\br)\right. \cr
 && \,\,\,\,\,\, -\int d^d r (\partial_k G)_\ell(\br) \delta f_i(\br) \int d^d r' G_\ell(r') \delta f_j(\br') \cr
 && \,\,\,\,\,\, -\left. \int d^d r G_\ell(\br) \delta f_i(\br) \int d^d r' (\partial_k G)_\ell(\br') \delta f_j(\br') \right\}.
\end{eqnarray*}
follows by differentiating both sides of (\ref{ident}) with respect to ${\textit {\textit {\textbf a}}}$ and then setting ${\textit {\textit {\textbf a}}}=\bzed.$
The same argument taking two derivatives with respect to ${\textit {\textit {\textbf a}}}$ yields formula (\ref{twice}).  




\bibliographystyle{apj}
\bibliography{TGGMR}

\providecommand{\noopsort}[1]{}\providecommand{\singleletter}[1]{#1}%
\begin{thebibliography}{}
\expandafter\ifx\csname natexlab\endcsname\relax\def\natexlab#1{#1}\fi

\bibitem[{Albright(1999)}]{Albright99}
Albright, B.~J. 1999, Phys. Plasmas, 6, 4222

\bibitem[{{Aluie}(2013)}]{Aluie13}
{Aluie}, H. 2013, Physica D, 247, 54

\bibitem[{{Aluie} \& {Eyink}(2010)}]{AluieEyink10}
{Aluie}, H., \& {Eyink}, G.~L. 2010, Phys. Rev. Lett., 104, 081101

\bibitem[{Axford(1984)}]{Axford84}
Axford, W.~I. 1984, in Geophys. Monogr. Ser., Vol.~30, Magnetic Reconnection in
  Space and Laboratory Plasmas, ed. J.~E.~W.~Hones (Washington, D.C.: AGU),
  1--8

\bibitem[{{Bellamy} {et~al.}(2005){Bellamy}, {Cairns}, \&
  {Smith}}]{Bellamyetal05}
{Bellamy}, B.~R., {Cairns}, I.~H., \& {Smith}, C.~W. 2005, J. Geophys. Res.
  (Space Phys.), 110, 10104

\bibitem[{{Bhattacharjee} {et~al.}(1999){Bhattacharjee}, {Ma}, \&
  {Wang}}]{Bhattacharjeeetal99}
{Bhattacharjee}, A., {Ma}, Z.~W., \& {Wang}, X. 1999, J. Geophys. Res., 104,
  14543

\bibitem[{{Bian} \& {Kontar}(2010)}]{BianKontar10}
{Bian}, N.~H., \& {Kontar}, E.~P. 2010, Phys. Plasmas, 17, 062308

\bibitem[{{Bian} {et~al.}(2010){Bian}, {Kontar}, \& {Brown}}]{Bianetal10}
{Bian}, N.~H., {Kontar}, E.~P., \& {Brown}, J.~C. 2010, Astron. \& Astrophys.,
  519, A114

\bibitem[{{Biskamp}(1986)}]{Biskamp86}
{Biskamp}, D. 1986, Phys. Fluids, 29, 1520

\bibitem[{Biskamp(2003)}]{Biskamp03}
Biskamp, D. 2003, Magnetohydrodynamic Turbulence (Cambridge University Press)

\bibitem[{Boldyrev(2005)}]{Boldyrev05}
Boldyrev, S. 2005, Astrophys. J., 626, L37

\bibitem[{Boldyrev(2006)}]{Boldyrev06}
---. 2006, Phys. Rev. Lett., 96, 115002

\bibitem[{{Bruno} \& {Carbone}(2013)}]{BrunoCarbone05}
{Bruno}, R., \& {Carbone}, V. 2013, Living Reviews in Solar Physics, 10

\bibitem[{{Burlaga} {et~al.}(1982){Burlaga}, {Lepping}, {Behannon}, {Klein}, \&
  {Neubauer}}]{Burlagaetal82}
{Burlaga}, L.~F., {Lepping}, R.~P., {Behannon}, K.~W., {Klein}, L.~W., \&
  {Neubauer}, F.~M. 1982, J. Geophys. Res., 87, 4345

\bibitem[{{Burlaga} \& {Ness}(2010)}]{BurlagaNess10}
{Burlaga}, L.~F., \& {Ness}, N.~F. 2010, Astrophys. J., 725, 1306

\bibitem[{{Burlaga} {et~al.}(2002){Burlaga}, {Ness}, {Wang}, \&
  {Sheeley}}]{Burlagaetal02}
{Burlaga}, L.~F., {Ness}, N.~F., {Wang}, Y.-M., \& {Sheeley}, N.~R. 2002, J.
  Geophys. Res. (Space Phys.), 107, 1410

\bibitem[{Caflisch {et~al.}(1997)Caflisch, Klapper, \& Steele}]{Caflischetal97}
Caflisch, R.~E., Klapper, I., \& Steele, G. 1997, Commun. Math. Phys., 184, 443

\bibitem[{{Chen} {et~al.}(2011){Chen}, {Bale}, {Salem}, \&
  {Mozer}}]{Chenetal11}
{Chen}, C.~H.~K., {Bale}, S.~D., {Salem}, C., \& {Mozer}, F.~S. 2011, Ap. J.
  Lett., 737, L41

\bibitem[{Collins(1984)}]{Collins84}
Collins, J. 1984, Renormalization: An Introduction to Renormalization, the
  Renormalization Group and the Operator-Product Expansion (Cambridge
  University Press)

\bibitem[{{Constantin} {et~al.}(1994){Constantin}, {E}, \&
  {Titi}}]{Constantinetal94}
{Constantin}, P., {E}, W., \& {Titi}, E.~S. 1994, Commun. Math. Phys., 165, 207

\bibitem[{{Craig} \& {Watson}(2003)}]{CraigWatson03}
{Craig}, I.~J.~D., \& {Watson}, P.~G. 2003, Sol. Phys., 214, 131

\bibitem[{{De Lellis} \& {Sz\'{e}kelyhidi Jr.}(2012)}]{DeLellisSzekelyhidiJr12}
{De Lellis}, C., \& {Sz\'{e}kelyhidi Jr.}, L. 2012, in 6th European Congress of
  Mathematics: 2-7 July Krak{\'o}w, ed. K.~Ciesielski, T.~Nadzieja, \&
  K.~Pawa{\l}owski, Wiadomo{\'s}ci Matematyczne (Polskie Towarzystwo
  Matematyczne), 13--29

\bibitem[{{Dmitruk} \& {Matthaeus}(2006)}]{DmitrukMatthaeus06}
{Dmitruk}, P., \& {Matthaeus}, W.~H. 2006, Phys. Plasmas, 13, 042307

\bibitem[{Evans(2010)}]{Evans10}
Evans, L. 2010, Partial Differential Equations, Graduate studies in mathematics
  (American Mathematical Society)

\bibitem[{Eyink {et~al.}(2013)Eyink, Vishniac, Lalescu, Aluie, Kanov, B\"urger,
  Burns, Meneveau, \& Szalay}]{Eyinketal13}
Eyink, G., Vishniac, E., Lalescu, C., {et~al.} 2013, Nature, 497, 466

\bibitem[{{Eyink}(2005)}]{Eyink05}
{Eyink}, G.~L. 2005, Physica D, 207, 91

\bibitem[{{Eyink}(2007)}]{Eyink-Notes}
---. 2007, {Turbulence Theory, Course Notes at Johns Hopkins University,
  Chapter II, section b}, {http://www.ams.jhu.edu/~eyink/Turbulence/notes.html}

\bibitem[{Eyink(2007)}]{Eyink07}
Eyink, G.~L. 2007, Phys. Lett. A, 368, 486

\bibitem[{{Eyink}(2008)}]{Eyink08}
{Eyink}, G.~L. 2008, Physica D, 237, 1956

\bibitem[{Eyink(2011)}]{Eyink11}
Eyink, G.~L. 2011, Phys. Rev. E, 83, 056405

\bibitem[{Eyink \& Aluie(2006)}]{EyinkAluie06}
Eyink, G.~L., \& Aluie, H. 2006, Physica D, 223, 82

\bibitem[{{Eyink} \& {Drivas}(2014)}]{EyinkDrivas14}
{Eyink}, G.~L., \& {Drivas}, T.~D. 2014, {J. Stat. Phys.}, {to appear},
  doi:10.1007/s10955-014-1135-3

\bibitem[{Eyink {et~al.}(2011)Eyink, Lazarian, \& Vishniac}]{Eyinketal11}
Eyink, G.~L., Lazarian, A.~L., \& Vishniac, E.~T. 2011, Astrophys. J., 743, 51

\bibitem[{{Eyink} \& {Sreenivasan}(2006)}]{EyinkSreenivasan06}
{Eyink}, G.~L., \& {Sreenivasan}, K.~R. 2006, Rev. Mod. Phys., 78, 87

\bibitem[{Favre(1969)}]{Favre69}
Favre, A. 1969, in Problems of Hydrodynamics and Continuum Mechanics:
  Contributions in Honor of the Sixtieth Birthday of Academician L. I. Sedov,
  14th Nov. 1967 ; English Edition, ed. L.~Sedov (SIAM), 231--266

\bibitem[{{Germano}(1992)}]{Germano92}
{Germano}, M. 1992, J. Fluid Mech., 238, 325

\bibitem[{Goldenfeld(1992)}]{Goldenfeld92}
Goldenfeld, N. 1992, Lectures on Phase Transitions and the Renormalization
  Group, Frontiers in Physics (Addison-Wesley, Advanced Book Program)

\bibitem[{Goldreich \& Sridhar(1995)}]{GoldreichSridhar95}
Goldreich, P., \& Sridhar, S. 1995, Astrophys. J., 438, 763

\bibitem[{Goldreich \& Sridhar(1997)}]{GoldreichSridhar97}
---. 1997, Astrophys. J., 485, 680

\bibitem[{{Gosling}(2007)}]{Gosling07}
{Gosling}, J.~T. 2007, Astrophys. J.:ett., 671, L73

\bibitem[{{Gosling}(2012)}]{Gosling12}
---. 2012, Space. Sci. Rev., 172, 187

\bibitem[{{Gosling} {et~al.}(2007){Gosling}, {Phan}, {Lin}, \&
  {Szabo}}]{Goslingetal07}
{Gosling}, J.~T., {Phan}, T.~D., {Lin}, R.~P., \& {Szabo}, A. 2007, Geophys.
  Res. Lett., 34, 15110

\bibitem[{{Gosling} \& {Szabo}(2008)}]{GoslingSzabo08}
{Gosling}, J.~T., \& {Szabo}, A. 2008, J. Geophys. Res. (Space Phys.), 113,
  10103

\bibitem[{{Grasso} {et~al.}(2000){Grasso}, {Califano}, \&
  {Pegoraro}}]{Grassoetal00}
{Grasso}, D., {Califano}, F., \& {Pegoraro}, F. 2000, Plasma Phys. Rep., 26,
  512

\bibitem[{Greene(1988)}]{Greene88}
Greene, J.~M. 1988, J. Geophys. Res., 93, 8583

\bibitem[{Hesse \& Schindler(1988)}]{HesseSchindler88}
Hesse, M., \& Schindler, K. 1988, J. Geophys. Res., 93, 5559

\bibitem[{{Hnat} {et~al.}(2011){Hnat}, {Chapman}, {Gogoberidze}, \&
  {Wicks}}]{Hnatetal11}
{Hnat}, B., {Chapman}, S.~C., {Gogoberidze}, G., \& {Wicks}, R.~T. 2011, Phys.
  Rev. E, 84, 065401

\bibitem[{{Howes} {et~al.}(2012){Howes}, {Bale}, {Klein}, {Chen}, {Salem}, \&
  {TenBarge}}]{Howesetal12}
{Howes}, G.~G., {Bale}, S.~D., {Klein}, K.~G., {et~al.} 2012, Astrophys. J.
  Lett., 753, L19

\bibitem[{Huang(1987)}]{Huang87}
Huang, K. 1987, Statistical Mechanics, 2nd edn. (Wiley)

\bibitem[{Iroshnikov(1964)}]{Iroshnikov64}
Iroshnikov, R.~S. 1964, Soviet Astron., 7, 566

\bibitem[{{Khabarova} \& {Obridko}(2012)}]{KhabarovaObridko12}
{Khabarova}, O., \& {Obridko}, V. 2012, Ap. J., 761, 82

\bibitem[{{Kleva} {et~al.}(1995){Kleva}, {Drake}, \&
  {Waelbroeck}}]{Klevaetal95}
{Kleva}, R.~G., {Drake}, J.~F., \& {Waelbroeck}, F.~L. 1995, Physics of
  Plasmas, 2, 23

\bibitem[{Kraichnan(1965)}]{Kraichnan65}
Kraichnan, R.~H. 1965, Phys. Fluids, 8, 1385

\bibitem[{Lau \& Finn(1990)}]{LauFinn90}
Lau, Y.~T., \& Finn, J.~M. 1990, Astrophys. J., 350, 672

\bibitem[{Lau \& Finn(1992)}]{LauFinn92}
---. 1992, Physica D, 57, 283

\bibitem[{Lazarian(2005)}]{Lazarian05}
Lazarian, A. 2005, EM de Gouveia Dal Pino, G. Lugones, \& A. Lazarian
  (Melville, NY: AIP), 42

\bibitem[{{Lazarian} \& {Opher}(2009)}]{LazarianOpher09}
{Lazarian}, A., \& {Opher}, M. 2009, Astrophys. J., 703, 8

\bibitem[{Lazarian \& Vishniac(1999)}]{LazarianVishniac99}
Lazarian, A., \& Vishniac, E. 1999, Astrophys. J., 517, 700

\bibitem[{LeVeque {et~al.}(1998)LeVeque, Steiner, \& Gautschy}]{Levequeetal98}
LeVeque, R., Steiner, O., \& Gautschy, A. 1998, Computational Methods for
  Astrophysical Fluid Flow: Saas-Fee Advanced Course 27. Lecture Notes 1997
  Swiss Society for Astrophysics and Astronomy, Lecture notes / Saas-Fee
  Advanced Course (Springer)

\bibitem[{{Loureiro} \& {Hammett}(2008)}]{LoureiroHammett08}
{Loureiro}, N.~F., \& {Hammett}, G.~W. 2008, J. Comp. Phys., 227, 4518

\bibitem[{{Matthaeus} \& {Goldstein}(1986)}]{MatthaeusGoldstein86}
{Matthaeus}, W.~H., \& {Goldstein}, M.~L. 1986, Phys. Rev. Lett., 57, 495

\bibitem[{{Matthaeus} \& {Lamkin}(1986)}]{MatthaeusLamkin86}
{Matthaeus}, W.~H., \& {Lamkin}, S.~L. 1986, Phys. Fluids, 29, 2513

\bibitem[{Mininni \& Pouquet(2009)}]{MininniPouquet09}
Mininni, P.~D., \& Pouquet, A. 2009, Phys. Rev. E, 80, 025401

\bibitem[{Newcomb(1958)}]{Newcomb58}
Newcomb, W.~A. 1958, Ann. Phys., 3, 347

\bibitem[{{Ohia} {et~al.}(2012){Ohia}, {Egedal}, {Lukin}, {Daughton}, \&
  {Le}}]{Ohiaetal12}
{Ohia}, O., {Egedal}, J., {Lukin}, V.~S., {Daughton}, W., \& {Le}, A. 2012,
  Phys. Rev. Lett., 109, 115004

\bibitem[{{Onsager}(1949)}]{Onsager49}
{Onsager}, L. 1949, Nuovo. Cim. Suppl., 6, 279

\bibitem[{{Parker}(1958)}]{Parker58}
{Parker}, E.~N. 1958, Astrophys. J., 128, 664

\bibitem[{{Parker}(1963)}]{Parker63}
---. 1963, Interplanetary Dynamical Processes (New York: Wiley Inter-Science)

\bibitem[{{Petschek}(1964)}]{Petschek64}
{Petschek}, H.~E. 1964, in AAS-NASA Symposium on the Physics of Solar Flares,
  NASA SP50, ed. W.~N. {Hess} (Washington, D.C.: NASA), 425

\bibitem[{{Phan} {et~al.}(2009){Phan}, {Gosling}, \& {Davis}}]{Phanetal09}
{Phan}, T.~D., {Gosling}, J.~T., \& {Davis}, M.~S. 2009, Geophys. Res. Lett.,
  36, 9108

\bibitem[{{Podesta} {et~al.}(2009){Podesta}, {Chandran}, {Bhattacharjee},
  {Roberts}, \& {Goldstein}}]{Podestaetal09}
{Podesta}, J.~J., {Chandran}, B.~D.~G., {Bhattacharjee}, A., {Roberts}, D.~A.,
  \& {Goldstein}, M.~L. 2009, J. Geophys. Res., 114, 1107

\bibitem[{{Podesta} {et~al.}(2007){Podesta}, {Roberts}, \&
  {Goldstein}}]{Podestaetal07}
{Podesta}, J.~J., {Roberts}, D.~A., \& {Goldstein}, M.~L. 2007, Astrophys. J.,
  664, 543

\bibitem[{Priest \& D\'emoulin(1995)}]{PriestDemoulin95}
Priest, E.~R., \& D\'emoulin, P. 1995, J. Geophys. Res., 100, 2344

\bibitem[{Priest \& Forbes(1992)}]{PriestForbes92}
Priest, E.~R., \& Forbes, T.~G. 1992, J. Geophys. Res., 97, 1521

\bibitem[{Priest {et~al.}(2003)Priest, Hornig, \& Pontin}]{Priestetal03}
Priest, E.~R., Hornig, G., \& Pontin, D.~I. 2003, J. Geophys. Res., 108, 1285

\bibitem[{{Richardson} {et~al.}(2013){Richardson}, {Burlaga}, {Decker},
  {Drake}, {Ness}, \& {Opher}}]{Richardsonetal13}
{Richardson}, J.~D., {Burlaga}, L.~F., {Decker}, R.~B., {et~al.} 2013, Ap. J.
  Lett., 762, L14

\bibitem[{{Roberts}(2010)}]{Roberts10}
{Roberts}, D.~A. 2010, J. Geophys. Res. (Space Phys.), 115, 12101

\bibitem[{{Roberts} {et~al.}(2005){Roberts}, {Keiter}, \&
  {Goldstein}}]{Robertsetal05}
{Roberts}, D.~A., {Keiter}, P.~A., \& {Goldstein}, M.~L. 2005, J. Geophys. Res.
  (Space Phys.), 110, 6102

\bibitem[{{Sahraoui} {et~al.}(2013){Sahraoui}, {Huang}, {Belmont}, {Goldstein},
  {R{\'e}tino}, {Robert}, \& {De Patoul}}]{Sahraouietal13}
{Sahraoui}, F., {Huang}, S.~Y., {Belmont}, G., {et~al.} 2013, Ap. J., 777, 15

\bibitem[{{Santos-Lima} {et~al.}(2010){Santos-Lima}, {Lazarian}, {de Gouveia
  Dal Pino}, \& {Cho}}]{Santos-Limaetal10}
{Santos-Lima}, R., {Lazarian}, A., {de Gouveia Dal Pino}, E.~M., \& {Cho}, J.
  2010, Astrophys. J., 714, 442

\bibitem[{{Schekochihin} {et~al.}(2009){Schekochihin}, {Cowley}, {Dorland},
  {Hammett}, {Howes}, {Quataert}, \& {Tatsuno}}]{Schekochihinetal09}
{Schekochihin}, A.~A., {Cowley}, S.~C., {Dorland}, W., {et~al.} 2009, Ap. J.
  Suppl., 182, 310

\bibitem[{Schindler {et~al.}(1988)Schindler, Hesse, \& Birn}]{Schindleretal88}
Schindler, K., Hesse, M., \& Birn, J. 1988, J. Geophys. Res., 93, 5547

\bibitem[{{Shay} {et~al.}(1998){Shay}, {Drake}, {Denton}, \&
  {Biskamp}}]{Shayetal98}
{Shay}, M.~A., {Drake}, J.~F., {Denton}, R.~E., \& {Biskamp}, D. 1998, J.
  Geophys. Res., 103, 9165

\bibitem[{{Stolovitzky} {et~al.}(1998){Stolovitzky}, {Meneveau}, \&
  {Sreenivasan}}]{Stolovitskyetal98}
{Stolovitzky}, G., {Meneveau}, C., \& {Sreenivasan}, K.~R. 1998, Phys. Rev.
  Lett., 80, 3883

\bibitem[{{Tadmor}(2004)}]{Tadmor04}
{Tadmor}, E. 2004, {Comm. Math. Sci.}, 2, 317

\bibitem[{{Uzdensky} \& {Kulsrud}(2000)}]{UzdenskyKulsrud00}
{Uzdensky}, D.~A., \& {Kulsrud}, R.~M. 2000, Phys. Plasmas, 7, 4018

\bibitem[{{Vasyliunas}(1975)}]{Vasyliunas75}
{Vasyliunas}, V.~M. 1975, Rev. Geophys. Space Phys., 13, 303

\bibitem[{{Wicks} {et~al.}(2013{\natexlab{a}}){Wicks}, {Mallet}, {Horbury},
  {Chen}, {Schekochihin}, \& {Mitchell}}]{Wicksetal13a}
{Wicks}, R.~T., {Mallet}, A., {Horbury}, T.~S., {et~al.} 2013{\natexlab{a}},
  Phys. Rev. Lett., 110, 025003

\bibitem[{{Wicks} {et~al.}(2013{\natexlab{b}}){Wicks}, {Roberts}, {Mallet},
  {Schekochihin}, {Horbury}, \& {Chen}}]{Wicksetal13b}
{Wicks}, R.~T., {Roberts}, D.~A., {Mallet}, A., {et~al.} 2013{\natexlab{b}},
  Ap. J., 778, 177

\end{thebibliography}

%% The following command ends your manuscript. LaTeX will ignore any text

\end{document}